\documentclass[10pt,journal,compsoc]{IEEEtran}

\pdfoutput=1

\ifCLASSOPTIONcompsoc
  
  \usepackage[nocompress]{cite}
\else
  \usepackage{cite}
\fi

\usepackage{amsmath,amsfonts}
\usepackage{csquotes}
\usepackage{algorithmic}
\usepackage{array}
\usepackage{xcolor}
\usepackage{subcaption}
\usepackage[utf8]{inputenc}
\usepackage[misc]{ifsym}
\usepackage{pifont}
\newcommand{\xmark}{\ding{55}}
\newcommand{\cmark}{\ding{51}}
\usepackage{textcomp}
\usepackage{stfloats}
\usepackage{url}
\usepackage{multirow}
\usepackage{verbatim}
\usepackage{graphics}
\usepackage{graphicx}
\usepackage{booktabs}
\usepackage{multicol}
\usepackage{caption} 
\usepackage{lipsum}
\ifCLASSINFOpdf
\else
\fi

\begin{document}

\title{rCamInspector: Building Reliability and Trust on IoT (Spy) Camera Detection using XAI}

\author{Priyanka Rushikesh Chaudhary,Manan Gupta, Jabez Christopher, Putrevu Venkata Sai Charan, Rajib Ranjan Maiti}

\IEEEtitleabstractindextext{%
\begin{abstract}
  The classification of network traffic using machine learning (ML) models is one of the primary mechanisms to address the security issues in IoT networks and/or IoT devices. 
However, the ML models often act as black-boxes that create a roadblock to take critical decision based on the model output. 
To address this problem, we design and develop a system, called \emph{rCamInspector}, that employs Explainable AI (XAI) to provide reliable and trustworthy explanations to model output. 
\emph{rCamInspector} adopts two classifiers, \emph{Flow Classifier} - categorizes a flow into one of four classes, \emph{IoTCam}, \emph{Conf}, \emph{Share} and \emph{Others}, and \emph{SmartCam Classifier} - classifies an IoTCam flow into one of six classes, Netatmo, Spy Clock, Canary, D3D, Ezviz, V380 Spy Bulb; both are IP address and transport port agnostic. 
\emph{rCamInspector} is evaluated using 38GB of network traffic and our results show that XGB achieves the highest accuracy of 92\% and 99\% in the Flow and  SmartCam classifiers respectively among eight supervised ML models.
We analytically show that the traditional mutual information (MI) based feature importance cannot provide enough reliability on the model output of XGB in either classifiers. 
Using SHAP and LIME, we show that a separate set of features can be picked up to explain a correct prediction of XGB. 
For example, the feature Init Bwd Win Byts turns out to have the highest SHAP values to support the correct prediction of both IoTCam in Flow Classifier and Netatmo class in SmartCam Classifier. 
To evaluate the faithfulness of the explainers on our dataset, we show that both SHAP and LIME have a consistency of more than 0.7 and a sufficiency of 1.0. 
Comparing with existing works, we show that \emph{rCamInspector} achieves a better accuracy (99\%), precision (99\%), and false negative rate (0.7\%).

\end{abstract}

\begin{IEEEkeywords}
Internet of Things (IoT), IoT Camera Detection, XAI SHAP and LIME, Classification of network flows using Flow-based Features,  Reliability and Trust
\end{IEEEkeywords}}

\maketitle

\IEEEdisplaynontitleabstractindextext

\IEEEpeerreviewmaketitle

\IEEEraisesectionheading{\section{Introduction}\label{sec:introduction}}

Rapid proliferation of Internet of Things (IoT) devices has raised significant privacy and security concerns, particularly with the increasing use of IoT cameras. While these cameras serve legitimate purposes such as surveillance and home automation, they can also be exploited as spy cameras, posing serious privacy threats, where face images, behaviors, gestures, personally identifiable information (PII), phone numbers, home addresses, or home assets can be directly exposed to attackers \cite{alharbi2018IET, Abdalla2020ISDFS}. 
Instead of solely addressing patching the firmware, security researchers have increasingly focused on identifying IoT devices through the monitoring of massive network traffic at strategic network points, like routers and/or gateways, to improve the security of an IoT network in recent times \cite{wang2020iot, saidi2020imc}. 
Therefore, machine learning (ML) based techniques have proven to be an effective way of classifying and identifying network flows and/or the IoT devices. 
For example, Naive Bayes or Random Forests models can analyze large volume of network traces(about 2GB) to distinguish legitimate and potentially malicious IoT devices \cite{sivanathan2017IEEEInfocom}, \cite{Arunan2018IEEE}, \cite{Salman2022Wiley}. 
While they can achieve a higher accuracy to detect and classify IoT devices, the models are often used as \enquote{black boxes} making it difficult to interpret the model outputs. 
In other words, the ML models lack transparency that raises concerns among the users of the corresponding systems creating a major roadblock in wider adoption of the systems. 
In this paper, we aim to address the problem of building reliability and trust on the outputs of the models that classifies and identifies the IoT (spy) cameras by using explainable AI (XAI) models.

The traditional mechanisms for building reliability and trust on the models output include the analysis of feature importance that provide a kind of global explanation of the overall performance of the model. 
The top features for any of the trained models can be identified relatively easily, but there is no explanation of their impact within the decision-making process of the model.
Recently, XAI models, like SHAP (Shapley Additive Explanations)\cite{scott2017NeurIPS} and LIME (Local Interpretable Model-Agnostic Explanations)\cite{dieber2020LIME}, we call these as explainers, have gain popularity in the domains like medical image analysis \cite{venkatesh2023IEEE}. 
Unlike traditional feature importance, XAI explainers can provide both global and local explanations for a specific output of the model. 
In general, such an explainer would take as input a trained model and a dataset with independent variables and produce the model output together with the feature-wise positive or negative supports causing the output. 
By using explainers, such as SHAP and LIME, one can visualize and comprehend the contribution of each feature to specific output, thus gaining a clearer understanding of feature significance and interactions. Explainability not only improves the model transparency but can also reveal the patterns in misclassification, helping to fine-tune the models and choose relevant features, which could address issues such as the high false-negative rate seen in some classifiers \cite{chinu2024Springer}. 
%
Like the other domains, security analysts and network administrators can gain confidence on ML-based defense systems by understanding, e.g., why a certain IoT device is flagged as a spy camera, and hence improve trust and reliability. 
Further, XAI-based security solutions that comply with privacy and cybersecurity regulations (i.e., Regulatory Compliance \cite{ENISA_AIStandard}, \cite{NIST2021_AIStandard}) are in high demand as they can provide interpretability to the prediction of IoT security incidents and hence builds up reliability and trust on the system.



In this paper, we have designed and implemented a system called \emph{rCamInspector} that addresses the problem of building reliability and trust in the classification of the network flows by using SHAP and LIME explainers. 
We present an in-depth analysis of both the traditional feature importance in search of an explanation of the model performance and the application of XAI explainers in two types of classifications performed by \emph{rCamInspector}, we denote them as \textbf{Flow classifier} - classifies the network flows into one of four broad categories, namely Conf, Share, IoTCam, and Others, and \textbf{SmartCam classifier} - classifies an IoTcam flow into one of six classes, namely Netatmo, Spy Clock, Canary, D3D, Ezviz, V380 Spy Bulb.  
We have used about 38GB of network trace that is first divided into four broad classes to be used by \textit{Flow Classifier}: \emph{Conf} - indicating the class of flows belonging to conferencing applications, like Skype, Meet, Zoom, and Teams, \emph{Share} - indicating the class of flows belonging to video sharing applications, like Youtube and Prime, \emph{IoTCam} - indicating the class of flows belonging to IoT (spy) cameras, and \emph{Others} - indicating the class of flows belonging to any application other than those involving video traffic. 
Then, the flows only in \emph{IoTCam} class are further divided into six classes, namely Netatmo, Spy Clock, Canary, D3D, Ezviz, V380 Spy Bulb, based on the IoT cameras that generates the flows to be used by \textit{SmartCam Classifier}.
Thus, once a flow is classified as IoTCam by \textit{Flow Classifier}, \emph{SmartCam Classifier} further classifies the flow to detect a particular camera out of 6 IoT cameras.
For example, a flow in \emph{IoTCam} category is detected as a flow of \emph{Ezviz} IoT camera, and hence the presence of the Ezviz IoT camera by \emph{rCamInspector}.
In this work, we have created our own network setup consisting of six commercially available IoT (spy) cameras and traditional computing devices like Desktop and Laptop, to collect the network traces (basically pcap files).  
The raw pcap files are then used to extract 77 flow-based features by considering a popular flow-based features extractor tool, called CICFlowmeter \cite{cicflowmeter}. 
\textit{rCamInspector} neither attempts to decrypt any network packet nor uses IP addresses or transport port numbers as features. 
In either of \emph{Flow Classifier} and \emph{SmartCam Classifier}, We have considered a total of eight supervised ML models, namely i.e. Decision Tree Classification (DT), k-Nearest Neighbour (kNN), Naive Bayes (NB), Linear Regression (LR), Random Forest (RF), Extreme Gradient Boosting (XGB), Extra Tree (ET) and AdaBoost (AB), for an investigation that forms a baseline for traditional feature importance to find most important features describing the global performance of each of the models. 
Next, we have consider the best-performing model and discover the most important features using  SHAP and LIME explainers. 
Finally, we have conducted a comparison study among the most important features using traditional feature importance analysis and by using the explainers. 
rCamInspector presents uniquely tailored SHAP and LIME explainers that give more interpretability for feature analysis specially designed for the complex task of IoT (spy) camera detection. 
We compared our results against notable state-of-the-art works on IoT device detection and/or classification. 
Our key contributions in this article are:
\begin{enumerate}
    \item \textbf{Development of rCamInspector System:}
     \begin{itemize}
       \item We design and develop \emph{rCamInspector}, a novel system capable of classifying network flows and and detect IoT (Spy) cameras using two classifiers, first, \textbf{Flow classifier:} classifies a flow into one of four classes of \emph{IoTCam}, \emph{Share}, \emph{Conf}, \emph{Others}, and second, \textbf{SmartCam classifier:} classifies a flow in \emph{IoTcam} into one of six classes of \emph{D3D}, \emph{Netatmo}, \emph{Spy Clock}, \emph{Canary}, \emph{Ezviz}, \emph{V380 Spy Bulb}. 
       Both the classifiers use only flow-based features to make the system IP address and transport port agnostic. 
       Either classifier uses eight distinct supervised machine learning models, namely, DT, kNN, NB, LR, RF, XGB, ET, and AB. Our results show that XGB achieves more than 92\% and 99\% accuracies in Flow and SmartCam Classifier respectively.  
       
       \item Our results show that features having higher \emph{mutual information} (MI) may not provide a good global explanation for the model output.  
       For example, two most common important features across eight models in SmartCam Classifier are \textit{Init Bwd Win Byts} and \textit{Pkt Len Min}, but do not have highest mutual information (MI values are 0.42 and 0.15 respectively). 
    \end{itemize}
    
    \item \textbf{Integration of XAI into rCamInspector for Reliability and Trust:}
    \begin{itemize}
       \item To enhance the reliability and trust on the output of rCamInspector, we employ two explainers, \textbf{SHAP} and \textbf{LIME}, each takes trained \textbf{XGB} as input, in both Flow and SmartCam classifier, and discovers the features that provides both global and local explanations for the model output. 
       For example, \textit{Init Bwd Win Byts} has the highest positive SHAP value to classify a flow as \emph{IoTCam} and  \emph{Netatmo} both. 
       
       \item Comparing the most important features in SHAP and LIME to explain any of the classes, our results reveal that there is a significant overlap in the most important features. 
       For instance, the feature \textit{Bwd Header Len} has a higher SHAP value and a higher LIME value to support the prediction of D3D camera. 
       Thus, rCamInspector provides reliable and trustworthy outputs at both Flow and SmartCam Classifier for a network administrator who can take actionable steps to protect IoT network. 
       
    \end{itemize}

\end{enumerate}

In the rest of the paper, 
Section \ref{sec:relwork} describes related works, 
Section \ref{sec:sysem-architecture} provides the system, threat model along with implementation of explainer system \textit{rCamInspector},
Section \ref{sec:dataset} states experimental setup, datasets, performance metrics, and feature importance metrics used in this paper.
Section \ref{sec:feat_engg_pruning} gives a classical analysis of the features and performance of the Flow Classifier, followed by Section \ref{sec:cam-detection}, which describes the classical analysis of SmartCam classifiers with a result summary. Section \ref{sec:xai} shows XAI interpretation with SHAP and LIME methods, followed by the conclusion in Section \ref{sec:conclusion}.

\section{Related Work}
\label{sec:relwork}
Classification of network traffic to detect online audio/video applications is an ongoing and attractive mechanism to defend against certain cyber threats, and broadly divide these works into three groups: classification of online audio/video applications, IoT device classification without and with using XAI. 

\subsection{Works on Classifying Online Audio/Video Applications}
Because any online audio/video application can pose a serious security and privacy threats, a number of early studies \cite{bonfiglio2007ACM, Radhakrishnan2015IEEE} has focused on detection of the network flows that carries \emph{Skype}, a popular online audio/video application, data at different stages of a call, the network location of the hosts, e.g., behind NAT or Firewall \cite{Adami2009Springer}. 
\emph{Skype} uses the RTP (Real-Time Transport Protocol) possibly to hide the presence of \emph{Skype} hosts \cite{SkypeRTP, FastRTPDetection}. 
Further, the feasibility of using flow-based features for the classification of \emph{Skype} and \emph{GTalk} has been investigated \cite{Skype_GtalkSkype}. 
The work in \cite{Priyanka2022ACM} proposes an initial idea to classify the video traffic and detect IoT (spy) camera using network traffic, that we consider to extend.

\subsection{Works on Classifying IoT Devices}
The lack of sufficient security and privacy features in IoT devices has attracted a large research community in this area. 
The classification of IoT devices using machine learning is one of those predominant works that aim to provide a kind of access control. 
For example, detection of hidden IoT cameras is possible by using wireless traffic \cite{wang2020iot, Cheng2018ASIACCS} or network traffic \cite{ SivanathanA2018IEEE, Gordon2021IEEE}. 
Meidan et. al. \cite{meidan2017arXiv} addressed the problem of IoT device classification so that a white-list based verification of the devices can be ensured. 
Their work relied exclusively on the features derived from TCP sessions and the RF classifier. 
Sivanathan et.al. \cite{sivanathan2017IEEEInfocom} have shown that high accuracy in IoT traffic classification can be achieved using 12 network traffic attributes and RF classifier. 
However, the system requires training data from every IoT device, limiting its practical usability. 
Miettinen et al. \cite{miettinen2017IEEEICDCS} proposed IoT device-type identification to enforce certain security policies. Their approach requires that all devices of the same type have identical firmware and hardware versions; this may not hold true in modern IoT networks. 
Arurnan et.al. \cite{Arunan2018IEEE} proposed a framework for classifying IoT devices in smart environments using network traffic characteristics. By leveraging Naive Bayes and Random Forest ML models and discriminative traffic features, their work enables device identification, which is crucial for security, management, and optimization in IoT ecosystems. 
Pratibha et. al. \cite{PratibhaIET2021} developed IoTHunter, a framework to  monitor and manage IoT devices, that can classify the IoT devices by using device-specific keywords and MAC address. 
Salman et. al. \cite{Salman2022Wiley} developed a classifier that detects a device based on behavioral patterns, whereas Bezawada et. al. \cite{bezawada2018ACM} demonstrated that  a minimal feature set, including the number of packets per session, can achieve a high accuracy (about 99\%) using gradient boosting models. 

Recently, Heo et. al. \cite{heoACSAC2022} proposed a system, called \emph{Spy Camera Finder (SCamF)}, that detects the presence of a streaming IoT camera by using Wi-Fi traffic. 
Also, Zhang et. al. \cite{Zhang2018ACMSIGSAC} have classified IoT devices via the decryption of encrypted Wi-Fi traffic from a previously known vulnerable IoT device. 
The use of application layer information to classify IoT devices has also been explored with the help of natural language processing(NLP) on application data by Thomsen et. al. \cite{thomsen2020smartlampsmartcam}.
Bai et. al. \cite{bai2018IEEELCN} have explored automatic IoT device classification through network traffic analysis using the LSTM-CNN cascade model.
Priyanka et. al. \cite{Priyanka2024BuildSec} have extended \cite{Priyanka2022ACM} to simply find the best performing models without having any explanation, either by using traditional feature importance or XAI.
However, the vast majority of existing methods to classify IoT devices have focused on the model performance either by reducing the number of features or by improving correctness in unseen scenarios. 
There is either very limited or no focus on the discovery of important features that can lead to enhance the reliability or trust on the output of these model. 

\subsection{Explainable AI in IoT Device Classification}
To enhance the reliability of the model output, explainable AI (XAI) has become effective in domains like image classification \cite{bird2024IEEEAccess}. 
The feasibility of using XAI in the domains of IoT and ICS (Industrial Control Systems) is shown by 
Yang et. al. \cite{yang2019Elsevier}, where a set of important features is used to fingerprint the communication protocols at ICS protocol layers.  
Augmenting this with automated rule generation can save a significant amount of human effort to label the training set. Lavrenovs et. al.\cite{lavrenovs2020IEEECyCon} trained classifier targeting interfaces based on generic HTTP protocols. 
The classifiers are commonly trained on labeled data either from a laboratory network \cite{YairSAC2017} or a campus network \cite{sivanathan2017IEEEInfocom}. 
Yadav et. al.  \cite{yadav2020ACM} provide a systematic categorization of ML-augmented techniques for fingerprinting IoT devices.
Multiple XAI frameworks, such as Local Interpretable Model-Agnostic Explanations (LIME) by Dieber et. al. \cite{dieber2020LIME} and Ignatieve et. al in \cite{ignatiev2020AAAI}, and SHapley Additive exPlanation (SHAP) have
been developed by Scott et. al. \cite{scott2017NeurIPS}, have been aiming to facilitate the transparency and trust in the underlying models. 
Tsakiridis et. al. \cite{tsakiridis2020Springer} have aimed to build XAI based detection support system for IoT-based sensor network. 
Garcia et. al. \cite{garcia2019IEEEAccess} developed  an approach within the human-centric AI for generating explanations by knowledge learned by a multilayer perceptron (MLP). 

Gummadi et al. \cite{gummadi2024xaiIEEEACCESS} have emphasized the dual benefits of XAI to improve the accuracy of the model for anomaly detection and increase trust among IoT stakeholders by making ML-based decision-making transparent. 
Das et.al. \cite{das2020arXivXAI} have shown that the full potential of XAI can bridge the gap between theoretical advancements and real-world applications. 
Further, Jagatheesaperumal et. al. \cite{jagatheesaperumal2022IEEEOpenJ} underscores the transformative role of XAI in IoT by addressing transparency and trust issues. 
The synthesis of explainability and IoT is essential for building next-generation intelligent and accountable IoT systems \cite{kok2023explainable}. 
By demonstrating its effectiveness in a practical case study, the Zolanvari et. al. \cite{zolanvari2021trustIEEEIoTJ} highlights the transformative potential of XAI in addressing the unique challenges of Industrial Internet of Things (IIoT) security to detect anomalous behaviors in IIoT devices, like IP cameras, sensors, and industrial control systems. 
Muna et. al. \cite{MUNA2023ScienceIoTJ} demonstrated the use of XAI to build transparency and interpretability of ML models for IoT attack detection in smart cities having large-scale IoT deployments. 
In this paper, we extend the state-of-the-art use of XAI, in particular SHAP and LIME, into the domain of IoT (spy) camera classification, allowing network administrators to understand which features contribute most to identifying a spy camera. 
We leverage advanced feature importance analysis at both global and local levels, ensuring the key flow-specific features for effective explanations. A summary of the comparison of selected existing works is shown in Table \ref{tab:comp_work}.

\begin{table*}[hbtp]
\caption{Comparison with Existing Work. (Note- Custom Dataset refers to traffic generated by own's lab.)}
\label{tab:comp_work}
\resizebox{\linewidth}{!}{
\begin{tabular}{c|c|c|c|c|c}
\hline
 Research Paper & Goal & Dataset & \#Features & Name of Features & XAI Techniques \\ \hline
 A. Sivanathan et. al.\cite{Arunan2018IEEE} & \begin{tabular}[c]{@{}c@{}}IoT Device\\ Classification\end{tabular} & UNSW & 4 & \begin{tabular}[c]{@{}c@{}}Flow Volume, Flow Duration,\\ Avg. Flow Rate, Device\\ Sleep Time\end{tabular} & \xmark \\
 
 P. Khandait et. al.\cite{PratibhaIET2021} & \begin{tabular}[c]{@{}c@{}}IoT Network Traffic\\ Classification\end{tabular} & UNSW & 1 & \begin{tabular}[c]{@{}c@{}}Device Keywords in Flows\end{tabular} & \xmark \\
 
 B. Bezawada et.al.\cite{bezawada2018ACM} & \begin{tabular}[c]{@{}c@{}}IoT devices Behavioral \\fingerprinting\end{tabular} & \begin{tabular}[c]{@{}c@{}}Custom\end{tabular} & 2 & \begin{tabular}[c]{@{}c@{}}Packet Header, Payload Features\\ (e.g. Entropy)\end{tabular} & \xmark \\

 M. Miettinen et. al.\cite{miettinen2017IEEEICDCS} & \begin{tabular}[c]{@{}c@{}}IoT Device\\ Type Identification\end{tabular} & \begin{tabular}[c]{@{}c@{}}Custom\end{tabular} & 23 & \begin{tabular}[c]{@{}c@{}}Packet features (ARP, LLC, IP,\\ TCP, UDP, etc.)\end{tabular} & \xmark \\
 
 A. Gummadi et.al.\cite{gummadi2024xaiIEEEACCESS} & \begin{tabular}[c]{@{}c@{}}Anomaly Detection\\ for IoT devices\end{tabular} & \begin{tabular}[c]{@{}c@{}}MEMS \& \\ N-BaIoT\end{tabular} & 12 & \begin{tabular}[c]{@{}c@{}}Stream Aggregation, Time-frame\\ (Lambda), Packet Stream Stats\end{tabular} & \begin{tabular}[c]{@{}c@{}}SHAP, LOCO,\\ CEM, PFI,\\ ProfWeight\end{tabular} \\
 
 Our Approach & \begin{tabular}[c]{@{}c@{}}IoT Device\\ Classification\end{tabular} & \begin{tabular}[c]{@{}c@{}}BITSPHC\\ (Custom)\end{tabular} & 77 & \begin{tabular}[c]{@{}c@{}}Flow Duration, Tot Fwd Pkts,\\ Flow IAT STD etc.\end{tabular} & SHAP, LIME \\ \hline
\end{tabular}}
\end{table*}

\section{System Design and Implementation}
\label{sec:sysem-architecture}

\subsection{System Model} 
\begin{figure*}[hbtp]
\centering
\includegraphics[width=\textwidth]{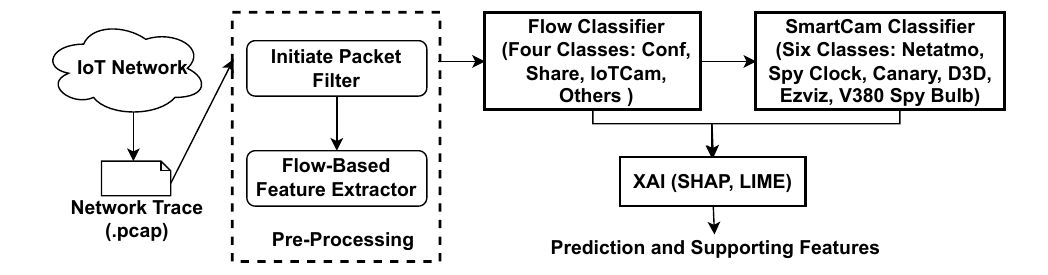}
\caption{Architecture of our proposed system (Note: VM: Vanilla Models and VMH: Vanilla Models with Hyperparameters).}
\label{fig:architecture}
\end{figure*}

Fig. \ref{fig:architecture} shows the architecture of our proposed system, called \emph{rCamInspector}, which has a simple component-based architecture. 
rCamInspector takes as input the network traces in the form of \emph{.pcap} files that can be collected at a router or a switch having a port-mirroring option.
The flow-based feature extractor module takes \emph{.pcap} files as input and extracts \emph{flow-based} features from each of the flows. 
Note that these features do not contain any IP address or transport port number.
Two classifier modules then use these features for training and prediction. 
The performance for individual classifiers in both the modules is calculated with different performance metrics, like accuracy, false positives, false negative, precision and recall. 
To get an explanation, \emph{rCamInspector} uses two XAI explainers, \enquote{SHAP} and \enquote{LIME}, to find global and local features that influence the prediction by the models in both Flow classifier and SmartCam Classifier. 
At first, a flow is classified by the \emph{Flow Classifier} into one of four classes, i.e., \emph{Conf}, \emph{Share}, \emph{IoTCam}, and \emph{Others} class. 
Next, if a flow is classified as \emph{IoTCam}, then the \emph{SmartCam Classifier} further classifies the flow into one of six classes, i.e., Netatmo, D3D, Canary, Ezviz, Spy Clock, and V380 Spy Bulb. 
Finally, the flow goes to the XAI explainer that then produces appropriate features that positively support the detection of the camera at both the \emph{Flow Classifier} and the \emph{SmartCam Classifier}. 
Thus, if any particular IoT camera becomes operational through the router, then the camera can be detected by \emph{rCamInspector} by classifying the flows generated by the camera, giving more insights into what features contribute to classifying the specific IoT camera.

\subsection{Threat Model}
We assume that an attacker can physically install an IoT camera using stolen credentials of a home/office Wi-Fi router, for instance. 
Alternatively, a remote attacker can also activate camera service in an IoT device that can offer multiple services, like smart plug-cum-camera or smart bulb-cum-camera.  
We consider an IoT device as a spy device if it offers IoT camera service along with other services like bulb or plug. 
Note that video traffic also arises out of online audio/video conferencing applications, but these applications run with the user's knowledge or permission, which is often missing in IoT cameras. 
Once an IoT camera is installed or activated, it can exfiltrate video stream to a remote attacker and directly invade the privacy of the targeted users. 
The attacker can also activate multiple IoT cameras simultaneously, change the focus area of the camera, zoom in/out for wider capture or better clarity, and so on. 
Alternatively, they can determine the timings at which the camera streams can be started or stopped without the knowledge of the owner. 
The attacker poses a security threat when it forces an IoT (spy) camera to start streaming.

\subsection{Defender Model}
A network administrator, say a defender, can execute \emph{rCamInspector} at an edge device or a gateway (similar to that in \cite{Sun2019AutomatedID, IbbadTNSM2020}), where the network and above layer traffic can be passively sniffed for the detection of the IoT camera. 
The devices in the network include traditional laptops, desktops, smartphones, and IoT devices, including IoT (spy) cameras. 
The administrator is interested in deploying a machine learning-based classifier and detector to reduce its huge manual effort in detecting the IoT (spy) cameras in its subnets. 
The traditional and advanced supervised ML models, like DT, kNN, NB, LR, RF, XGB, ET and AB, can used and the defender looks for the best-performing ones in both the classifications. 
The defender further looks for the explanations by using XAI explainers, like SHAP and LIME, in terms of class-specific features that positively support the model's output. 
The ultimate intent is to reduce false alarms and increase reliability and trust on the ML models. 
We propose \emph{rCamInspector} as an effective solution for such a defender.

\subsection{Implementation of rCamInspector}
\label{sec:imple_rCamInspector}
We implement \emph{rCamInspector} using freely available tools that are compatible with Python3 and Python packages like sklearn, pandas, SHAP and LIME libraries.
Well-known \emph{CICFlowmeter} \cite{cicflowmeter} extracts 77 flow-based features from each of the flows in every pcap file (Please refer Appendix \ref{subsec:appendixA}). 
In CICFlowmeter, a flow is a sequence of packets with a maximum time window of 600 sec, where all the packets have the same source and destination IP addresses and transport protocol. 
It generates a \emph{csv} file containing 77 flow-based features (i.e., none of these features include IP address or port number) in a row corresponding to each flow.
The use of flow-based features is advocated by the fact that the collective behavior of the flow can be more prevalent than using packet-specific features, like header information at the network layer. 
Further, the solution based on flow-based features can be highly generalizable to a large variety of IoT cameras.
The flow-based features extracted by CICFlowmeter are generic features, i.e., not optimized for a particular application.
We develop a complete toolset for analyzing feature importance based on SHAP and LIME explainers using Python's shap and lime with lime\_tabular packages. 
In our feature importance analysis, we extract the model-specific features (i.e., top important features for each ML model) and IoT-specific features (i.e., top features for each IoT camera) for all eight models in both classifications.

Our goal is to get a good interpretation of the outcomes and tune model parameters so that a set of baseline results can be gathered. 
Unless specified otherwise, we will employ the ML model that achieves top performance in the experiments. 
The selection of an ensemble model is based on the ability to accurately classify the network flows.

\section{Experimental Setup and Data Set}
\label{sec:dataset}
\subsection{Network Setup and Data Collection}
In our laboratory setup, we have collected 38GB of network traces. 
We have configured a Wi-Fi router using a Raspberry Pi 3 Model B to provide Internet connectivity to the IoT and non-IoT devices.
Table \ref{tab:camera-data-set} shows a summary of the dataset. 
Four broad categories of data are collected, namely \emph{Conf} - containing the traces of four  (Meet, Teams, Skype and Zoom) online audio/video conferencing applications, \emph{Share} - containing the traces of two (Youtube and Prime) video sharing applications, \emph{IoTCam} - containing the traces of six (Netatmo, Spy Clock, Canary, D3D, Ezviz, V380 Spy Bulb) commercially available IoT cameras, and  \emph{Others} - containing the traces of a set of randomly selected applications that do not produce any video traffic. 
While collecting traffic, various services, e.g., pan, tilt, 360-degree moving, and switching audio on and off mode, in IoT cameras are activated to include a wider range of payloads. 
The traffic in \emph{Others} is collected by surfing websites like newspapers, blogs, research paper reading, etc.

\begin{table*}[hbtp] 
\centering
\caption{Dataset Summary with flow counts and labels. }
\label{tab:camera-data-set}
\begin{tabular}{|c|c|c|c|c|c|c|}
\hline
Set  & Applications     & Size                                                                & \begin{tabular}[c]{@{}c@{}}Packet Count\end{tabular} & \begin{tabular}[c]{@{}c@{}}\#Flows, Percentage\\ /App\end{tabular} & \begin{tabular}[c]{@{}c@{}}\#Flows, Percentage\\ /Label\end{tabular} & Label                   \\ \hline \hline
\multirow{4}{*}{\begin{tabular}[c]{@{}c@{}}Set I\end{tabular}}   & Meet      & \multirow{4}{*}{\begin{tabular}[c]{@{}c@{}}10.06\\ GB\end{tabular}} & \multirow{4}{*}{22429473}   & 31154, 77.83\%                                                  & \multirow{4}{*}{40025(39.75\%)}                                   & \multirow{4}{*}{Conf}   \\ \cline{2-2} \cline{5-5}
 & Teams     &   &   & 6840, 17.08\%     &    &                         \\ \cline{2-2} \cline{5-5}
   & Skype     &   &   & 943, 2.36\%   &  &       \\ \cline{2-2} \cline{5-5}
 & Zoom      & &     & 1088, 2.72\%    &  &    \\ \hline
\multirow{2}{*}{\begin{tabular}[c]{@{}c@{}}Set II\end{tabular}}  & Prime     & \multirow{2}{*}{\begin{tabular}[c]{@{}c@{}}11.81\\ GB\end{tabular}} & \multirow{2}{*}{8026748}   & 3574, 45.36\%  & \multirow{2}{*}{7880 (7.82\%)}  & \multirow{2}{*}{Share}  \\ \cline{2-2} \cline{5-5}
 & YouTube   &     &   & 4306, 54.64\%     &     &     \\ \hline
\multirow{6}{*}{\begin{tabular}[c]{@{}c@{}}Set III\end{tabular}} & Netatmo   & \multirow{6}{*}{\begin{tabular}[c]{@{}c@{}}14.96\\ GB\end{tabular}} & \multirow{6}{*}{25675111}   & 943, 9.43\%      & \multirow{6}{*}{10000 (9.93\%)}   & \multirow{6}{*}{IoTCam} \\ \cline{2-2} \cline{5-5}
& Spy Clock &    &       & 395, 3.95\%  &      &   \\ \cline{2-2} \cline{5-5}
    & Canary    &    & & 421, 4.21\%   & &   \\ \cline{2-2} \cline{5-5} 
    & D3D       &    &  & 1733, 17.33\%    &   &   \\ \cline{2-2} \cline{5-5}
  & Ezviz     &   &    & 2586, 25.86\%  &  &  \\ \cline{2-2} \cline{5-5}
   & Spy Bulb  &   &    & 3922, 39.22\%   & & \\ \hline
Set IV    & Others & \begin{tabular}[c]{@{}c@{}}500\\ MB\end{tabular}                    & 738067   & 42764  & 42764 (42.47\%) & Others  \\ \hline
\end{tabular}
\end{table*}

\subsection{Experimental Setup}
We have used the Python-based \emph{sklearn} library for our classification tasks. 
Like any other library, this library has a default set of hyper-parameters for each of the models, and it provides additional scopes for the programmers to tune these parameters to suit a specific classification task. 
For example, the tree depth in CART model by default is 3, but it can be changed other values to obtain a better results in a particular data set. 
Unless specified otherwise, we have shown the result in this paper with default and tuned parameters. 

\subsection{Performance Metrics}
Let us denote by $FP_i$, $TN_i$, $TP_i$ and $FN_i$ the false positive, true negatives, true positive and false negative cases of a particular class of $L_i$ in either flow classifier or SmartCam classifier.
For example, in Flow classifier, we have $l=4$ classes as $L_1=IoTCam$, $L_2 = Conf$, $L_3 = Share$ and $L_4 = Others$, and the corresponding confusion matrix $C[1..4][1..4]$.

\begin{itemize}
\item Accuracy: $AC_{i}$  = $\frac{TP_{i} + TN_{i}} {TP_{i} + TN_{i} + FN_{i} + FP_{i}}$  and $AC_{\mu}$  = $\frac{\sum AC_i} {l}$ -- to measure the rate of correct classification of $L_i$.\\

\item Precision: $PR_{i}$  = $\frac{TP_{i}} {TP_{i} + FP_{i}}$ and $PR_{\mu}$  = $\frac{\sum PR_i} {l}$ -- to measure the rate of correct prediction of $L_i$.  \\

\item Recall: $RC_{i}$  = $\frac{TP_{i}} {TP_{i} + FN_{i}}$ and $RC_{\mu}$  = $\frac{\sum RC_i} {l}$ -- to measure the rate of correct prediction of $L_i$. \\

\item F1-Score: $FS_{i}$  = $\frac{PR_{i} \times RC_{i}} {PR_{i} + RC_{i}}$ and $FS_{\mu}$  = $\frac{\sum FS_i} {l}$ -- to measure a combination of both precision and recall.\\
\end{itemize}
where $TP_{i} = C[i,i]$, 
$TN_{i} = \sum_{j\neq i, k \neq i} ^ {j = 4, k = 4} C[j,k]$,  
$FP_{i} = \sum_{j \ neq i} ^{j = 4} C[i,j]$,  
$FN_{i} = \sum_{j\neq i} ^{j = 4}C[j,i]$. 

\subsection{Feature Importance Metrics}
Feature importance metrics help understand the impact of individual features on a model’s predictions, enhancing interpretability and explainability. We use two statistical measures and two explainer values for feature importance:
\begin{itemize}
\item \textbf{Pearson Correlation} measures the linear relationship between a feature and the target variable, providing insights into direct dependencies. Pearson Correlation $r$ of a pair of features $f_i$ and $f_j$ with n data point:
\begin{center}
    $r = \frac{n(\sum f_i f_j) - (\sum f_i) (\sum f_j)}{\sqrt{[n(\sum f_i^2) - (\sum f_i)^2][n(\sum f_j^2) - (\sum f_j)^2]}}$
\end{center}

\item \textbf{Mutual Information} captures nonlinear dependencies by quantifying the reduction in uncertainty about the target given a feature. Mutual Information $I(f_i;L_j)$ of a feature $f_i$ w.r.t. target class $L_j$: 
\begin{center}
$I(f_i;L_j) = \sum_{f_i,L_j} \Big[ p_{(F,L)}(f_i, L_j)\log(\frac{p_{(F,L)}(f_i, L_j)}{(p_F(f_i))(p_L(L_j))}\Big]$
\end{center}
Where, \begin{itemize} 
\item $p_{(F,L)}(f_i, L_j)$ is joint probability distribution of $(F,L)$.
\item $(p_F(f_i))$, $(p_L(L_j))$ is marginal probability distribution of $(F)$ and $(L)$ respectively.
\end{itemize}

\item \textbf{SHAP} assigns importance scores to features based on their contribution to model predictions using cooperative game theory principles. SHAP value $\phi_i$ for feature $f_i$ with respect to a model $g$ is given by: 
\begin{center}$\phi_i = \sum_{S \subseteq F \setminus \{f_i\}} \frac{|S|!(|F| - |S| - 1)!}{|F|!} \Big[g(S \cup \{f_i\}) - g(S)\Big]$
\end{center}
Where,\begin{itemize} 
\item $F$ is the set of all features, $S$ is subset of features excluding $i$,
\item $g(S)$ is the model's prediction given the feature subset $S$,
\item $|S|$ is the number of features in the subset $S$, $|F|$ is the total number of features, 
\item The summation iterates over all subsets $S$ that do not include feature $i$, whereas the fraction is a weighting factor based on the number of features. \end{itemize}

\item \textbf{LIME} approximates model behavior locally by training interpretable models on perturbed instances. LIME local interpretable model $\hat{g}$ for perturbed instance $z_i$ with respect to a model $g$ is given by: 
\begin{center}
$\hat{g}$ = $\arg\min_{g \in G} \sum_{i=1}^{N} \pi_x(z_i) (f(z_i) - g(z_i))^2 + \Omega(g)$ 
\end{center}
 Where,  \begin{itemize}
  \item $g$ is the interpretable model (e.g., decision tree) from the family of models $G$, $\Omega(g)$ is complexity penalty for $g$,
    \item $\pi_x(z_i)$ denotes proximity function that assigns higher weight to samples closer to x,
    \item $f(z_i)$ denotes black-box model's prediction for $(z_i)$,  $g(z_i)$ denotes  surrogate model's prediction for $(z_i)$. 
\end{itemize} 
\end{itemize}

\section{Classical Analysis of Features and Performance of Flow Classifier}
\label{sec:feat_engg_pruning}

\subsection{Classical Analysis of Features using Correlation}
\begin{figure}[hbtp]
\centering
\includegraphics[width=\linewidth]{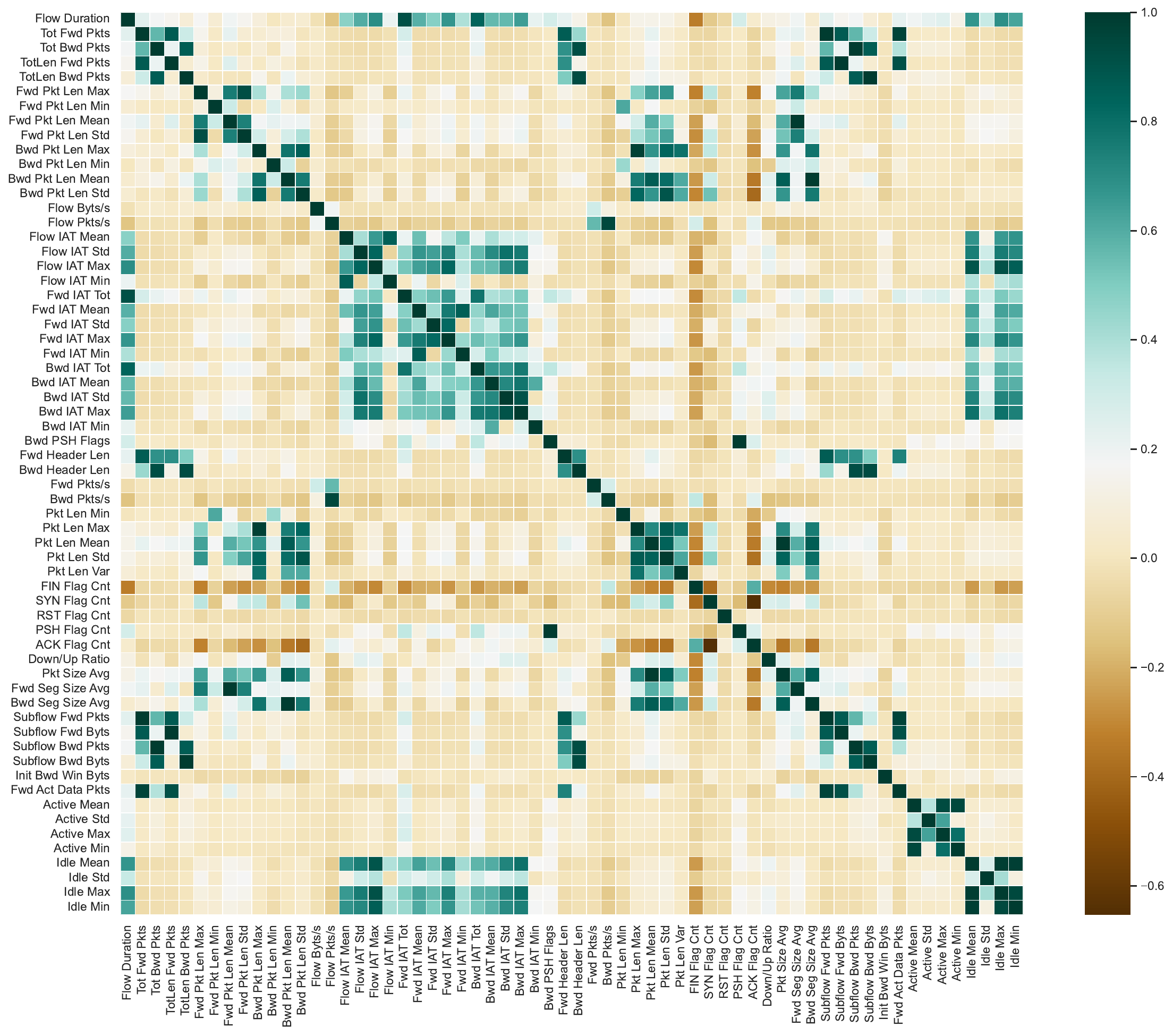}
\caption{Corelation Matrix for 62 features.}
\label{fig:corr_matrix1}
\end{figure}

Our feature extractor module extracts 77 generic flow-based features (detailed description of features is in Appendix \ref{subsec:appendixA}) and 62/77 have non-zero standard deviation that we consider for the work in this paper. 
Figure \ref{fig:corr_matrix1} shows the Pearson Correlation matrix of all 62 features. 
The correlation varies between the range [-0.6531, 0.9999]. 
We consider a correlation threshold as $\pm0.9$, i.e., we arbitrarily choose one feature from a pair of features having a correlation greater than $|\pm0.9|$. 
For example, \textit{\{Subflow Fwd Bytes, Totlen Fwd Pkts\}} pair has a correlation of 0.99, and hence we discard \textit{\enquote{Subflow Fwd Bytes}} feature. Thus, 20/62 features may be discarded based on high correlation, but we keep all for further analysis.

\subsection{Classical Analysis of Features using Mutual Information}
\label{sec:detection-vconf-vshare}
\begin{figure*}[hbt]
\includegraphics[width=\linewidth]{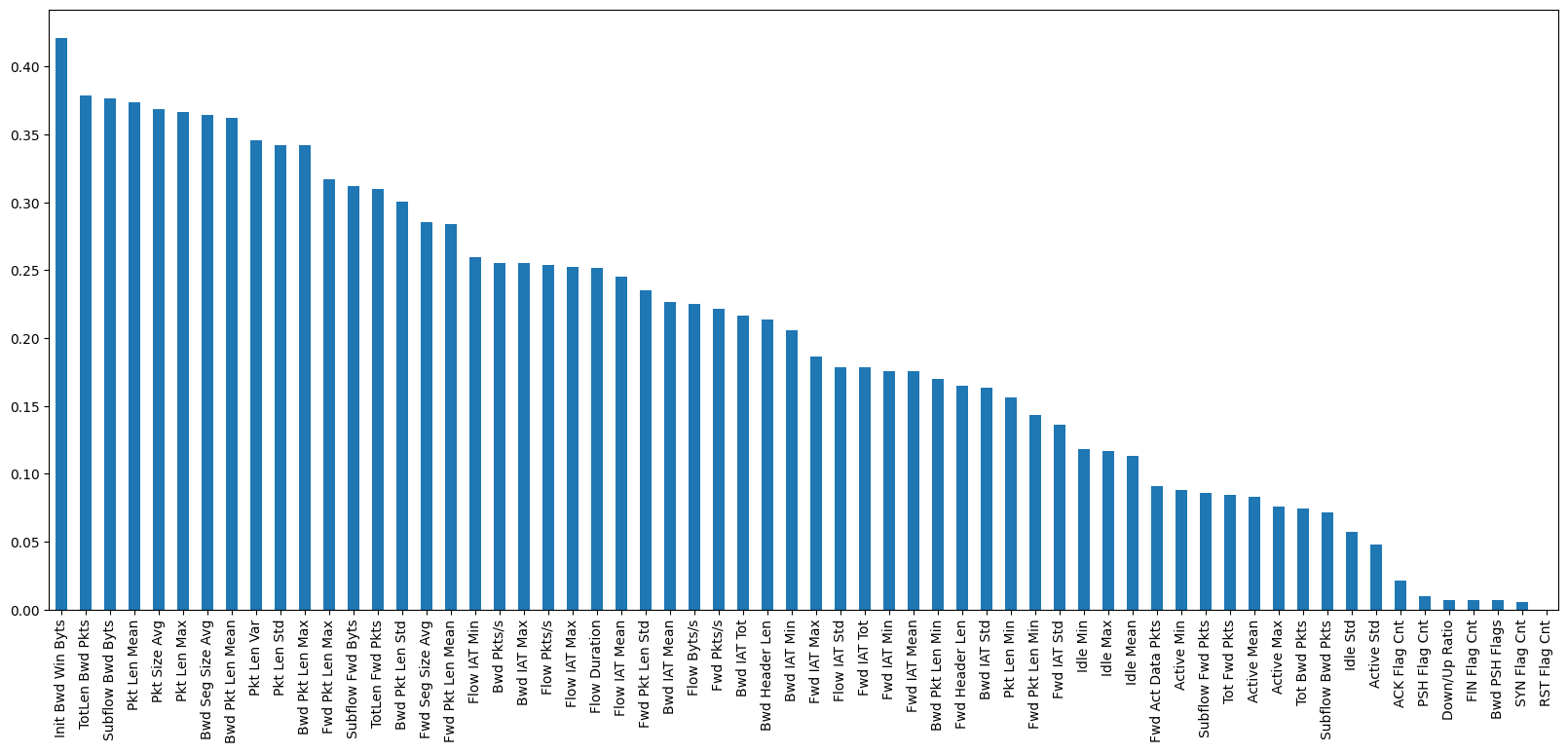}
\caption{Mutual information (MI) of all 62 features in flow classifier.} 
\label{fig:mutualInfo4class}
\end{figure*}
Figure \ref{fig:mutualInfo4class} shows the mutual information of each of the 62 features. It turns out that \textit{Init Bwd Win Byts} feature has the highest mutual information, i.e., 0.42, indicating that this feature can highly contribute to a better classification result. 
One can notice that a pair of features having a high correlation have almost same mutual information, and hence, one can discard any one of the two. 
Alternately, six features, namely \textit{RST Flag Cnt}, \textit{Down/Up Ratio}, \textit{SYN Flag Cnt}, \textit{PSH Flag Cnt}, \textit{FIN Flag Cnt} and \textit{Bwd PSH Flags} have less than \textit{0.02} mutual information, indicating that these feature may not contribute to achieve a better classification result. 

\begin{table}[htb!]
\caption{Model specific hyperparameters for best results.}
\label{tab:hyperparams}
\begin{tabular}{cc}
\hline
\textbf{Model} & \textbf{Hyperparameters} \\ \hline
DT & Max Depth = 30 \\
kNN & Nearest neighbours = 8 \\
LR & Max Iterations = 1000 \\
RF & Max Depth = 30, Number of Estimators = 300 \\
XGB & Max Depth = 30, Number of Estimators = 200 \\
ET & Max Depth = 50, Number of Estimators = 300 \\
AB & Max Depth = 30, Number of Estimators = 100 \\ \hline
\end{tabular}
\end{table}

\subsection{Performance of Flow Classifier without and with Parameter Tuning}

We have conducted a comprehensive set of experiments to find an efficient flow classifier out of 8 models,i.e., DT, kNN, NB, LR, RF, XGB, ET, and AB. 
Our aim is to find the best performing models without or with model-specific hyper-parameter tuning. 
Table \ref{tab:performance_VM_VMHy} shows the results in terms of average accuracy, recall, precision, F1 score, and false negative using two settings, default (VM) and tuned hyper-parameters (VMH).  The model-specific hyper-parameters are shown in Table \ref{tab:hyperparams}. 
We specifically show the false negatives for the IoTCam class, because it is a crucial metric that highlights the proportion of actual IoTCam instances that were incorrectly classified as the other three classes. 
DT using tuned hyper-parameters achieves about 13\% better accuracy compared to its default parameters, whereas XGB shows the precision to remains almost the same (92\%) with default and tuned hyper-parameters.
In fact, XGB achieves the best precision and recall. 
Alternately, if false negative is more important then AB with tuned hyper-parameters performs slightly better compared to XGB. 
Because the difference in the performance is not very significant, any of AB, RF and XGB can be used for this classification; we consider XGB among these three. 

\begin{table*}[hbt]
\centering
\caption{Performance of different Classifier using Vanilla Model (VM) and Vanilla Model with Hyperparameters (VMH)}
\label{tab:performance_VM_VMHy}
\begin{tabular}{c|cc|cc|cc|cc|cc}
\hline
\multirow{2}{*}{\textbf{Classifier}} & \multicolumn{2}{c|}{\textbf{Accuracy}} & \multicolumn{2}{c|}{\textbf{F1 Score}} & \multicolumn{2}{c|}{\textbf{Recall}} & \multicolumn{2}{c|}{\textbf{Precision}} & \multicolumn{2}{c}{\textbf{FN (IoTCam)}} \\ \cline{2-11} 
 & \textbf{VM} & \textbf{VMH} & \textbf{VM} & \textbf{VMH} & \textbf{VM} & \textbf{VMH} & \textbf{VM} & \textbf{VMH} & \textbf{VM} & \textbf{VMH} \\ \hline
\textbf{DT} & 0.7829 & 0.9172 & 0.7605 & 0.9167 & 0.7829 & 0.9172 & 0.7859 & 0.9167 & 0.5458 & 0.0489 \\
\textbf{kNN} & 0.8110 & 0.7957 & 0.8079 & 0.7915 & 0.8110 & 0.7957 & 0.8094 & 0.7958 & 0.2145 & 0.2288 \\
\textbf{NB} & 0.4998 & 0.4998 & 0.4134 & 0.4134 & 0.4998 & 0.4998 & 0.5885 & 0.5885 & 0.6096 & 0.6096 \\
\textbf{LR} & 0.5649 & 0.5649 & 0.5536 & 0.5536 & 0.5649 & 0.5649 & 0.5791 & 0.5791 & 0.6728 & 0.6728 \\
\textbf{RF} & 0.7930 & \textbf{0.9218} & 0.7756 & \textbf{0.9210} & 0.7930 & \textbf{0.9218} & 0.8120 & \textbf{0.9212} & 0.5268 & \textbf{0.0883} \\
\textbf{XGB} & \textbf{0.9239} & \textbf{0.9298} & \textbf{0.9227} & \textbf{0.9293} & \textbf{0.9239} & \textbf{0.9298} & \textbf{0.9236} & \textbf{0.9294} & \textbf{0.0550} & \textbf{0.0360} \\
\textbf{ET} & 0.6026 & 0.9176 & 0.5665 & 0.9169 & 0.6026 & 0.9176 & 0.6629 & 0.9170 & 0.7237 & 0.0971 \\
\textbf{AB} & 0.6704 & \textbf{0.9271} & 0.6650 & \textbf{0.9267} & 0.6704 & \textbf{0.9271} & 0.7148 & \textbf{0.9267} & 0.2288 & \textbf{0.0441} \\ \hline
\end{tabular}
\end{table*}

\subsection{Performance of Flow Classifier using PCA features with 95\% and 99\% Variance}

\begin{table*}[hbtp]
\centering
\caption{Dimensionality Reduction using PCA with 95\% and 99\% Variance.}
\label{tab:pca_var}
\begin{tabular}{c|cc|cc|cc|cc|cc}
\hline
\multirow{2}{*}{\textbf{Classifier}} & \multicolumn{2}{c|}{\textbf{Accuracy}} & \multicolumn{2}{c|}{\textbf{F1 Score}} & \multicolumn{2}{c|}{\textbf{Recall}} & \multicolumn{2}{c|}{\textbf{Precision}} & \multicolumn{2}{c}{\textbf{FNR (IoTCam)}} \\ \cline{2-11} 
 & \textbf{95\%V} & \textbf{99\%V} & \textbf{95\%V} & \textbf{99\%V} & \textbf{95\%V} & \textbf{99\%V} & \textbf{95\%V} & \textbf{99\%V} & \textbf{95\%V} & \textbf{99\%V} \\ \hline
\textbf{DT} & 0.7829 & 0.7979 & 0.7818 & 0.7969 & 0.7829 & 0.7979 & 0.7816 & 0.7965 & 0.2858 & 0.2716 \\
\textbf{KNN} & 0.7638 & 0.7699 & 0.7587 & 0.7652 & 0.7638 & 0.7699 & 0.7645 & 0.7709 & 0.3388 & 0.3177 \\
\textbf{NB} & 0.5262 & 0.5217 & 0.4754 & 0.4764 & 0.5262 & 0.5217 & 0.5052 & 0.5260 & 0.6999 & 0.6748 \\
\textbf{LR} & 0.5126 & 0.5259 & 0.4568 & 0.4696 & 0.5126 & 0.5259 & 0.5074 & 0.5547 & 0.6219 & 0.6056 \\
\textbf{RF} & 0.8103 & 0.8248 & 0.8082 & 0.8229 & 0.8103 & 0.8248 & 0.8090 & 0.8237 & 0.2858 & 0.2709 \\
\textbf{XGB} & 0.7955 & 0.8152 & 0.7926 & 0.8131 & 0.7955 & 0.8152 & 0.7945 & 0.8139 & 0.3164 & 0.2709 \\
\textbf{ET} & 0.8127 & 0.8250 & 0.8106 & 0.8229 & 0.8127 & 0.8250 & 0.8110 & 0.8237 & 0.2865 & 0.2763 \\
\textbf{AB} & 0.8038 & 0.8239 & 0.8023 & 0.8224 & 0.8038 & 0.8239 & 0.8021 & 0.8225 & 0.2817 & 0.2580 \\ \hline
\end{tabular}
\end{table*}

We investigate if reducing feature dimension can help in achieving a better model performance undermining the high feature correlation. 
Unlike mutual information, Principal component analysis (PCA) is one of the effective unsupervised mechanisms to understand inter-class relationships by reducing the feature dimension and can help in classifying the flows effectively. In particular, we consider PCA with 95\% and 99\% variance, i.e., by retaining 95\% and 99\% of complete information with all 62 features. 
Having a higher variance offers a balance between feature reduction and data retention, providing a more detailed feature set that may yield higher accuracy for certain classifiers while still achieving dimensionality reduction.
With 95\% and 99\% variance, PCA results in 5 and 11 features. 
Table \ref{tab:pca_var} shows the performances of all eight models. 
For example, XGB achieves about 80\% and 81\% accuracies using 95\% and 99\% variance, i.e., using 5 and 11 features, respectively. 
This is about 11\% less accuracy compared to that without using PCA (Table \ref{tab:performance_VM_VMHy}).
While a reduced feature set can lead to better computational efficiency, the models are significantly reducing the performance, potentially indicating the flip side of PCA in this classification. 

Figure \ref{fig:Conf_matr4Class} shows the confusion matrix using XGB for the flow classification having 8:2 train and test split ratio. 
It is important to note that only about 0.05\%, 0.08\%, and 0.7\% of Others, Conf, and Share flows are falsely detected as IoTCam flows, respectively.
Similarly, only about 1.07\%, 1.3\%, and 1.13\% for IoTCam flows are falsely detected as Others, Conf, and Share flows, respectively. 
In other words, the misclassification rate is relatively higher within the three classes of \emph{Others}, \emph{Conf}, and \emph{Share}, which causes the overall decay in accuracy. 

\begin{table*}[hb]
	\noindent\begin{minipage}{0.5\linewidth}
\caption{Feature Importance Across Classifiers for flow classifier.} 
    \label{tab:top_across_all_models_4Class}
\resizebox{\columnwidth}{!}{\begin{tabular}{llll}
        \toprule
        \textbf{Feature} & \textbf{Classifiers} & \textbf{Count} & \textbf{MI Value}\\ 
         \midrule
        Init Bwd Win Byts & ET, AB, RF, DT, NB & 5 & 0.4197\\
        Flow IAT Min & ET, AB, RF, DT & 4 & 0.2611 \\
        Bwd IAT Min & AB, RF, DT, NB & 4 & 0.2070\\
        Flow Duration & ET, RF, DT, KNN & 4 & 0.2507\\
        Bwd IAT Tot & ET, RF, DT, KNN & 4 & 0.2121\\
        Bwd Pkts/s & ET, AB, RF & 3 & 0.2576\\
        Fwd Pkts/s & AB, DT, NB & 3 & 0.2226\\
        Bwd Header Len & RF, DT, NB & 3 & 0.2155\\
        \bottomrule
    \end{tabular}}
\end{minipage}
\begin{minipage}{0.45\linewidth}
\includegraphics[width=75mm, height=50mm]{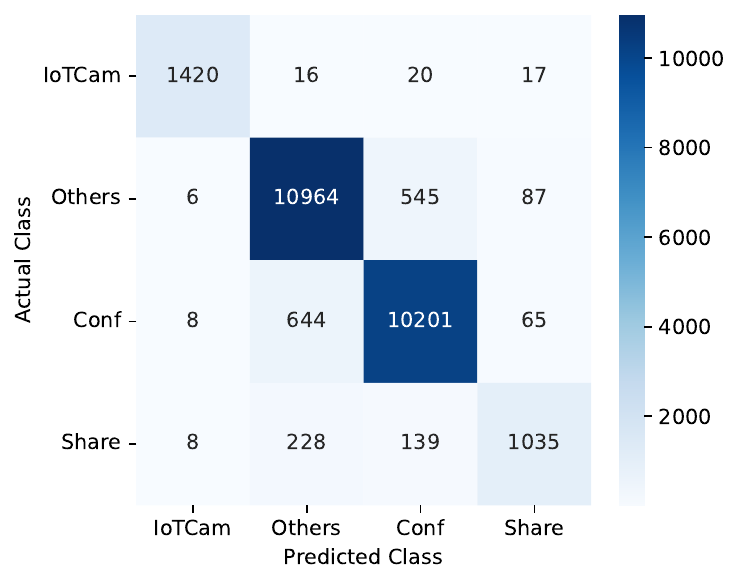}
\captionof{figure}{Confusion Matrix using XGB for Flow Classifier.}
\label{fig:Conf_matr4Class}
\end{minipage}
\end{table*}

\subsection{Model Specific Important Features }
Keeping the focus on explainability, we explore the top ten most important features of each of eight models. 
This step is crucial to know the attributes significantly influencing model performance. 
Table \ref{tab:top10feats} shows top ten features along with their feature importance. 
Though overlaps, the sets of important features in any pair of models have exclusive features. 
For example, \textit{flow IAT Min} is the third most important feature in DT, but  not present in top ten most important features in XGB. 
Note that \textit{flow IAT Min} has $18^{th}$ highest mutual information, i.e., about 0.27, and it has the highest correlation with \textit{Fwd IAT Total}, i.e., about 0.8. 
In other words, the contribution of this feature cannot be substantiated uniformly across the models. 
AB has no common features with XGB, even when both are achieving almost similar accuracy. 
Though not surprising, one can find a set of specific features to achieve a better performance in a particular model, like XGB, DT, or AB, but a uniform explanation to classify a particular class may not be possible using traditional feature analysis.

\begin{table*}[hbtp]
\centering
\caption{Top 10 Features with Importance Values for flow Classifier using Different ML Models.}
\label{tab:top10feats}
\resizebox{\linewidth}{!}{
\begin{tabular}{cc|cc|cc|cc}
\hline
\multicolumn{2}{c|}{\textbf{Decision Tree}} & \multicolumn{2}{c|}{\textbf{Naive Bayes}} & \multicolumn{2}{c|}{\textbf{Random Forest}} & \multicolumn{2}{c}{\textbf{XGB}} \\ \hline 
\textbf{Feature} & \textbf{Importance} & \textbf{Feature} & \textbf{Importance} & \textbf{Feature} & \textbf{Importance} & \textbf{Feature} & \textbf{Importance} \\ \hline

Init Bwd Win Byts & 0.2499 & Bwd IAT Min & 0.0160 & Init Bwd Win Byts & 0.1359 & Bwd Pkt Len Min & 0.1187 \\
Bwd IAT Min & 0.1283 & Fwd IAT Min & 0.0086 & Bwd IAT Min & 0.0627 & Fwd Pkt Len Min & 0.0770\\
Flow IAT Min & 0.0504 & Subflow Fwd Byts & 0.0077 & Flow IAT Min & 0.0576 & Fwd Pkt Len Mean & 0.0541\\
Bwd IAT Tot & 0.0476 & TotLen Fwd Pkts & 0.0077 & Bwd Header Len & 0.0342 & Bwd PSH Flags & 0.0476\\
Fwd Pkt Len Std & 0.0434 & Fwd Header Len & 0.0074 & Flow Duration & 0.0324 & ACK Flag Cnt & 0.0470\\
Bwd Header Len & 0.0349 & Bwd Header Len & 0.0053 & Bwd Pkts/s & 0.0272 & Pkt Len Max & 0.0425\\
Flow Duration & 0.0305 & Init Bwd Win Byts & 0.0030 & Flow Pkts/s & 0.0269 & RST Flag Cnt & 0.0362\\
Fwd Pkts/s & 0.0299 & Fwd Pkts/s & 0.0027 & Flow IAT Max & 0.0262 & Pkt Len Min & 0.0349 \\
Bwd Pkt Len Min & 0.0246 & Fwd IAT Mean & 0.0015 & Bwd IAT Tot & 0.0248 & Bwd Pkt Len Max & 0.0276\\
Flow Byts/s & 0.0201 & Tot Fwd Pkts & 0.0014 & Flow IAT Mean & 0.0228 & Fwd Pkt Len Std & 0.0255\\  \hline \hline
\end{tabular}}

\resizebox{0.7\linewidth}{!}{\begin{tabular}{cc|cc|cc}
\hline
\multicolumn{2}{c|}{\textbf{ET}} & \multicolumn{2}{c|}{\textbf{AB}} & \multicolumn{2}{c}{\textbf{KNN}} \\ \hline
\textbf{Feature} & \textbf{Importance} & \textbf{Feature} & \textbf{Importance} & \textbf{Feature} & \textbf{Importance}  \\ \hline

Init Bwd Win Byts & 0.1073 & Init Bwd Win Byts & 0.4596 & Flow Duration & 0.2332 \\
ACK Flag Cnt & 0.0353 & FIN Flag Cnt & 0.1085 & Bwd IAT Tot & 0.2150 \\
Flow Duration & 0.0342 & Bwd Pkts/s & 0.0845 & Fwd IAT Tot & 0.2048 \\
Flow IAT Min & 0.0321 & Flow IAT Mean & 0.0467 & Flow IAT Max & 0.1916 \\
Fwd Pkt Len Max & 0.0312 & Fwd Pkts/s & 0.0380 & Fwd IAT Max & 0.1224 \\
Bwd IAT Tot & 0.0280 & Flow Byts/s & 0.0370 & Bwd IAT Max & 0.1207 \\
Bwd Pkts/s & 0.0261 & Flow IAT Min & 0.0260 & Flow IAT Std & 0.1206 \\
Flow Pkts/s & 0.0257 & Bwd IAT Min & 0.0254 & Fwd IAT Mean & 0.1192 \\
Bwd Pkt Len Max & 0.0250 & Idle Max & 0.0224 & Fwd IAT Std & 0.1170 \\
SYN Flag Cnt & 0.0236 & Fwd IAT Min & 0.0166 & Flow IAT Mean & 0.1021 \\ \hline
\end{tabular}}
\end{table*}

We extend our analysis to find the most important and common features across the models. 
Table \ref{tab:top_across_all_models_4Class} shows the features that have commonly appeared in top ten features sets. 
For example, \textit{Init Bwd Win Byts} has appeared in top ten features in 5 out of 8 models, namely ET, AB, RF, DT, and NB. 
This particular feature has the highest mutual information (MI), i.e., 0.42.  
We consider a threshold as 3, i.e., a feature that appeared in at least three models, then it can be considered as very important to achieve better accuracy. 
Such a threshold resulted in eight features that are important across any model. 
The remaining seven features in this list have much less MI compared to the first one. 
Each of these eight features has a higher mutual information i.e., greater than or equal to 0.23 (Figure \ref{fig:mutualInfo4class}). 

\section{Classical Analysis of SmartCam Classifier}
\label{sec:cam-detection}
\subsection{Feature Analysis based on Mutual Information}

\begin{figure*}[hbtp]
\includegraphics[width=\linewidth]{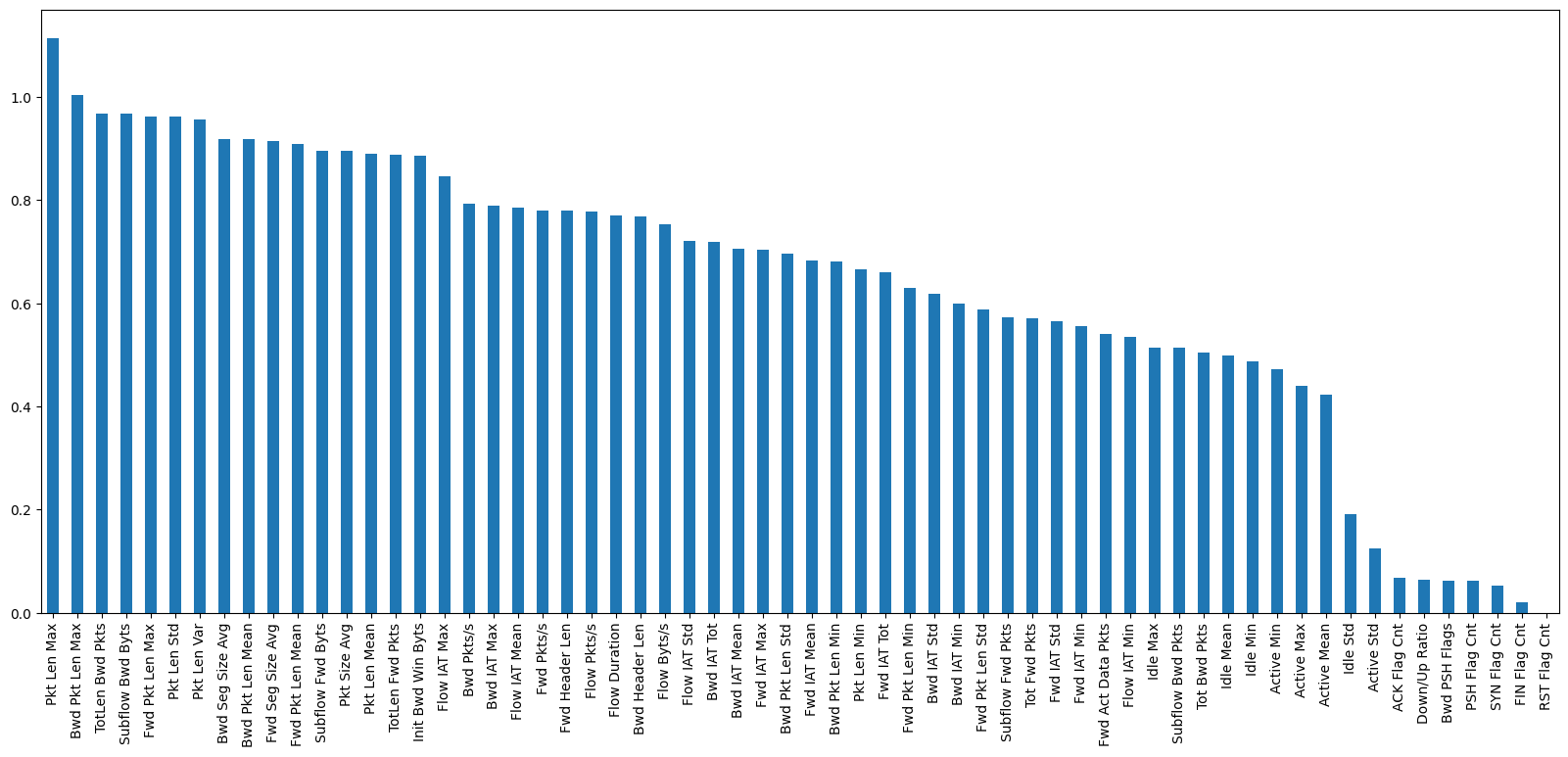}
\caption{Mutual information (MI) of all 62 features in SmartCam classifier.}
\label{fig:mutualInfo6Class}
\end{figure*}

Figure \ref{fig:mutualInfo6Class} shows the mutual information of all 62 features using this dataset having six classes, namely SpyClock, Canary, D3D, Exviz, Netatmo, and SpyBulb. 
Unlike flow classifier, \textit{Pkt Len Max} has the highest mutual information (MI), i.e., about 1.10, in SmartCam classifier, whereas \textit{Init Bwd Win Byts} has the $16^{th}$ highest MI. 
In other words, the MI values in SmartCam classifier can vary significantly in deciding the model performances. 
\begin{table*}[hbtp]
    \centering
    \caption{Classwise and Overall Accuracy for SmartCam Classifier}
    \resizebox{\linewidth}{!}{\begin{tabular}{lcccccccc}
        \toprule
        \textbf{Classifier} & \textbf{AlarmClock} & \textbf{Canary} & \textbf{D3D} & \textbf{Ezviz} & \textbf{Netatmo} & \textbf{V380} & \textbf{Overall Accuracy} & \textbf{Overall FN\%} \\ 
        \midrule
        DT & 0.9645 & 0.9497 & 0.9982 & 0.9918 & 0.9741 & 0.9906 & 0.9876 & 1.24\% \\ 
        KNN          & 0.7943 & 0.8428 & 0.9110 & 0.9544 & 0.9482 & 0.8925 & 0.9103 & 8.97\% \\ 
        NB  & 0.2979 & 0.3962 & 0.7313 & 0.3310 & 0.9871 & 0.2190 & 0.4191 & 58.09\% \\ 
        LR & 0.4043 & 0.6541 & 0.5125 & 0.6047 & 0.5307 & 0.8391 & 0.6664 & 33.36\% \\ 
        RF & 0.9149 & 0.9308 & 0.9911 & 0.9860 & 0.9579 & 0.9906 & 0.9803 & 1.97\% \\ 
        \textbf{XGB}      & \textbf{0.9716} & \textbf{0.9434} & \textbf{1.0000} & \textbf{0.9942} & \textbf{0.9871} & \textbf{0.9969} & \textbf{0.9921} & \textbf{0.79\%} \\ 
        ET  & 0.8936 & 0.9182 & 0.9893 & 0.9801 & 0.9676 & 0.9922 & 0.9785 & 2.15\% \\ 
        AB     & 0.9645 & 0.9560 & 0.9982 & 0.9930 & 0.9741 & 0.9843 & 0.9858 & 1.42\% \\ 
        \bottomrule
    \end{tabular}}
    \label{tab:PerformanceMatrixSixClass}
\end{table*}

Table \ref{tab:PerformanceMatrixSixClass} shows the performance of the classifier using eight models for each of the six classes. 
Our aim is to find best performing model(s) along with the important features to identify a given IoT camera. 
For each classifier, we report the class-wise accuracies, which indicate how well each model performs on the individual categories. 
This explanatory analysis allows us to identify which classifiers excel in certain subclasses and where they may struggle.
For example, DT achieves an average accuracy of about 98\%, whereas it achieves more than 99\% accuracy to identify D3D, Ezviz and V380 cameras. 
Similarly, XGB achieves an average accuracy of more than 99\%, whereas it finds no mistakes to identify D3D camera and achieves more than 99\% of accuracy to identify Ezviz and V380 cameras. 
Looking at false negatives, we find that XGB achieve the least false negatives, and the next least false negatives can be seen in DT, AB and RF. 
In other words, XGB can be the most efficient model.
Looking at a particular run of XGB (Figure \ref{fig:Conf_matr6Class}), none of the D3D getting misclassified as any other camera, whereas SpyClock and Canary suffer most misclassifications, i.e., 2.8\% and 5.7\% respectively. 

\begin{table*}[htb!]
\noindent\begin{minipage}{0.5\linewidth}
\caption{Feature Importance Across Classifiers for SmartCam Classifier.} 
    \label{tab:mostCommonFeats6Class}
\resizebox{\columnwidth}{!}{\begin{tabular}{llll}
        \toprule
        \textbf{Feature} & \textbf{Classifiers} & \textbf{Count} & \textbf{MI Value}\\ 
        \midrule
        Init Bwd Win Byts  & AB, ET, RF, NB, DT & 5  & 0.4206\\
        Pkt Len Min  & AB, ET, XGB, RF, DT & 5 & 0.1545\\
        Bwd IAT Tot  & DT, AB, ET, NB , KNN & 5 & 0.2108\\
        Bwd Header Len  & AB, ET, RF, NB, DT & 5  & 0.2150\\
        Bwd IAT Min  & DT, RF, AB, NB & 4 & 0.2067\\
        Pkt Len Max  & AB, ET, RF, DT & 4 & 0.3631 \\
        Bwd Pkts/s  & AB, RF, DT & 3 & 0.2584 \\
        Fwd IAT Tot  & ET, NB, KNN & 3 & 0.1757 \\
        Flow Duration & KNN, NB, ET & 3 & 0.2519 \\
        Bwd IAT Mean & RF, NB , KNN & 3 & 0.2267\\
        \bottomrule
    \end{tabular}}
\end{minipage}
\begin{minipage}{0.5\linewidth}
\centering
\includegraphics[width=75mm, height=60mm]{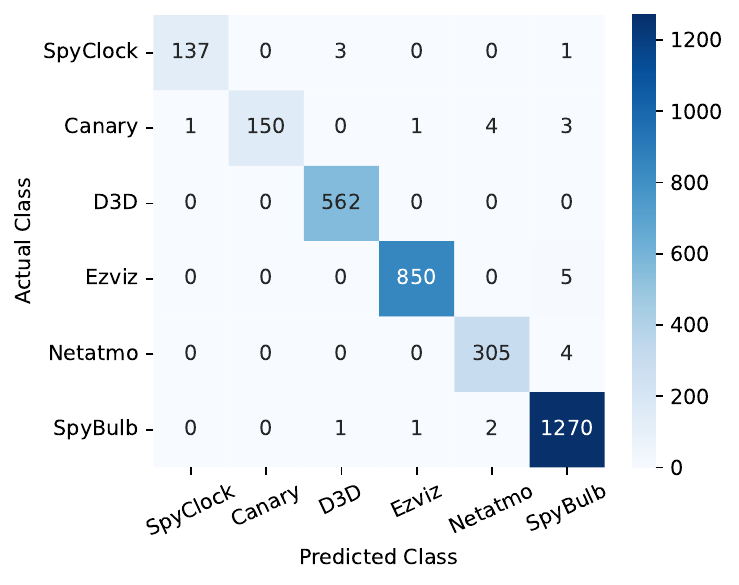}
\captionof{figure}{Confusion Matrix using XGB for SmartCam Classifier.}
\label{fig:Conf_matr6Class}
\end{minipage}
\end{table*}

\subsection{Classical Analysis of Important Features}
Similar to the Flow Classifier, we look into the top ten important model-specific features. 
Table \ref{tab:top10sixclass} shows the eight sets of top ten features. 
Looking at the top two models of XGB and DT in terms of average accuracy, there is only one common feature, i.e., \textit{Pkt Len Min}, and the rest are exclusive. 
Comparing the sets in the same model across the Flow and SmartCam classifiers, we find that five features are common in DT, whereas two features are common in XGB. 
Hence, like Flow classifier, the models have rather an exclusive set of features to achieve their respective performance.  

\begin{table*}[h]
\centering
\caption{Top 10 Features with Importance Values for SmartCam Classifier using Different ML Models.}
\label{tab:top10sixclass}
\resizebox{\linewidth}{!}{
\begin{tabular}{cc|cc|cc|cc}
\hline
\multicolumn{2}{c|}{\textbf{Decision Tree}} & \multicolumn{2}{c|}{\textbf{Naive Bayes}} & \multicolumn{2}{c|}{\textbf{Random Forest}} & \multicolumn{2}{c}{\textbf{XGB}}\\ \hline 
\textbf{Feature} & \textbf{Importance} & \textbf{Feature} & \textbf{Importance} & \textbf{Feature} & \textbf{Importance} & \textbf{Feature} & \textbf{Importance} \\ \hline
Init Bwd Win Byts & 0.3103 & Fwd IAT Min & 0.1168 & Init Bwd Win Byts & 0.0767 & Bwd PSH Flags & 0.1397 \\
Pkt Len Min & 0.2363 & Bwd IAT Min & 0.0694 & Pkt Len Min & 0.0440 & Idle Max & 0.0901 \\
Bwd IAT Min & 0.1005 & Init Bwd Win Byts & 0.0552 & Pkt Len Max & 0.0393 & Fwd IAT Std & 0.0712\\
Bwd IAT Tot & 0.0840 & Pkt Len Var & 0.0198 & Bwd Header Len & 0.0371 & Pkt Len Min & 0.0705 \\
Bwd Pkts/s & 0.0375 & Fwd IAT Tot & 0.0171 & Bwd Pkt Len Min & 0.0344 & Pkt Len Std & 0.0409\\
Pkt Len Max & 0.0335 & Bwd IAT Tot & 0.0144 & Bwd IAT Min & 0.0312 & TotLen Fwd Pkts & 0.0358\\
Bwd Header Len & 0.0253 & Bwd Header Len & 0.0143 & Bwd Pkts/s & 0.0304 & RST Flag Cnt & 0.0333\\
Flow IAT Min & 0.0245 & Bwd IAT Mean & 0.0135 & Pkt Len Var & 0.0276 & Active Std & 0.0330\\
Fwd Pkt Len Min & 0.0234 & Fwd Header Len & 0.0132 & Bwd IAT Mean & 0.0254 & Fwd Pkt Len Max & 0.0330\\
Bwd Pkt Len Max & 0.0176 & Flow Duration & 0.0122 & Flow Pkts/s & 0.0248 & Fwd IAT Mean & 0.0312\\  \hline  \hline
\end{tabular}}

\resizebox{0.7\linewidth}{!}{
\begin{tabular}{cc|cc|cc}
\hline
 \multicolumn{2}{c|}{\textbf{ET}} & \multicolumn{2}{c|}{\textbf{AB}} & \multicolumn{2}{c}{\textbf{KNN}} \\ \hline
\textbf{Feature} & \textbf{Importance} & \textbf{Feature} & \textbf{Importance} & \textbf{Feature} & \textbf{Importance} \\ \hline
Init Bwd Win Byts & 0.0656 & Init Bwd Win Byts & 0.3092 & Fwd IAT Tot & 0.3570 \\
ACK Flag Cnt & 0.0395 & Pkt Len Min & 0.2366 & Bwd IAT Tot & 0.3352 \\
Bwd Header Len & 0.0287 & Bwd IAT Min & 0.1028 & Flow Duration & 0.2730 \\
Pkt Len Min & 0.0287 & Bwd IAT Tot & 0.0836 & Flow IAT Max & 0.1967 \\
Flow Duration & 0.0266 & Bwd Pkts/s & 0.0378 & Bwd IAT Max & 0.1808 \\
Bwd Pkt Len Max & 0.0256 & Pkt Len Max & 0.0346 & Fwd IAT Max & 0.1755 \\
Fwd IAT Tot & 0.0252 & Fwd Pkt Len Min & 0.0312 & Fwd IAT Mean & 0.1725 \\
Pkt Len Max & 0.0252 & Bwd Header Len & 0.0269 & Bwd IAT Mean & 0.1671 \\
Flow IAT Max & 0.0242 & Flow IAT Min & 0.0238 & Flow IAT Mean & 0.1424 \\
Bwd IAT Tot & 0.0224 & Bwd IAT Max & 0.0174 & Flow IAT Std & 0.1335 \\ \hline
\end{tabular}}
\end{table*}

Next, we look at the most common features in the sets of top ten features, having a threshold of 3, i.e., a feature is present in at least three of the sets.
Table \ref{tab:mostCommonFeats6Class} shows that 14 features are most common across the eight models.
For example, \textit{Init Bwd Win Byts} is present in top ten sets of five models, namely, ET, AdaBoost, RF, DT, and NB, whereas \textit{Bwd header Len} is present in three models, namely RF, DT and NB. 
Interestingly, the former feature has the highest mutual information (MI) among 14 features, and this is the same as that in the Flow Classifier.  
Comparing the Flow and SmartCam classifiers, six features are common in the most common feature lists. 
This clearly shows the detection of the IoT cameras requires certain additional features. 
However, it is still challenging to discover a comprehensive explanation of the classification results for the detection of a specific camera by classical analysis of features.

\subsection{Summary of Results}
In summary, NB (41.91\%) and LR (66.64\%) exhibited subpar performance, struggling to achieve high accuracy in the SmartCam classifier. 
In contrast, XGB (99.21\%) and AB (98.70\%,) emerge as the top-performing models in our analysis. 
These findings suggest that ensemble methods, like XGB and AB, significantly outperform simpler models like NB and LR for IoT camera detection, underscoring the importance of selecting the right model based on the dataset characteristics and classification objectives.

\section{Improving Reliability and Trust on Model Performance using XAI}
\label{sec:xai}

\subsection{XAI Settings}
To empower a network administrator for decision-making based on the results of an ML model, \emph{rCamInspector} uses two XAI explainers, namely \enquote{SHAP} and \enquote{LIME}. 
Any explainer takes as input a trained model and a test dataset.
For every test sample, an explainer provides the corresponding values, i.e., SHAP and LIME values, for each of the features to support the prediction obtained by the trained model. 
For example, the SHAP values can be positive and negative indicating the positive and negative supports to the prediction for the given data point. 
Note that the classical analysis of features does not indicate their shares of contributions to a specific prediction.
Hence, a network administrator often considers an ML-based model as a black box and blindly relies on the results produced by the model. 

XAI explainer can build a level of trust in the model prediction. 
An explainer can visualize the contribution of each feature to specific predictions, thus gaining a clearer understanding of significant features and model prediction. 
Explainability not only improves model transparency but also reveals the patterns in misclassifications, helping to fine-tune the models and choose the relevant features, which could address issues such as the high false-negative rate seen in some models. 
SHAP values can be used to describe the model both globally (for overall model performance) and locally (for specific classes). 
LIME values can be used to cross-validate the interplay between the features for a correct prediction by the model. Specifically, we use the LIME to obtain a better local explanation.

\subsection{SHAP-Features for Flow Classifier}

Using SHAP explainer, we analyze the top features contributing positively and negatively to IoTCam predictions in the Flow Classifier using XGB. 
This dual analysis provides a clearer understanding of the factors driving the model behavior for IoTCam class.
Note that XGB model is trained with the dataset having four classes. 
SHAP explainer then takes as inputs the trained XGB model and the test dataset or train dataset having only the independent variables, i.e., the X values.
The explainer computes the SHAP values for each of the features for every data point.

\begin{figure*}[htbp]
\begin{tabular}{cc}
\includegraphics[width=0.48\linewidth]{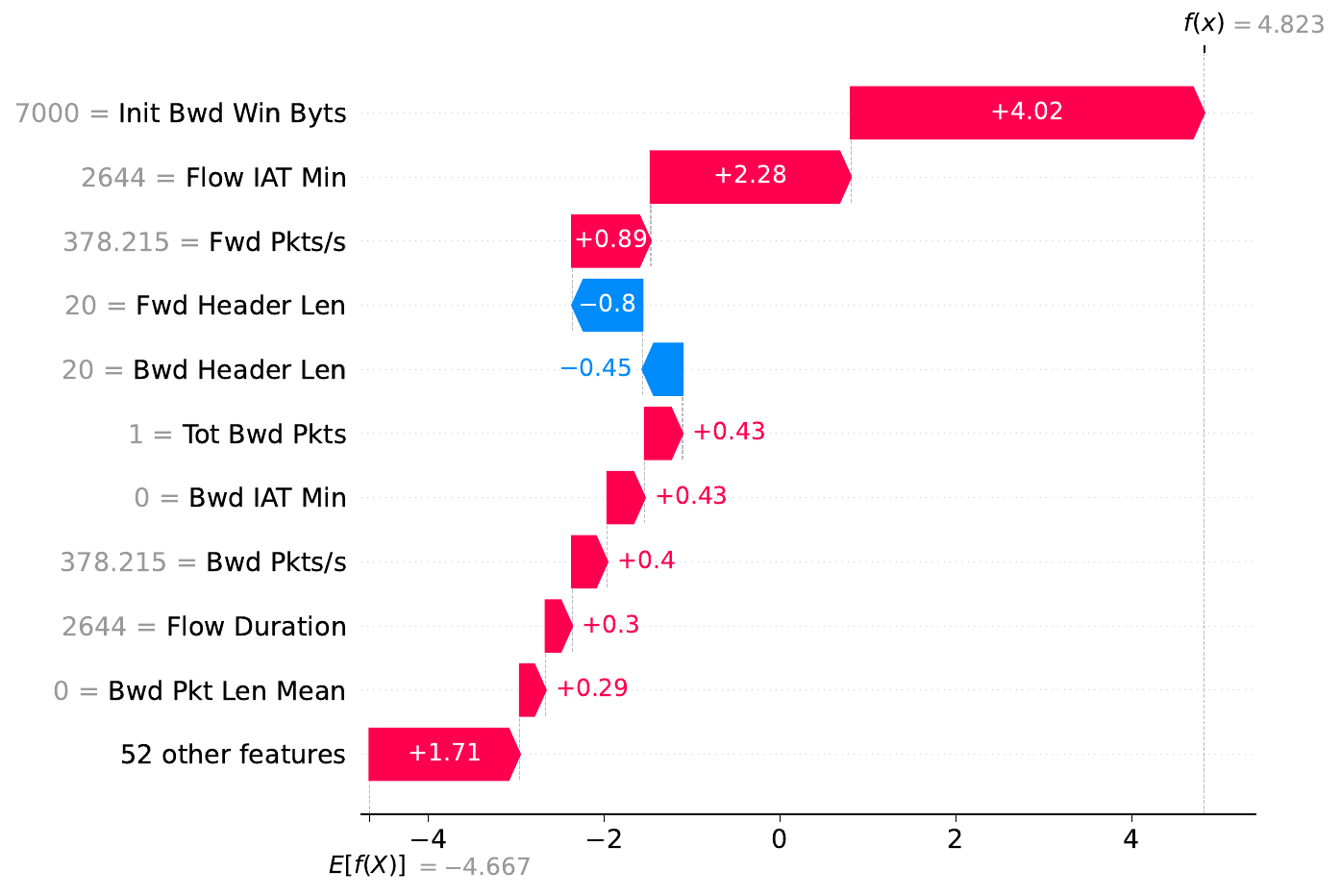} &
\includegraphics[width=0.48\linewidth]{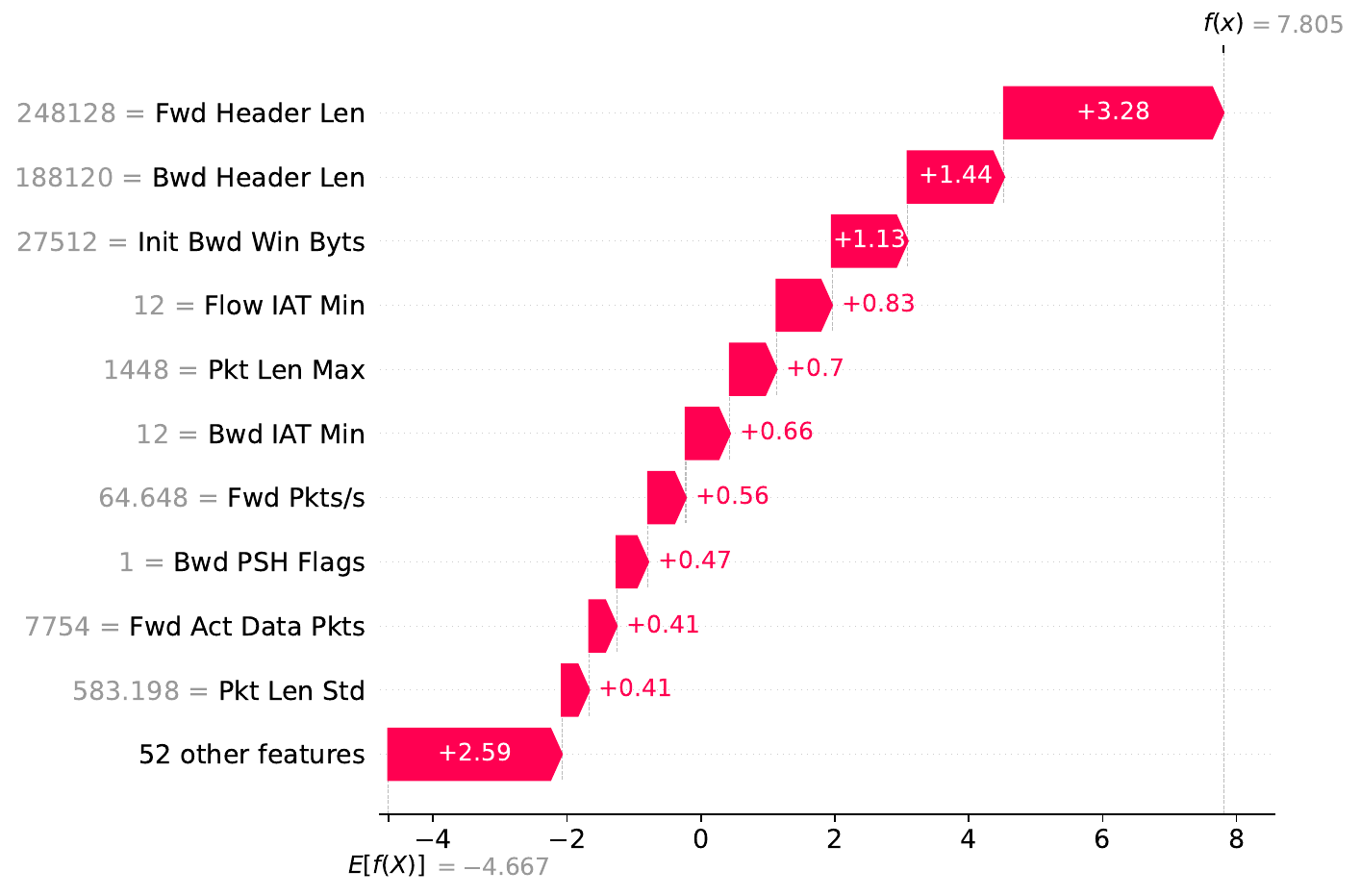} \\ 
(a) True class IoTCam in Training.   & (b) True class IoTCam in Testing. \\
\end{tabular}
\caption{SHAP values using XGB model while (a) training and (b) testing for IoTCam while flow classifier.}
\label{fig:SHAP_traintestplots_4Class}
\end{figure*}

Figure \ref{fig:SHAP_traintestplots_4Class} shows the SHAP values for two data points taken, one each from training and testing samples; both belong to the IoTCam class. 
It turns out that the correct classification for this particular data point in training is positively supported by features like \textit{Init Bwd Win Byts}, \textit{Flow IAT Min} and \textit{Bwd IAT Min} in the top ten features list, whereas \textit{Bwd Header Len} and \textit{Fwd Header Len} are the ones that negatively support for the correct classification. 
Looking at the test sample, i.e., Figure \ref{fig:SHAP_traintestplots_4Class}(b), it turns out that several other features, like \textit{Fwd Pkt Len Max} and \textit{ACK Flag Cnt}, positively supports the correct prediction, compared to that in the training sample. 
In fact, all the top ten features positively support the correct prediction of the test sample.

We have investigated the SHAP values of the features that positively supports the incorrect predictions of this same samples to be classified as other three classes, i.e., \enquote{Others}, \enquote{Conf} and \enquote{Share} (the plots for training sample are shown in Figure \ref{fig:SHAP_traintestplots_4Class_appendix} (a), (c) and (e) and testing in Figure \ref{fig:SHAP_traintestplots_4Class_appendix} (b), (d) and (f) in Appendix \ref{subsec:appendixB}). 
We found that \textit{Init Bwd Win Byts} feature negatively supports to classify the sample as \enquote{Others} and \enquote{Conf}, whereas it positively supports the prediction as \enquote{Share}. 
Further, though \textit{Bwd IAT Min} feature has positive support for   \enquote{Others}, it has negative support for \enquote{Conf}. 
While \textit{Bwd Header Len} has positive support for both \enquote{Others} and \enquote{Conf}, it does not belong to 10 important features for \enquote{Share}.  Further, it turns out that \textit{Bwd Header Len} feature negatively supports for IoTCam in training (Figure \ref{fig:SHAP_traintestplots_4Class}(a)), but positively supports for both \enquote{Others} and \enquote{Conf}, whereas it does not belong to the list of top ten features to support \enquote{Share}. 
While \textit{Fwd Header Len} appears to positively support \enquote{Others}, \textit{Flow IAT Min} negatively supports other three of \enquote{Others}, \enquote{Conf} and \enquote{Share}. 
Also, \textit{Bwd Pkt Len Std} negatively supports \enquote{Others} and it positively supports \enquote{Conf}, and it does not appear in the top ten list in  \enquote{Share}.
Overall, the features have complex interdependencies to classify any data point into its respective classes. 
Though such complex relationships may not be unexpected, they provide an initial basis for understanding the contributions of individual features during training and testing. 

\begin{figure*}[hbt!]
\centering
\includegraphics[width=\linewidth]{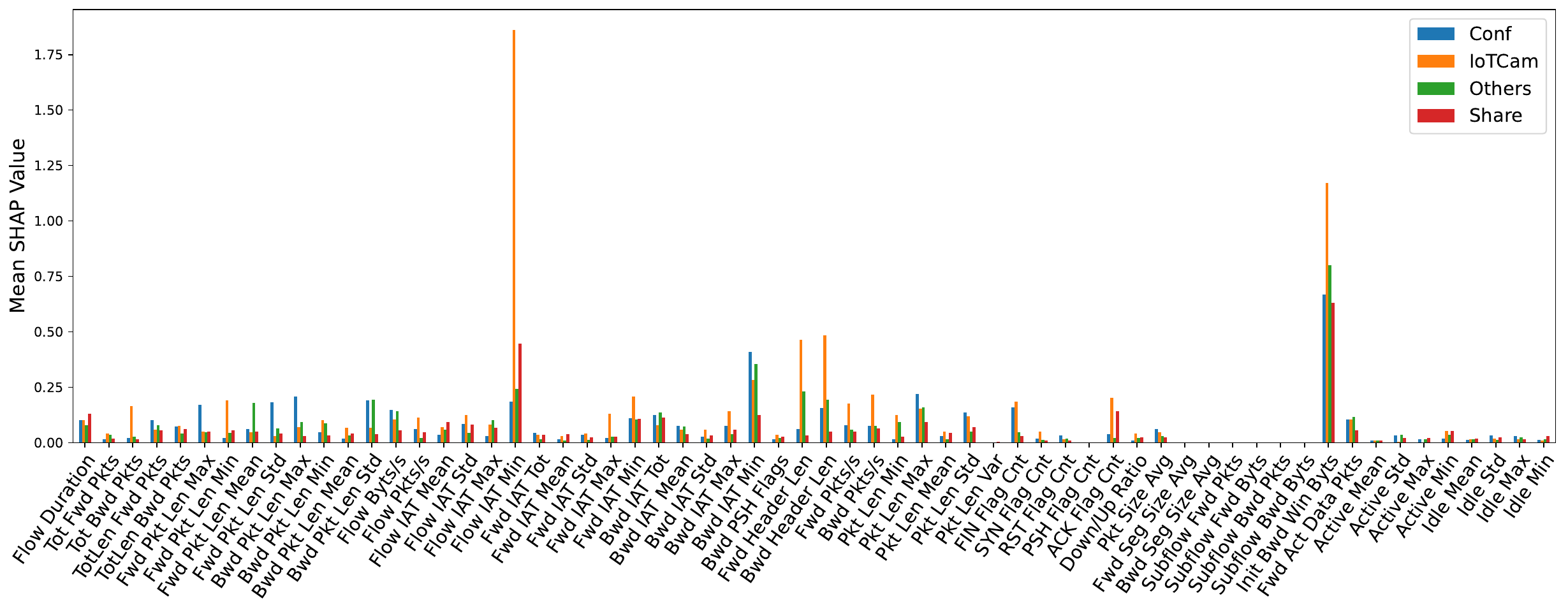}
\caption{Mean SHAP Values using XGB for Flow classifier.}
\label{fig:Mean_4Class}
\end{figure*}

Finally, we have investigated the mean SHAP values, i.e., the mean is obtained by computing the average of SHAP values across all the data samples for a given class for a complete dataset to get an overall idea of the important features supporting any particular class. 
Figure \ref{fig:Mean_4Class} shows the mean SHAP values for the flow classification. 
It turns out that \textit{Init Bwd Win Byts} is the top feature that positively supports the classification of both \enquote{IoTCam} and \enquote{Share}. 
The other most important features are \textit{Bwd Header Len} and \textit{ACK Flag Cnt} that distinguish two classes, i.e., IoTCam and Share.  
Similarly, \textit{Init Bwd Win Byts} and \textit{Bwd IAT Min} are the two most important features positively supporting both \emph{Others} and \emph{Conf}. 
The other most important features for \emph{Others} and \emph{Conf} are \textit{Flow IAT Min} and \textit{Fwd Header Len}, and \textit{Pkt Len Max} and \textit{Bwd Pkt Len Std} respectively that can classify the flows properly.


\subsection{SHAP-Features for SmartCam Classifier}
\begin{figure*}[hbtp]
\centering
\begin{tabular}{cc}
\includegraphics[width=0.48\linewidth]{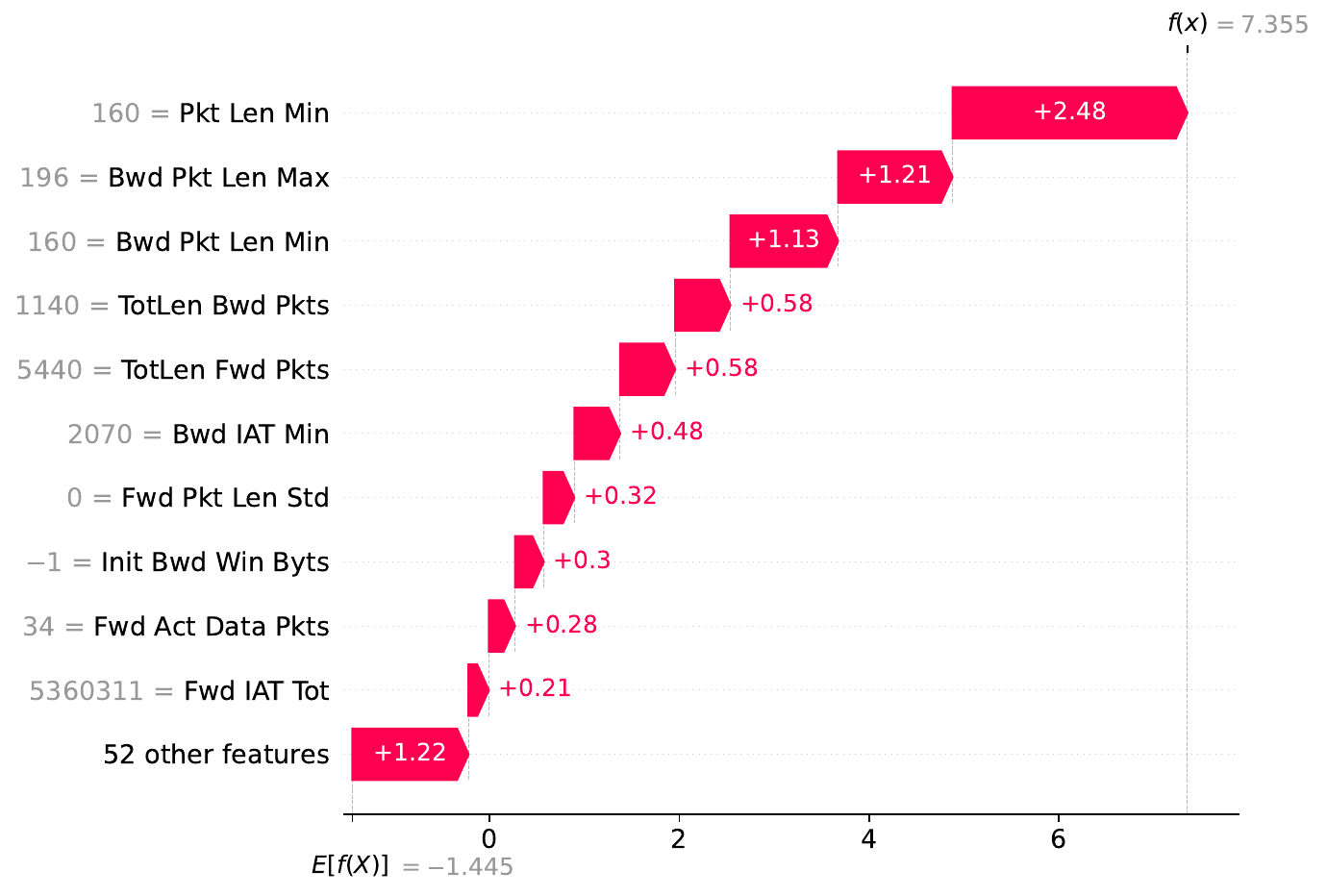} &
\includegraphics[width=0.48\linewidth]{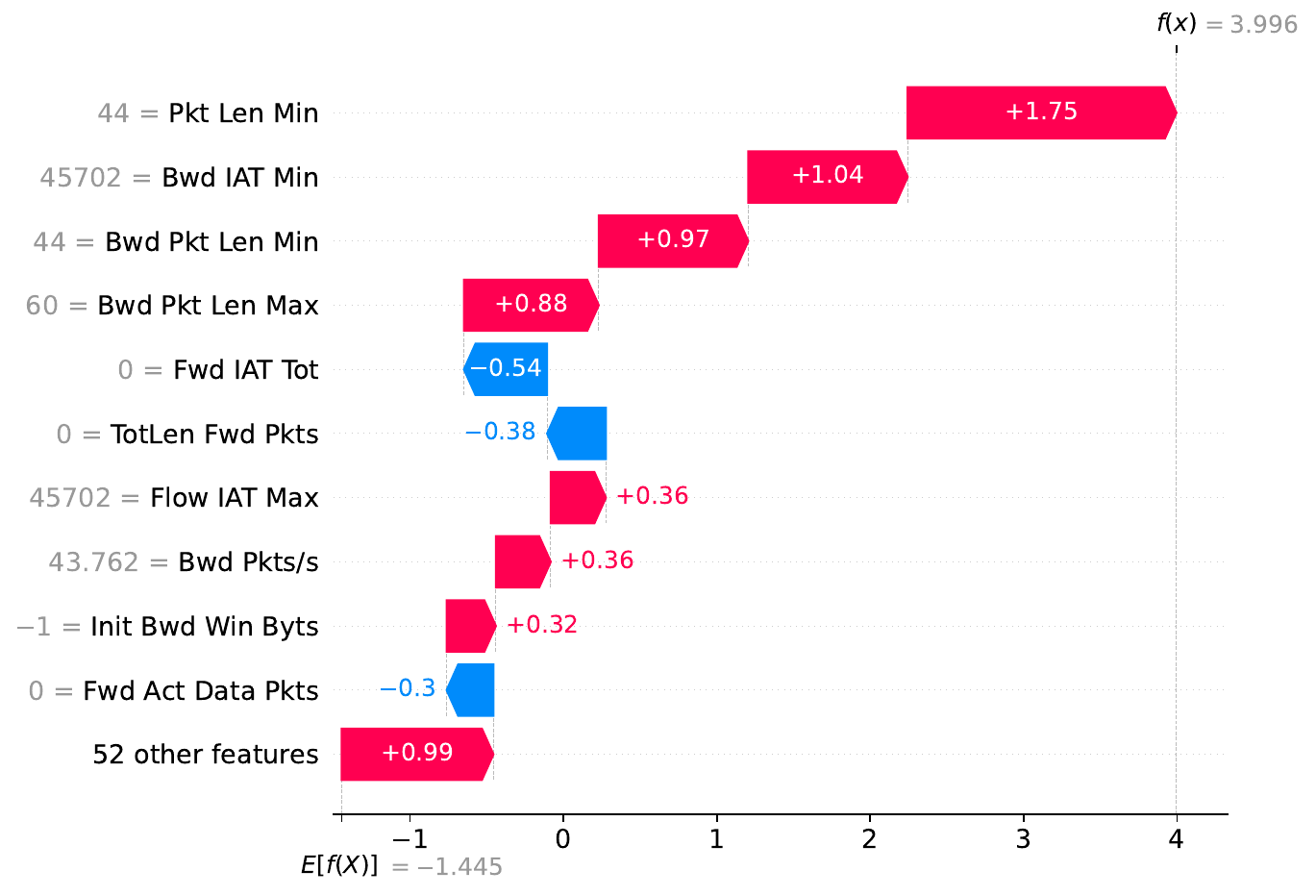} \\ 
(a) Training sample belong to Ezviz.   & (b) Testing sample belong to Ezviz. \\
\end{tabular}
\caption{SHAP values for (a) a training sample and (b) a testing sample of Ezviz using trained XGB in SmartCam Classifier.}
\label{fig:SHAP_Ezviz_traintestplots_6Class}
\end{figure*}

Now, we look at the SmartCam Classifier in search of explanations.  
XGB is trained with the dataset having six classes, i.e., AlarmClock, Canary, D3D, Ezviz, Netatmo, and V380. 
The trained model and the data samples (i.e., only X values) are provided as input to the SHAP explainer. 
We consider one data point from each of training (shown in Figure \ref{fig:SHAP_Ezviz_traintestplots_6Class} (a)) and testing (Figure \ref{fig:SHAP_Ezviz_traintestplots_6Class} (b)) in Ezviz class, where both the samples are correctly predicted as Ezviz.
The split ratio of dataset is 8:2.
Figure \ref{fig:SHAP_Ezviz_traintestplots_6Class} (a) shows that \textit{Pkt Len Min}, \textit{Bwd Pkt Len Max}, and \textit{Bwd Pkt Len Min} are among the most important features that positively support the correct classification in training. 
In fact, all the top ten features have positive SHAP values supporting the correct prediction.
Figure \ref{fig:SHAP_Ezviz_traintestplots_6Class} (b) shows that only seven out of the top ten features positively support the correct prediction in the test, and these features do not completely overlap with those in the training (Figure \ref{fig:SHAP_Ezviz_traintestplots_6Class} (a)).
Three features, namely \textit{Fwd IAT Tot}, \textit{TotLen Fwd Pkts}, and \textit{Fwd Act Data pkts}, show negative SHAP values.

We further investigate the SHAP values for the prediction of the other five classes for the same training sample as that in Figure \ref{fig:SHAP_Ezviz_traintestplots_6Class} (a); all these predictions are incorrect, yet we look for the feature overlap 
(The plots are not shown due to space constraints). 
We have observed the following in these cases:
\begin{enumerate}
    \item \textit{Fwd IAT Max} and \textit{Pkt Len Min} are top two features positively supporting the prediction for \enquote{AlarmClock}.

    \item \textit{TotLen Bwd Pkts} and \textit{Bwd Pkt Len Mean} are top two features positively supporting for  \enquote{Canary}.
    \item \textit{Init Bwd Win Byts}, and \textit{Bwd IAT Min} are top two features positively supporting for \enquote{D3D}.
    \item \textit{Bwd IAT Min} is the only one among the top ten features positively supporting for \enquote{Netatmo}.
    \item \textit{Fwd Header Len} and \textit{Pkt Len Min} are top two features positively supporting for  \enquote{V380}.
\end{enumerate}  
Further, each of these incorrect predictions has a significant number of features that have negative SHAP values, indicating negative support for the predictions. 
For example, \textit{Bwd Pkt Len Mean} and \textit{Fwd Pkt Len Max} have negative SHAP values for the prediction as AlarmClock.

Extending this study, we look into the SHAP values for the other five classes in case of the test sample (plots are not shown due to space constraint). 
This study basically provides a parity check of the features retracting the correct prediction. 
It turns out that,
\begin{enumerate}
    \item \textit{Bwd IAT Mean} and \textit{Flow IAT Min}, are two most positively supporting features for the prediction of \enquote{V380}.
    \item \textit{Pkt Len Min} and \textit{Flow Duration} for the prediction of \enquote{AlarmClock}. 
    \item \textit{Fwd Pkts/s} and \textit{Fwd IAT Mean}  for the prediction of  \enquote{Canary}. 
    \item \textit{Init Bwd Win Byts} and \textit{Bwd IAT Mean} for the prediction of \enquote{D3D}. 
     \item  \textit{Fwd Pkt Len Max} and \textit{Fwd Pkts/s}  for the prediction of \enquote{Netatmo}.
\end{enumerate}


\begin{figure*}[hbtp]
\centering
\includegraphics[width=\linewidth]{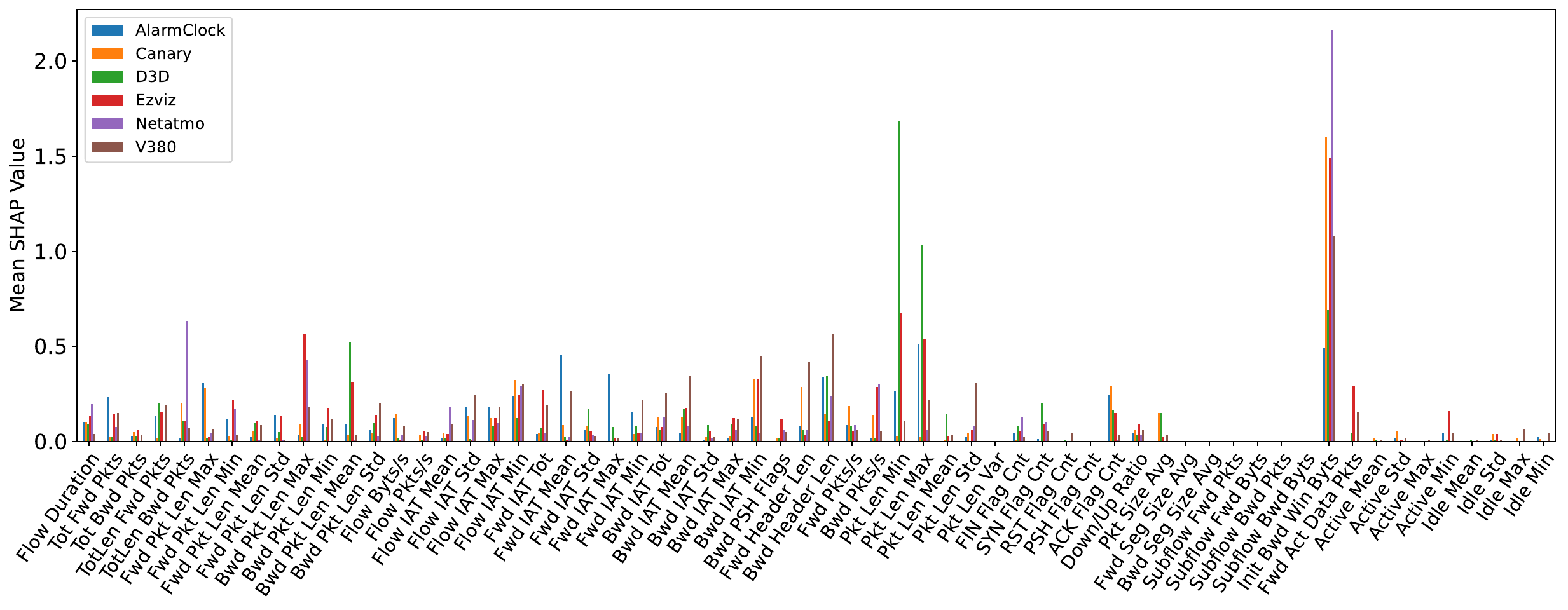}
\caption{Mean SHAP Values using XGB for SmartCam classifier.}
\label{fig:Mean_6Class}
\end{figure*}

Figure \ref{fig:Mean_6Class} depicts the mean SHAP values computed across all the training samples for each of the features for each of the six classes. 
Top two features having highest mean SHAP values are  \textit{Init Bwd Win Byts} and \textit{Pkt Len Min} that support the prediction of Ezviz; this is in line with the SHAP values in Figure \ref{fig:SHAP_Ezviz_traintestplots_6Class} (a) and Figure \ref{fig:SHAP_Ezviz_traintestplots_6Class} (b). 
Similarly, \textit{Pkt Len Min} and \textit{Pkt Len Max} have the highest mean SHAP values supporting the prediction of D3D, whereas  \textit{Pkt Len Max} and \textit{Init Bwd Win Byts} are the top features supporting the AlarmClock. 
The prediction of Canary has the highest support from \textit{Init Bwd Win Byts} and \textit{Bwd IAT Min}. 
Overall, it turns out that having only a limited number of features, e.g., top two, may not be sufficient to build enough explanation for the predictions of each the six cameras. 
Thus, we conjecture that as the number of IoT cameras increases, the number of features required to explain the prediction will be higher.

\subsection{LIME-Features for Flow Classifier}

\begin{table*}[hbtp]
\caption{Comparison of top ten SHAP- and LIME-features for Flow Classifier (\colorbox{green!30}{Green}- same and \colorbox{red!30}{Red}- differing features) for the same training sample as that in Figure \ref{fig:SHAP_traintestplots_4Class}. }
\label{tab:SHAPLIME_impfeature}
\resizebox{\linewidth}{!}{
\begin{tabular}{cc|cc|cc|cc}
\hline
\multicolumn{2}{c|}{\textbf{IoTCam}}           & \multicolumn{2}{c|}{\textbf{Others}}          & \multicolumn{2}{c|}{\textbf{Conf}}             & \multicolumn{2}{c}{\textbf{Share}}       \\ \hline
\textbf{SHAP}             & \textbf{LIME}              & \textbf{SHAP}             & \textbf{LIME}              & \textbf{SHAP}              & \textbf{LIME}              & \textbf{SHAP}          &   \textbf{LIME}            \\ \hline

\colorbox{green!30}{Init Bwd Win Byts} & \colorbox{red!30}{RST Flag Cnt} & \colorbox{green!30}{Init Bwd Win Byts} &\colorbox{red!30}{ RST Flag Cnt} & \colorbox{green!30}{Init Bwd Win Byts} & \colorbox{red!30}{RST Flag Cnt} & 
\colorbox{red!30}{Init Bwd Win Byts} & \colorbox{red!30}{RST Flag Cnt} \\

\colorbox{red!30}{Flow IAT Min}     & \colorbox{green!30}{Bwd Header Len}    & \colorbox{green!30}{Fwd Header Len}   & \colorbox{green!30}{Bwd Header Len}    &  \colorbox{red!30}{Bwd IAT Min}       & \colorbox{red!30}{Bwd PSH Flags}     & 
\colorbox{red!30}{Flow IAT Min}  & \colorbox{red!30}{Fwd Header Len}  \\

\colorbox{green!30}{Fwd Pkts/s}       & \colorbox{red!30}{Bwd PSH Flags}     & \colorbox{green!30}{Bwd IAT Min}      & \colorbox{red!30}{Bwd PSH Flags}     & \colorbox{red!30}{Flow IAT Min}      & \colorbox{red!30}{Fwd Header Len}    & 
\colorbox{red!30}{ACK Flag Cnt}  & \colorbox{red!30}{Bwd PSH Flags}   \\

\colorbox{green!30}{Fwd Header Len}   & \colorbox{green!30}{Fwd Header Len}    & \colorbox{green!30}{Bwd Pkt Len Std}  & \colorbox{green!30}{Fwd Header Len}    & \colorbox{red!30}{Fwd Pkt Len Std}   & \colorbox{green!30}{Bwd Header Len}    & \colorbox{red!30}{Flow Duration} & \colorbox{red!30}{Bwd IAT Min}     \\

\colorbox{green!30}{Bwd Header Len}   & \colorbox{green!30}{Init Bwd Win Byts} & \colorbox{green!30}{Bwd Header Len}   & \colorbox{green!30}{Bwd IAT Min}       & \colorbox{green!30}{Bwd Pkt Len Std}   & \colorbox{red!30}{Fwd Header Len}    & \colorbox{green!30}{Flow Byts/s}   & \colorbox{red!30}{Bwd Header Len}  \\

\colorbox{red!30}{Tot Bwd Pkts}     & \colorbox{red!30}{FIN Flag Cnt}      & \colorbox{red!30}{Pkt Len Max}      & \colorbox{red!30}{Fwd Header Len}    & \colorbox{green!30}{Bwd Header Len}    & \colorbox{red!30}{Fwd Pkt Len Mean}  & 
\colorbox{red!30}{Fwd Pkts/s}    & \colorbox{red!30}{Bwd Header Len}  \\

\colorbox{red!30}{Bwd IAT Min}      & \colorbox{red!30}{Bwd Pkt Len Std}   & \colorbox{red!30}{Bwd Pkt Len Max}  & \colorbox{red!30}{FIN Flag Cnt}      & \colorbox{red!30}{Bwd Pkt Len Max}   & \colorbox{green!30}{Bwd Pkt Len Std}   & \colorbox{red!30}{Flow IAT Mean} & \colorbox{red!30}{FIN Flag Cnt}    \\

\colorbox{red!30}{Bwd Pkts/s}       & \colorbox{green!30}{Bwd Header Len}    & \colorbox{red!30}{Flow IAT Min}     & \colorbox{green!30}{Init Bwd Win Byts} & \colorbox{red!30}{Fwd IAT Min}       & \colorbox{red!30}{FIN Flag Cnt}      & 
\colorbox{red!30}{Pkt Len Mean}  & \colorbox{red!30}{Bwd Pkt Len Std} \\

\colorbox{red!30}{Flow Duration}    & \colorbox{green!30}{Fwd Pkts/s}        & \colorbox{green!30}{Fwd Pkt Len Mean} & \colorbox{green!30}{Fwd Pkt Len Mean}  & \colorbox{red!30}{Bwd Pkts/s}        & \colorbox{green!30}{Init Bwd Win Byts} & 
\colorbox{red!30}{Bwd IAT Tot}   & \colorbox{red!30}{Pkt Len Max}     \\

\colorbox{red!30}{Bwd Pkt Len Mean} & \colorbox{green!30}{Fwd Header Len}    & \colorbox{red!30}{Pkt Len Min}      & \colorbox{green!30}{Bwd Pkt Len Std}   & \colorbox{red!30}{Fwd Act Data Pkts} & \colorbox{green!30}{Bwd Header Len}    &
\colorbox{red!30}{Bwd IAT Max}   & \colorbox{green!30}{Flow Byts/s}     \\ \hline
\end{tabular}}
\end{table*}

\begin{table*}[hbtp]
\centering
\caption{Comparison of top ten SHAP- and LIME-features for SmartCam classifier  (\colorbox{green!30}{Green}- same and \colorbox{red!30}{Red}- differing features).}
\label{tab:6Class_SHAPLIME_impfeature}
\resizebox{\linewidth}{!}{
\begin{tabular}{cc|cc|cc}
\hline
\multicolumn{2}{c|}{\textbf{AlarmClock}}      & \multicolumn{2}{c|}{\textbf{Canary}}           & \multicolumn{2}{c}{\textbf{D3D}}   \\ \hline 
\textbf{SHAP}             & \textbf{LIME}              & \textbf{SHAP}             & \textbf{LIME}           & \textbf{SHAP}             & \textbf{LIME}   \\ \hline
\colorbox{red!30}{Fwd IAT Max}      & \colorbox{red!30}{Fwd Header Len}    & \colorbox{red!30}{TotLen Bwd Pkts}   & \colorbox{green!30}{Init Bwd Win Byts} & \colorbox{red!30}{Bwd Pkt Len Mean}  & \colorbox{green!30}{Init Bwd Win Byts} \\
\colorbox{red!30}{Pkt Len Min}      & \colorbox{red!30}{Init Bwd Win Byts} & \colorbox{green!30}{Init Bwd Win Byts} & \colorbox{red!30}{Flow IAT Std}      & \colorbox{red!30}{Pkt Len Max}       & \colorbox{red!30}{Fwd Header Len} \\
\colorbox{red!30}{Fwd Pkt Len Min}  & \colorbox{red!30}{Active Std}        & \colorbox{red!30}{Bwd Pkt Len Mean}  & \colorbox{red!30}{Active Std}        & \colorbox{red!30}{Pkt Len Min}       & \colorbox{red!30}{Active Std}   \\
\colorbox{red!30}{Bwd Pkt Len Mean} & \colorbox{red!30}{ACK Flag Cnt}      & \colorbox{green!30}{ACK Flag Cnt}      & \colorbox{green!30}{Bwd IAT Min}       & \colorbox{red!30}{Pkt Len Mean}      & \colorbox{red!30}{ACK Flag Cnt}  \\
\colorbox{red!30}{Fwd Pkt Len Max}  & \colorbox{red!30}{Bwd IAT Min}       & \colorbox{green!30}{Bwd IAT Min}       & \colorbox{red!30}{RST Flag Cnt}      & \colorbox{green!30}{Init Bwd Win Byts} & \colorbox{red!30}{RST Flag Cnt}  \\
\colorbox{green!30}{Flow IAT Std}     & \colorbox{red!30}{RST Flag Cnt}      & \colorbox{red!30}{Fwd Pkt Len Min}   & \colorbox{green!30}{ACK Flag Cnt}      & \colorbox{red!30}{TotLen Fwd Pkts}   & \colorbox{green!30}{Bwd Header Len} \\
\colorbox{red!30}{Fwd IAT Tot}      & \colorbox{red!30}{Init Bwd Win Byts} &  \colorbox{red!30}{Bwd IAT Mean}      & \colorbox{red!30}{Flow IAT Min}      & \colorbox{red!30}{TotLen Bwd Pkts}   & \colorbox{red!30}{Flow IAT Std}  \\
\colorbox{red!30}{Flow Duration}    & \colorbox{red!30}{Bwd Header Len}    & \colorbox{red!30}{Flow Duration}     & \colorbox{red!30}{Flow IAT Max}      & \colorbox{green!30}{Bwd Header Len}    & \colorbox{green!30}{Bwd IAT Min}  \\
\colorbox{red!30}{Bwd Pkt Len Min}  & \colorbox{green!30}{Flow IAT Std}      & \colorbox{red!30}{Bwd Header Len}    & \colorbox{red!30}{Active Std}        & \colorbox{green!30}{Bwd IAT Min}       & \colorbox{red!30}{Flow IAT Max}  \\
\colorbox{red!30}{Flow Byts/s}      & \colorbox{red!30}{Active Std}        & \colorbox{red!30}{Fwd IAT Mean}      & \colorbox{red!30}{Fwd Header Len}    & \colorbox{red!30}{Fwd Pkt Len Mean}  & \colorbox{red!30}{Flow IAT Min}  \\ \hline \hline 
\end{tabular}}

\resizebox{\linewidth}{!}{
\begin{tabular}{cc|cc|cc}
\hline
\multicolumn{2}{c|}{\textbf{Ezviz}}            & \multicolumn{2}{c|}{\textbf{Netatmo}}          & \multicolumn{2}{c}{\textbf{V380}}              \\ \hline
\textbf{SHAP}             & \textbf{LIME}             & \textbf{SHAP}             & \textbf{LIME}             & \textbf{SHAP}             & \textbf{LIME}   \\ \hline
\colorbox{red!30}{Pkt Len Min}       & \colorbox{green!30}{Init Bwd Win Byts} & \colorbox{green!30}{Init Bwd Win Byts} & \colorbox{green!30}{Init Bwd Win Byts} & \colorbox{green!30}{Bwd IAT Min}       & \colorbox{green!30}{Bwd IAT Min}       \\
\colorbox{red!30}{Bwd Pkt Len Max}   & \colorbox{green!30}{Init Bwd Win Byts} & \colorbox{green!30}{TotLen Bwd Pkts}   & \colorbox{red!30}{Active Std}        & \colorbox{red!30}{Pkt Len Std}       & \colorbox{green!30}{Init Bwd Win Byts} \\
\colorbox{red!30}{Bwd Pkt Len Min}   & \colorbox{red!30}{RST Flag Cnt}      & \colorbox{red!30}{Flow IAT Mean}     & \colorbox{red!30}{RST Flag Cnt}      & \colorbox{green!30}{Fwd Header Len }   & \colorbox{red!30}{RST Flag Cnt}      \\
\colorbox{red!30}{TotLen Bwd Pkts}   & \colorbox{red!30}{Active Std}       & \colorbox{green!30}{Flow IAT Min}      & \colorbox{red!30}{Fwd Header Len}    & \colorbox{red!30}{Flow IAT Max}      & \colorbox{red!30}{Active Std}       \\
\colorbox{red!30}{TotLen Fwd Pkts}   & \colorbox{red!30}{ACK Flag Cnt}      & \colorbox{red!30}{Fwd Pkt Len Min}   & \colorbox{green!30}{Flow IAT Min}      & \colorbox{red!30}{Bwd Pkt Len Std}   & \colorbox{green!30}{Init Bwd Win Byts} \\
\colorbox{red!30}{Bwd IAT Min}       & \colorbox{red!30}{Flow IAT Std}      & \colorbox{red!30}{Fwd Pkts/s}        & \colorbox{red!30}{ACK Flag Cnt}      & \colorbox{green!30}{Init Bwd Win Byts} & \colorbox{red!30}{ACK Flag Cnt}      \\
\colorbox{red!30}{Fwd Pkt Len Std}   & \colorbox{red!30}{Active Std}        & \colorbox{red!30}{Bwd IAT Min}       & \colorbox{red!30}{Fwd Header Len}    & \colorbox{red!30}{Bwd IAT Mean}      & \colorbox{red!30}{Bwd Header Len}    \\
\colorbox{green!30}{Init Bwd Win Byts} & \colorbox{red!30}{Flow IAT Max}      & \colorbox{red!30}{Bwd Pkts/s}        &\colorbox{red!30}{ACK Flag Cnt}      & \colorbox{red!30}{Bwd Pkt Len Max}   & \colorbox{red!30}{ACK Flag Cnt}      \\
\colorbox{red!30}{Fwd Act Data Pkts} & \colorbox{red!30}{Flow IAT Min}      & \colorbox{red!30}{SYN Flag Cnt}      & \colorbox{red!30}{Bwd Header Len}    & \colorbox{red!30}{Pkt Len Min}       & \colorbox{red!30}{Active Std}        \\
\colorbox{red!30}{Fwd IAT Tot}       & \colorbox{red!30}{Fwd Header Len}    & \colorbox{red!30}{Fwd Pkt Len Max}   & \colorbox{green!30}{TotLen Bwd Pkts}   & \colorbox{red!30}{Bwd Pkt Len Mean}  & \colorbox{green!30}{Fwd Header len}    \\ \hline
\end{tabular}}
\end{table*}

\begin{figure*}[hbt!]
\centering
\begin{tabular}{cc}
\includegraphics[width=0.48\linewidth]{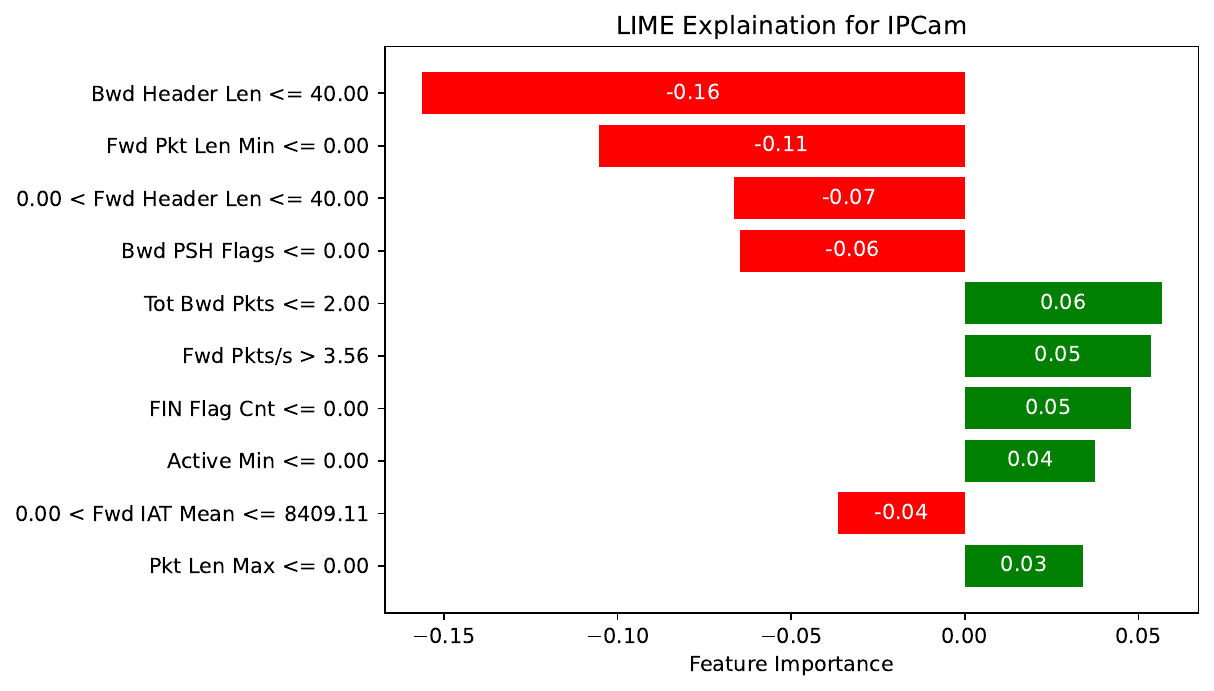} &
\includegraphics[width=0.48\linewidth]{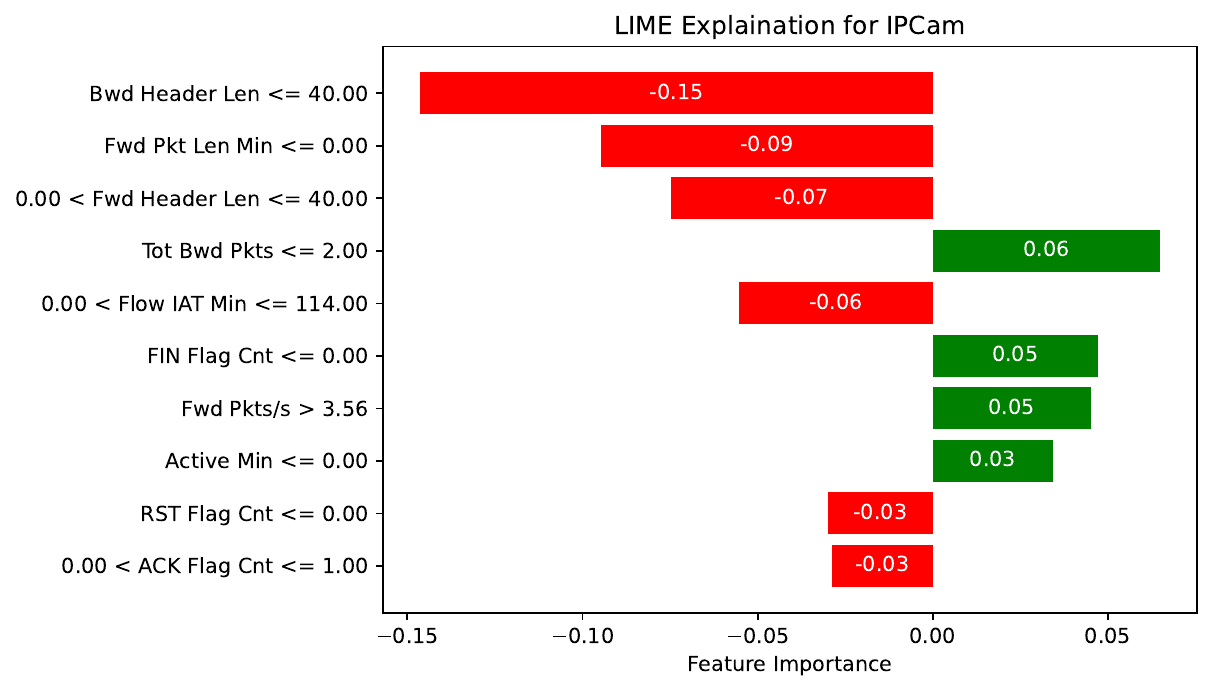} \\ 
(a) Training IoTCam & (b) Testing IoTCam\\
\end{tabular}
\caption{LIME explanation using XGB model in Flow Classifier.}
\label{fig:Lime_plots_4Class_appendix}
\end{figure*}

We extend our study to include a LIME explainer for both Flow and SmartCam classifiers. 
Our aim is to find additional explanations for the classifiers and contrast the important features. 
To maintain uniformity, we consider the trained XGB model and the same datasets for training and testing in both classifications. 
Basically, LIME explainer perturbs input data samples and observes changes in predictions, providing insights into model behavior. 
Similar to SHAP, an IoTCam sample is considered from each of train an test datasets to maintain uniformity. 
Table \ref{tab:SHAPLIME_impfeature} shows the top features in SHAP and LIME for the correct prediction of IoTCam sample in training.  
Features like \textit{RST Flag Cnt}, \textit{Bwd Header Len}, \textit{Bwd PSH Flags}, and \textit{Fwd Header Len} show the highest support in LIME. 
A feature may appear twice in the list of LIME-features because the feature may have two range of values that supports the prediction.
For example, \textit{Fwd Header Len} appears twice in the list to support IoTCam. 
Basically, a feature and its specific range of values form a condition, indicating the support for the prediction of a class. 
Figure \ref{fig:Lime_plots_4Class_appendix} (a) shows the top features along with their corresponding range to support the prediction.
For example, when the \textit{Bwd Header Len} is less than or equal to 40.00, it negatively contributes about 49.184 units to support IoTCam.
Similarly, when the feature\textit{RST Flag Cnt} is less than or equal to 0.00, it contributes approximately 70.196 units to support IoTCam in the testing (Figure \ref{fig:Lime_plots_4Class_appendix} (b)). 

Further, \textit{Bwd Header Len} appears twice with two different ranges of values to negatively support for IoTCam, one as \textit{Bwd Header Len $\leq$ 40.00: 49.18} and another as \textit{Bwd Header Len $>$ 184.00: 18.28}. 
It indicates that\textit{Bwd Header Len} having a value less than or equal to 40 bytes positively contributes approximately 49.18 units for IoTCam. 
And, \textit{Bwd Header Len} having a value greater than 184.00 positively contributes approximately 18.28 units for  IoTCam. 
In other words, both very small and very large values of \textit{Bwd Header Len} have a positive support on the likelihood of the instance being IoTCam. 
Overall, features like \textit{Tot Bwd Pkts}, \textit{Fwd Pkts/s}, \textit{FIN Flag Cnt},\textit{Active Min}, and \textit{Pkt Len Max} positively support the prediction of IoTCam, whereas the features like  \textit{Fwd Pkt Len Min}, \textit{Fwd Header Len}, \textit{Bwd PSH Flags}, negatively supports the prediction of IoTCam.
Thus, this explainer provides a more nuanced understanding of the features impact across its spectrum of values. 
We have analyzed the LIME features for the prediction of the same train and test samples for the other three classes (Plots are not reported due to space constraints). 
A summary of the features in these cases of predictions is shown in right most three columns in Table \ref{tab:SHAPLIME_impfeature}.

\subsection{LIME-Features for SmartCam Classifier}
\begin{figure*}[hbt!]
\centering
\begin{tabular}{cc}
\includegraphics[width=0.48\linewidth]{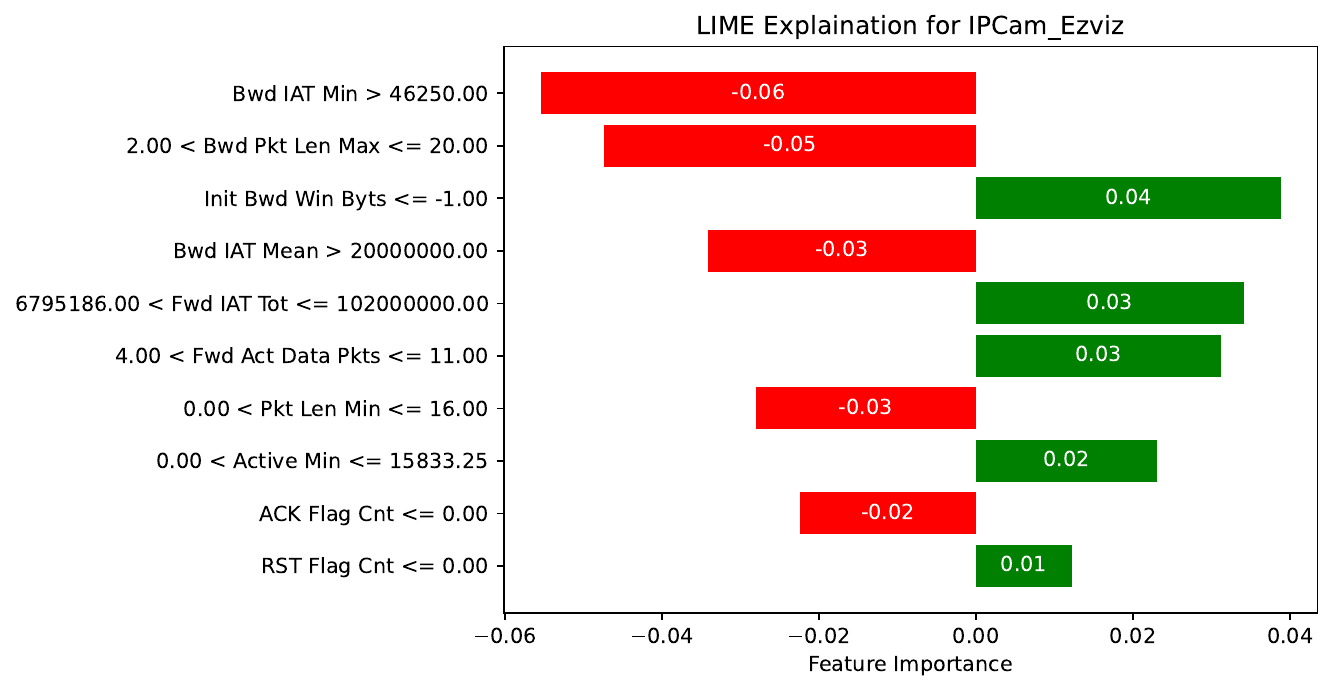} &
\includegraphics[width=0.48\linewidth]{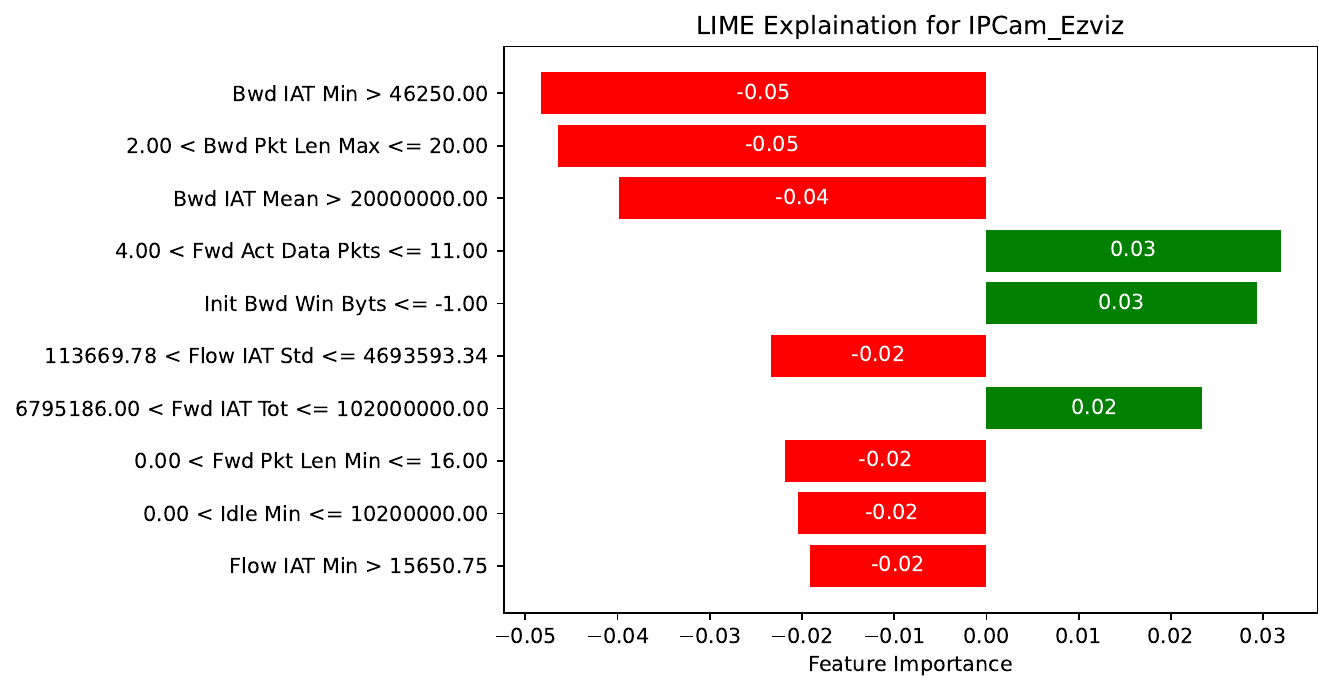} \\ 
(a) Training Ezviz & (b) Testing Ezviz\\
\end{tabular}
\caption{LIME explanation using XGB model in SmartCam Classifier.}
\label{fig:Lime_plots_6Class_appendix}
\end{figure*}

Table \ref{tab:6Class_SHAPLIME_impfeature} shows the top features based on the LIME explainer for the SmartCam Classifier. 
The top features include \textit{Fwd Header Len}, \textit{Init Bwd Win Byts}, \textit{Active Std} and \textit{ACK Flag Cnt} across all six classes. 
Similar to LIME in Flow Classifier, a same feature appears twice or more indicating more than one value range positively supporting a class, e.g., \textit{Init Bwd Win Byts} appears twice to support AlarmClock with \textit{Init Bwd Win Byts $\leq$ -1.00: 1.680} and \enquote{23.00 $<$ Init Bwd Win Byts $\leq$ 1383.00: 1.124}. 
Figure \ref{fig:Lime_plots_6Class_appendix} shows the top LIME-features along with their value range to support the prediction. 
For example, when the \textit{Bwd IAT Min} is greater than 46250.00, it negatively contributes about 1.17 units to support Ezviz in training (Figure \ref{fig:Lime_plots_6Class_appendix}a). 
Similarly, when the feature \textit{Flow IAT Min} is greater than 15650.75, it negatively contributes 2.10 units to support Ezviz in testing (Figure \ref{fig:Lime_plots_6Class_appendix}b). 
(The plots for LIME values to support the prediction as other five classes for the Ezviz sample are in Figure \ref{fig:LIME_traintestplots_6Class_appendix} in Appendix \ref{subsec:appendixB}).
Essentially, the discretization process in the LIME explainer can result in the same feature appearing multiple times with different condition values, and this helps to better interpret the nuanced ways in which specific value ranges of a feature influence the model's predictions. 
In general, compared to LIME, the SHAP explainer is computationally intensive, yet it often provides more consistent explanations. 

\subsection{Summary of SHAP- and LIME Features}
Table \ref{tab:SHAPLIME_impfeature} and Table \ref{tab:6Class_SHAPLIME_impfeature} show the top ten SHAP- and LIME-features for Flow and SmartCam classifiers respectively.
\begin{itemize}
    \item \textbf{Flow Classifier :}  
    One of the important features is \textit{Init Bwd Win Byts} that has a higher mean SHAP value as 1.26 to classify \enquote{IoTCam}, 0.8066 to classify  \enquote{Others}, 0.6961 to classify \enquote{Conf}, and 0.6810 to classify  \enquote{Share}. 
    Other important features include  \textit{Flow IAT Min} having mean SHAP value as 1.82 for \enquote{IoTCam},  \textit{Bwd IAT Min} with mean SHAP values as 0.40 and 0.33 support  \enquote{Conf} and \enquote{Others} respectively. 
    Alternately, there are some features, like \textit{Pkt Len Min} and \textit{TotLen Bwd Pkts}, that have a higher mean SHAP value to classify only classes, like D3D and Netatmo resp., and have very low values to support the other classes. 

    LIME explainer provides average absolute weights across the instances. 
    The top most feature is \textit{RST Flag Cnt} having LIME value $(\leq 0.00)$ with weights as 70.12,  563.83, 519.24, and 49.71 to support \enquote{IoTCam}, \enquote{Others}, \enquote{Conf} and \enquote{Share} respectively. 
    Another important feature is \textit{Bwd Header Len} having LIME value $(\leq 40.00)$ with weights as 49.18, 428.22, 294.41 and 24.32 to classify  \enquote{IoTCam}, \enquote{Others},  \enquote{Conf} and  \enquote{Share} respectively.
    Among others, \textit{Bwd PSH Flags} positively supports IoTCam, together with \textit{Fwd Header Len} and \textit{FIN Flag Cnt}. 
    
    \item \textbf{SmartCam Classifier :}  
    Similar to flow classifier,\textit{Init Bwd Win Byts} is one of most important features that has a mean SHAP value of about 2.20 for the classification of \enquote{Netatmo}, 1.5872 for \enquote{Canary}, 1.4355 for \enquote{Ezviz}, 1.06 for \enquote{V380}, 0.75 for \enquote{D3D}, and 0.4850 for \enquote{AlarmClock}.
    Another feature that has a higher mean SHAP value (1.65) to classify D3D is \textit{Pkt Len Min}; similarly, \textit{Pkt Len Max} has 0.50 and 0.42 for \enquote{Ezviz} and  \enquote{AlarmClock}, and \textit{Bwd IAT Min} has 0.42 for \enquote{V380}. 
    
    The most important LIME-feature is \textit{Init Bwd Win Byts} with varying condition, like $(\leq -1.00)$ with weight as 13.8198 to support V380, and 12.8005 to support D3D, 1.68 to support AlarmClock, whereas $(23.00 < Init Bwd Win Byts \leq 1383.00)$ supports Canary, Ezviz, Netatmo and V380 with weights as 3.20, 11.97, 5.53 and 12.93 respectively. 
    Another important feature is \textit{Fwd Header Len}, having $(\leq 24.00)$ with weights as 1.93, 9.67, 3.47 and 3.25 supports AlarmClock, D3D, Ezviz, and Netatmo respectively. 
    The features of \textit{Active Std}, \textit{ACK Flag Cnt}, and \textit{Flow IAT Std} have positive supports only for IoTCam.
    
\end{itemize}

\subsection{Faithfulness of Explainers}
To assess the usability of SHAP and LIME in the classification of flows and the detection of IoT (spy) cameras can be achieved through an analysis of faithfulness, i.e., using consistency and sufficiency \cite{Dasgupta2022PMLR}, of the explainers. 

\begin{itemize}
    \item \textbf{Consistency:} An XAI explainer is considered consistent if the ranks of the features based on the explainer and a consistency measurement metric, e.g., Spearman's Correlation Interpretation (SCI), has similar even in the presence of data perturbations \cite{silva2024arXiv}. 
    In general, SCI value ranges in [-1, 1], where a value for an explainer approaching 1 is considered as more consistent explainer. 
    In the Flow Classifier, overall consistency turns out to be 0.8111 and 0.7413 using SHAP and LIME, respectively, whereas the SmartCam Classifier has consistency as 0.7101 and 0.8584 in SHAP and LIME, respectively. 
    Further, to assess the consistency of individual features, we have found that the features like \textit{Fwd Header Len} and \textit{Bwd Header Len} have a score as 1.0, whereas the features like \textit{Init Bwd Win Byts}, and \textit{Fwd Pkts/s} have scores about 0.8 in Flow classifier using SHAP. 
    Similarly \textit{Flow IAT Std}, and  \textit{Fwd IAT Tot} have scores as 1.0, whereas \textit{Init Bwd Win Byts} and  \textit{Bwd Pkt Len Max} have about 0.94 using SHAP in the SmartCam classifier.
    Thus, consistency indicates a strong agreement in feature importance rankings, meaning the explanations are fairly stable despite perturbations in the data. 
    
    \item \textbf{Sufficiency :} Sufficiency provides a measure of how well a subset of important features can predict the original output of the model. 
    A higher sufficiency indicates that the most influential features identified by SHAP or LIME are enough to explain the performance of a model. 
    The sufficiency score turns out to be  1.00 in both SHAP and LIME explainers for both Flow and SmartCam classifiers.
    Hence, we consider that the explanations produced by the results in this paper are reliable and trustworthy. 
\end{itemize}

\subsection{Comparative Analysis with Prior Work}
\begin{table*}[h]
\caption{A comparative analysis of the flow-based features between the prior related works in IoT systems and our framework.}
\label{tab:comp_analysis}
\resizebox{\linewidth}{!}{\begin{tabular}{c|c|c|c|c|c|cccc|c}
\hline
\multirow{2}{*}{\begin{tabular}[c]{@{}c@{}}Research\\ Paper\end{tabular}} & \multirow{2}{*}{\begin{tabular}[c]{@{}c@{}}IoT (Spy)\\ Camera Detection\end{tabular}} & \multirow{2}{*}{\begin{tabular}[c]{@{}c@{}}Flow-based \\ Features\end{tabular}} & \multirow{2}{*}{\begin{tabular}[c]{@{}c@{}}XAI\\ support\end{tabular}} & \multirow{2}{*}{Classification} & \multirow{2}{*}{\begin{tabular}[c]{@{}c@{}}ML \\ Techniques\end{tabular}} & \multicolumn{4}{c|}{Performance Metrics (\%)} & \multirow{2}{*}{\begin{tabular}[c]{@{}c@{}}Faithfullness\\ of Explainer\end{tabular}} \\ \cline{7-10}
 &  &  &  &  &  & \multicolumn{1}{c|}{Accuracy} & \multicolumn{1}{c|}{Precision} & \multicolumn{1}{c|}{Recall} & FNR &  \\ \hline \hline
 
\cite{Arunan2018IEEE} & \xmark & \cmark & \xmark & 28-Classes & \begin{tabular}[c]{@{}c@{}}NB, RF\end{tabular} & \multicolumn{1}{c|}{99} & \multicolumn{1}{c|}{\xmark} & \multicolumn{1}{c|}{\xmark} & \xmark & \xmark \\ 
\cite{PratibhaIET2021} & \xmark & \cmark & \xmark & \xmark & \xmark & \multicolumn{1}{c|}{\xmark} & \multicolumn{1}{c|}{\xmark} & \multicolumn{1}{c|}{98} & \xmark & \xmark\\ 
\cite{bezawada2018ACM} & \xmark & \cmark & \xmark & \begin{tabular}[c]{@{}c@{}}7-Device Categories\end{tabular} & \begin{tabular}[c]{@{}c@{}}KNN, DT, GBRT, \\ Majority Voting\end{tabular} & \multicolumn{1}{c|}{99} & \multicolumn{1}{c|}{99} & \multicolumn{1}{c|}{99} & \xmark & \xmark\\ 
\cite{gummadi2024xaiIEEEACCESS} & \xmark & \cmark & \cmark & \begin{tabular}[c]{@{}c@{}}3-Class, 6-Class\end{tabular} & \begin{tabular}[c]{@{}c@{}}DT, DNN,ADA, \\ SVM, MLP, RF\end{tabular} & \multicolumn{1}{c|}{72} & \multicolumn{1}{c|}{71} & \multicolumn{1}{c|}{72} & \xmark & \xmark\\ 
Our Approach & \cmark & \cmark & \cmark & \begin{tabular}[c]{@{}c@{}}4-Class, 6-Class\end{tabular} & \begin{tabular}[c]{@{}c@{}}DT, KNN, ET, NB,\\ LR,RF,XGB,ADA\end{tabular} & \multicolumn{1}{c|}{99.21} & \multicolumn{1}{c|}{99.22} & \multicolumn{1}{c|}{99.21} & 0.7 & \cmark\\ \hline
\end{tabular}}
\end{table*}
Existing works on IoT device classification have primarily focused on identifying the most effective ML model for classification. In this section, we provide a comparative analysis between our proposed approach and previously developed techniques for IoT camera classification. 
Table \ref{tab:comp_analysis} summarizes the key differences in types of features, types of IoT devices, number of IoT devices/classes, ML models, and their performances and the use of XAI techniques with the faithfulness of explainers.

The work in \cite{Arunan2018IEEE} has focused on the classification of a variety of IoT devices, 28 in total, including IoT cameras, Smart Bulbs, and plugs, using simpler models like Naive Bayes and Random Forest, and have achieved about 99\% accuracy. The work in \cite{bezawada2018ACM} have used flow-based features, but a reduced set compared to \cite{Arunan2018IEEE}, to perform classification of 7 IoT devices using a wider range of ML models, including gradient boosting based models, and has achieved about 99\% accuracy. 
The work in \cite{PratibhaIET2021} has used flow-based features for device classification for the 18 devices as in \cite{Arunan2018IEEE}, but without using any traditional ML models, yet achieves about 98\% recall. 
None of these three works has considered the use of XAI or similar to enhance the usability of their results.  
Among others, the most relevant work to compare with our work is in \cite{gummadi2024xaiIEEEACCESS}, which has used XAI for anomaly detection in IoT networks. The work has used the XAI framework to explain the performance of models like decision trees, deep neural networks, and multilayer perceptrons, where the models could achieve about 72\% accuracy. Further, the work has considered nine classes in N-BaIoT dataset \cite{meidan2018IEEE}.

Compared with the state-of-the-art, we have mainly focused on explainability both with and without the XAI framework. 
In particular, we have shown the global explanations of each of the traditional models in terms of the most important features and then contrasted them with the features using XAI, in a variety of traditional and advanced models. 
Further, we have discovered the features that explain the local performance of the most efficient model, i.e., the features that contribute, both positively and negatively, to predict a specific class of IoTCam or a specific IoT camera device, like D3D, Netatmo, and Spy bulb.

\section{Conclusion}
\label{sec:conclusion}
In this paper, we address the problem of establishing reliability and trust on the classification of flow and detecting of IoT (spy) cameras using advanced supervised machine learning (ML) models. 
We have designed and developed a system, called rCamInspector, that utilizes Flow and SmartCam classifiers to classify the flows and IoT (spy) cameras respectively, and employ Explainable AI (XAI) to provide a reliable and trustworthy explanation on the model output. 
Each of the classifiers uses eight supervised models, namely, DT, kNN, NB, LR, RF, XGB, ET, and AB, to find the best performing model for the classifier.
Flow classifier classifies a flow into one of four classes, namely IoTCam, Share, Conf and Others. 
SmartCam Classifier classifies a IoTCam flow into one of six IoT (spy) cameras, namely, D3D, Netatmo, Spy Clock, Canary, Ezviz, V380 Spy Bulb. 
Using a dataset of about 38GB, we have shown that XGB performs best in both the classifiers with about 92\% and 99\% accuracies respectively. 
Using two XAI explainers, namely SHAP and LIME, we have shown that a distinctive set of features can support the classification in either Flow or SmartCam classifier. 
For example, \textit{Init Bwd Win Byts} and \textit{Fwd Pkts/s} features positively support the IoTCam class in both SHAP and LIME, and \textit{Init Bwd Win Byts} and \textit{Fwd IAT Tot} features positively supports the Ezviz class in both SHAP and LIME. 
Further, our results show that the set of most important SHAP- and LIME-features need not be exactly same to support any particular class. 
Then, using two faithfulness metrics, namely, consistency and sufficiency, we show that both SHAP and LIME have produced trustworthy explanations for Flow and SmartCam Classifiers, where consistency values are 0.8111 and 0.7101 in SHAP and 0.7413 and 0.8584 in LIME. 
Finally, we have compared our work with state-of-the-art and shown that rCamInspector achieved a significantly better precision  (about 99\%) and very less false negative rate (about 0.7\%) in SmartCam classifier.
In future, we plan to extend this work to include a wider range of IoT devices and open-sourced datasets to generalize the applicability of rCamInspector.


\bibliographystyle{IEEEtran}
\bibliography{references}


\appendix
\begin{appendices}
\label{sec:appendix}
\renewcommand{\thesubsection}{\Alph{subsection}}

\section{Flow-based Features and Description}
\label{sec:appendixA}
\begin{table}[h]
\caption{CICFlowMeter Feature and description (A total of 63 Features were used in rCamInspector, whereas a total of 14 features were discarded while preprocessing (Mentioned in \textcolor{red}{Red} color)).}
\scalebox{0.82}{
\begin{tabular}{|p{0.5cm}|p{2.5cm}|p{6.0cm}|}
\hline
S.N. & Name & Description\\
 \hline
 \hline
1 & Flow Duration & Duration of flow in microsecond\\
2 & Tot Fwd Pkts & Number of packets in forward direction \\
3 & Tot Bwd Pkts & Number of packets in backward direction \\
4 & TotLen Fwd Pkts & Total size (in bytes) of packets in forward direction \\
5 & TotLen Bwd Pkts & Total size (in bytes) of packets in backward direction \\ 
6 & Fwd Pkt Len Max & Maximum size (in bytes) of packet in forward direction \\
7 & Fwd Pkt Len Min & Minimum size (in bytes) of packet in forward direction \\
8 & Fwd Pkt Len Mean & Mean size (in bytes) of packet in forward direction \\ 
9 & Fwd Pkt Len Std & Standard deviation size of packet in forward direction \\
10 & Bwd Pkt Len Max & Maximum size (in bytes) of packet in backward direction \\
11 & Bwd Pkt Len Min & Minimum size (in bytes) of packet in backward direction \\
12 & Bwd Pkt Len Mean & Mean size (in bytes) of packet in backward direction \\ 
13 & Bwd Pkt Len Std	& Standard deviation size of packet in backward direction \\ 
14 & Flow Byte/s & Number of flow bytes per second \\
15 & Flow Packets/s & Number of flow packets per second \\ 
16 & Flow IAT Mean & Mean time between two packets sent in the flow \\ 
17 & Flow IAT Std & Standard deviation time between two packets sent in the flow \\
18 & Flow IAT Max & Maximum time between two packets sent in the flow \\
19 & Flow IAT Min & Minimum time between two packets sent in the flow \\ 
20 & Fwd IAT Max & Maximum time between two packets sent in the forward direction \\
21 & Fwd IAT Min & Minimum time between two packets sent in the forward direction \\
22 & Fwd IAT Mean & Mean time between two packets sent in the forward direction \\ 
23 & Fwd IAT Std & Standard deviation time between two packets sent in the forward direction \\
24 & Fwd IAT Total & Total time between two packets sent in the forward direction \\ 
25 & Bwd IAT Min & Minimum time between two packets sent in the backward direction \\ 
26 & Bwd IAT Max & Maximum time between two packets sent in the backward direction \\
27 & Bwd IAT Mean & Mean time between two packets sent in the backward direction \\
28 & Bwd IAT Std & Standard deviation time between two packets sent in the backward direction \\
29 & Bwd IAT Total & Total time between two packets sent in the backward direction \\ 
30 & Fwd Header Length & Total bytes used for headers in the forward direction \\
31 & Bwd Header Length & Total bytes used for headers in the backward direction \\
32 & FWD Packets/s & Number of forward packets per second \\
33 & Bwd Packets/s & Number of backward packets per second \\
34 & Min Packet Length & Minimum length of a packet \\ 
35 & Max Packet Length & Maximum length of a packet \\ 
36 & Packet Length Mean & Mean length of a packet \\ 
37 & Packet Length Std & Standard deviation length of a packet \\
38 & Packet Length Variance & Variance length of a packet \\ 
39 & Down/Up Ratio	& Download and upload ratio \\ \hline
\end{tabular}}
\label{tab:cicflowmeter-feature}
\end{table}

\begin{table}[h]
\scalebox{0.82}{
\begin{tabular}{|p{0.5cm}|p{3.2cm}|p{6.0cm}|} 
\hline

40 & Average Packet Size & Average size of packet \\ 
41 & Fwd Header Len & Total bytes used for headers in the forward direction \\
42 & Avg Fwd Segment Size & Average size observed in the forward direction \\
43 & AVG Bwd Segment Size & Average number of bytes bulk rate in the backward direction \\ 
44 & Bwd PSH Flag & Number of times the PSH flag was set in packets travelling in the backward direction (0 for UDP)\\ 
45 & FIN Flag Count & Number of packets with FIN \\
46 & SYN Flag Count & Number of packets with SYN \\
47 & RST Flag Count & Number of packets with RST \\
48 & PSH Flag Count & Number of packets with PUSH \\
49 & ACK Flag Count & Number of packets with ACK \\
50 & Subflow Fwd Packets & The average number of packets in a sub-flow in the forward direction \\
51 & Subflow Fwd Bytes & The average number of bytes in a sub-flow in the forward direction \\
52 & Subflow Bwd Packets & The average number of packets in a sub-flow in the backward direction \\
53 &  Subflow Bwd Bytes &  The average number of bytes in a sub-flow in the backward direction  \\
54 &  Init\_Win\_bytes\_backward &  The total number of bytes sent in initial window in the backward direction \\
55& Act\_data\_pkt\_forward & Count of packets with at least 1 byte of TCP data payload in the forward direction \\
56 & Active Min	& Minimum time a flow was active before becoming idle\\
57 & Active Mean & Mean time a flow was active before becoming idle \\
58 & Active Max & Maximum time a flow was active before becoming idle \\
59 & Active Std	& Standard deviation time a flow was active before becoming idle \\
60 & Idle Min &	Minimum time a flow was idle before becoming active \\
61 & Idle Mean & Mean time a flow was idle before becoming active \\ 
62 & Idle Max & Maximum time a flow was idle before becoming active \\
63 & Idle Std & Standard deviation time a flow was idle before becoming active \\
64 & \textcolor{red}{Fwd PSH flag} & \textcolor{red}{Number of times the PSH flag was set in packets travelling in the forward direction (0 for UDP)} \\
65 & \textcolor{red}{Fwd URG Flag} & \textcolor{red}{Number of times the URG flag was set in packets travelling in the forward direction (0 for UDP)} \\ 
66 & \textcolor{red}{Bwd URG Flag} & \textcolor{red}{Number of times the URG flag was set in packets travelling in the backward direction (0 for UDP)} \\
67 & \textcolor{red}{URG Flag Count} & \textcolor{red}{Number of packets with URG} \\
68 & \textcolor{red}{CWR Flag Count} & \textcolor{red}{Number of packets with CWE} \\
69 & \textcolor{red}{ECE Flag Count} & \textcolor{red}{Number of packets with ECE} \\ 
70 & \textcolor{red}{Fwd Avg Bytes/Bulk} & \textcolor{red}{Average number of bytes bulk rate in the forward direction} \\
71 & \textcolor{red}{Fwd AVG Packet/Bulk} & \textcolor{red}{Average number of packets bulk rate in the forward direction} \\
72 & \textcolor{red}{Fwd AVG Bulk Rate} & \textcolor{red}{Average number of bulk rate in the forward direction} \\
73 & \textcolor{red}{Bwd Avg Bytes/Bulk} & \textcolor{red}{Average number of bytes bulk rate in the backward direction} \\
74 & \textcolor{red}{Bwd AVG Packet/Bulk} & \textcolor{red}{Average number of packets bulk rate in the backward direction} \\
75 & \textcolor{red}{Bwd AVG Bulk Rate} & \textcolor{red}{Average number of bulk rate in the backward direction} \\ 
76 & \textcolor{red}{Init\_Win\_bytes\_forward} & \textcolor{red}{The total number of bytes sent in initial window in the forward direction} \\
77 & \textcolor{red}{Min\_seg\_size\_forward} & \textcolor{red}{Minimum segment size observed in the forward direction} \\ 

\hline
\end{tabular}}
\end{table}

\newpage
\section{SHAP and LIME Figures}
\label{sec:appendixB}
\begin{figure*}[htbp]
\begin{tabular}{cc}
\includegraphics[width=0.5\linewidth]{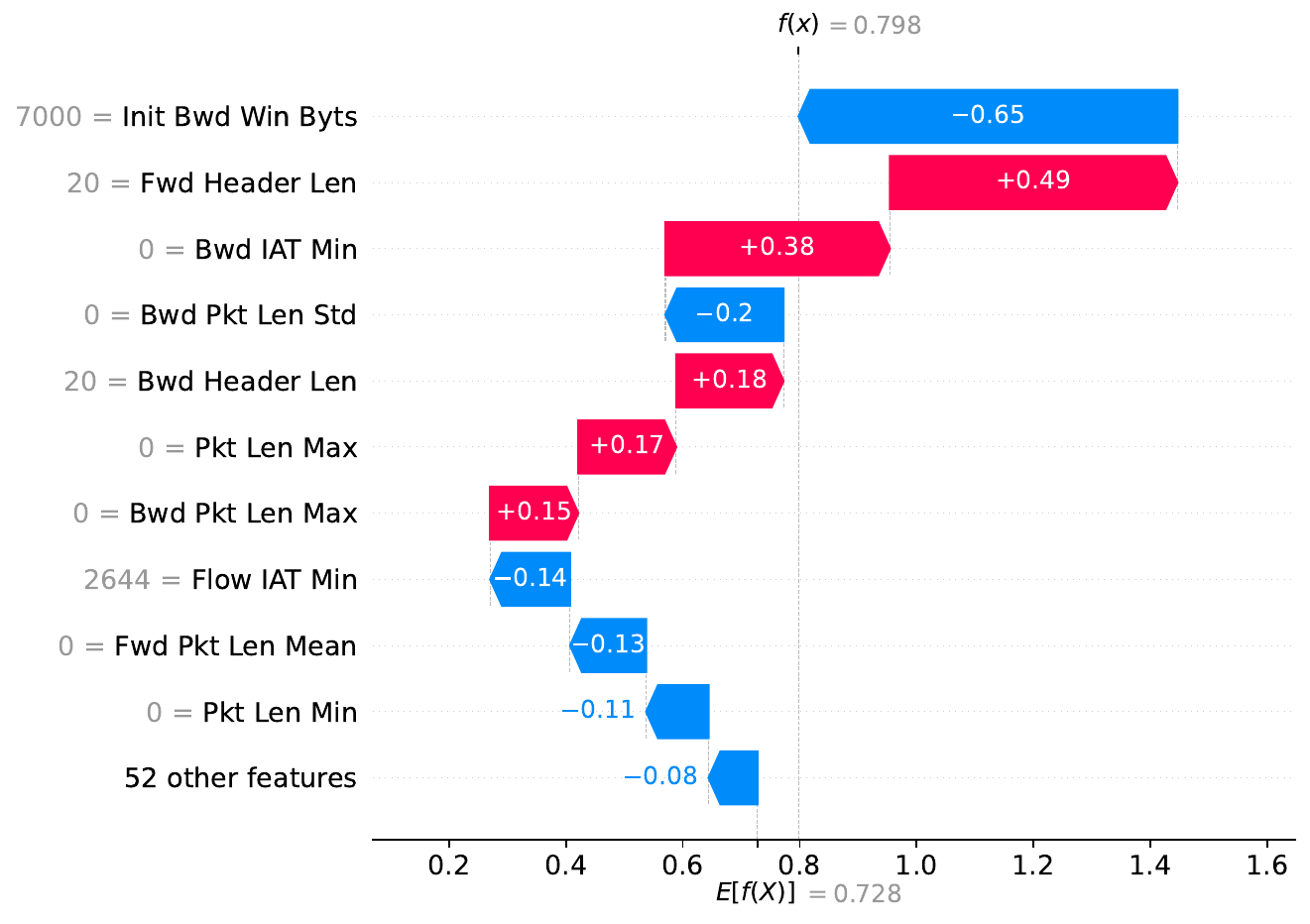} &
\includegraphics[width=0.5\linewidth]{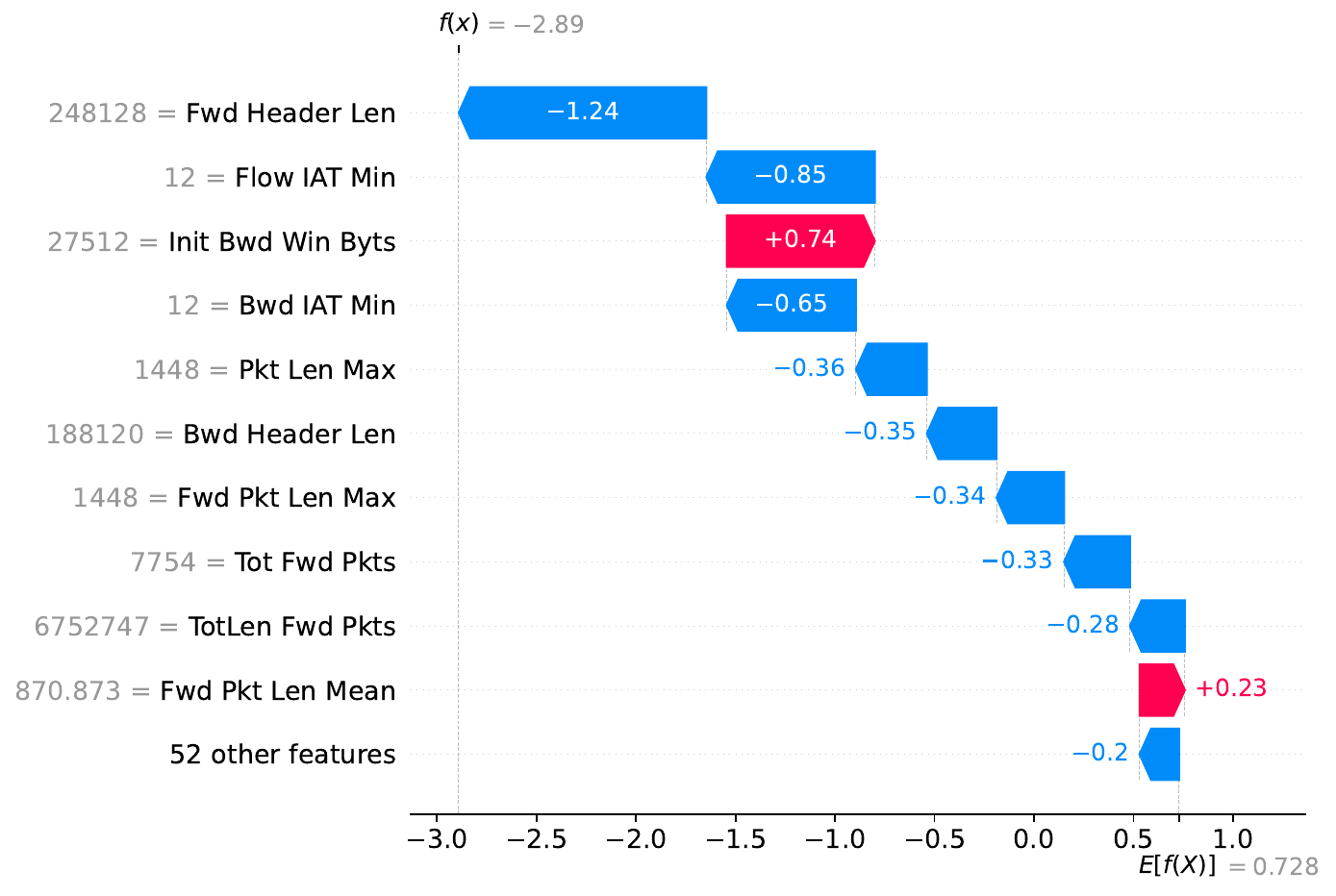} \\ 
(a) Others in Training.   & (b) Others in Testing. \\

\includegraphics[width=0.5\linewidth]{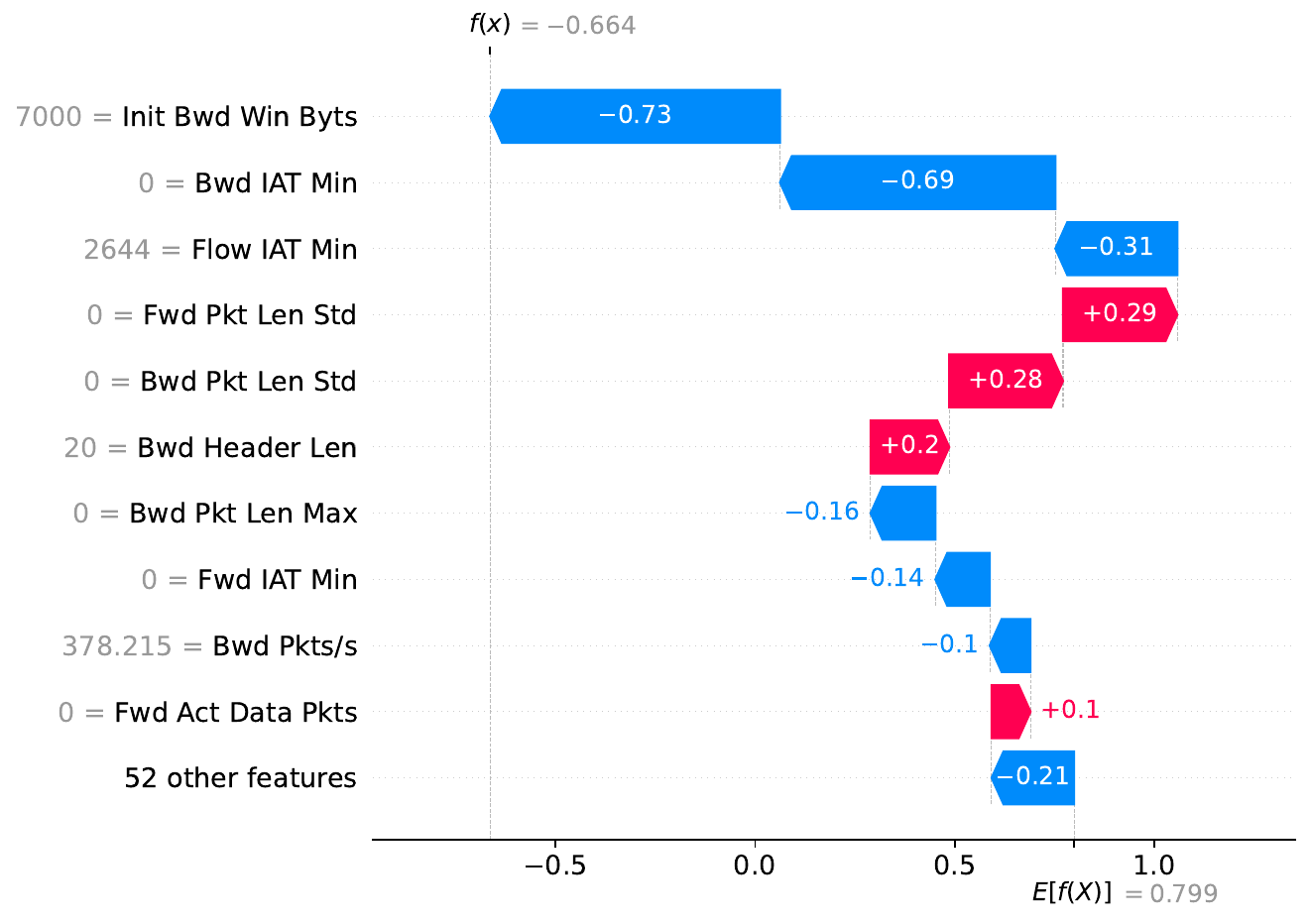} &
\includegraphics[width=0.5\linewidth]{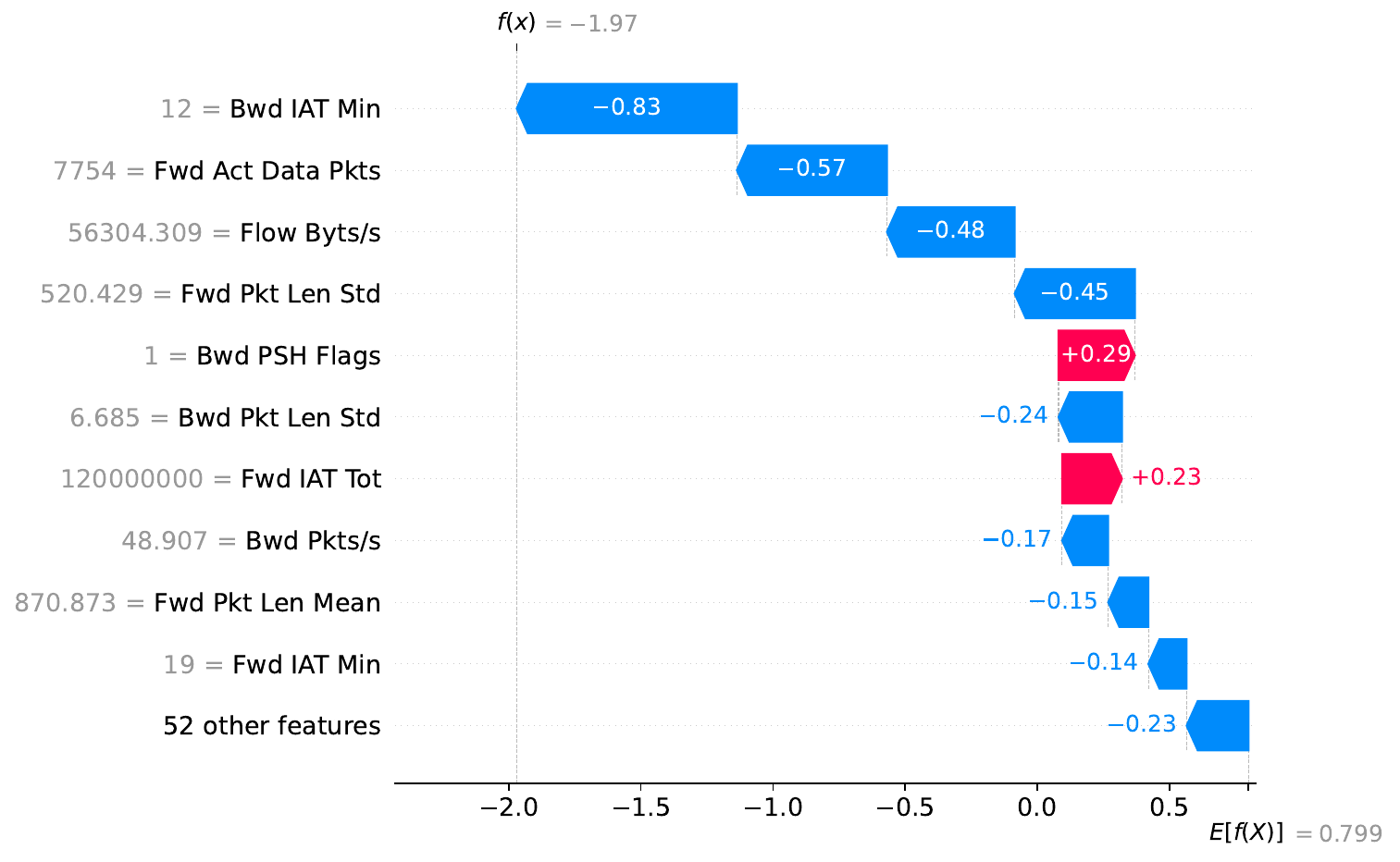} \\ 
(c) Conf in Training.   & (d) Conf in Testing. \\

\includegraphics[width=0.5\linewidth]{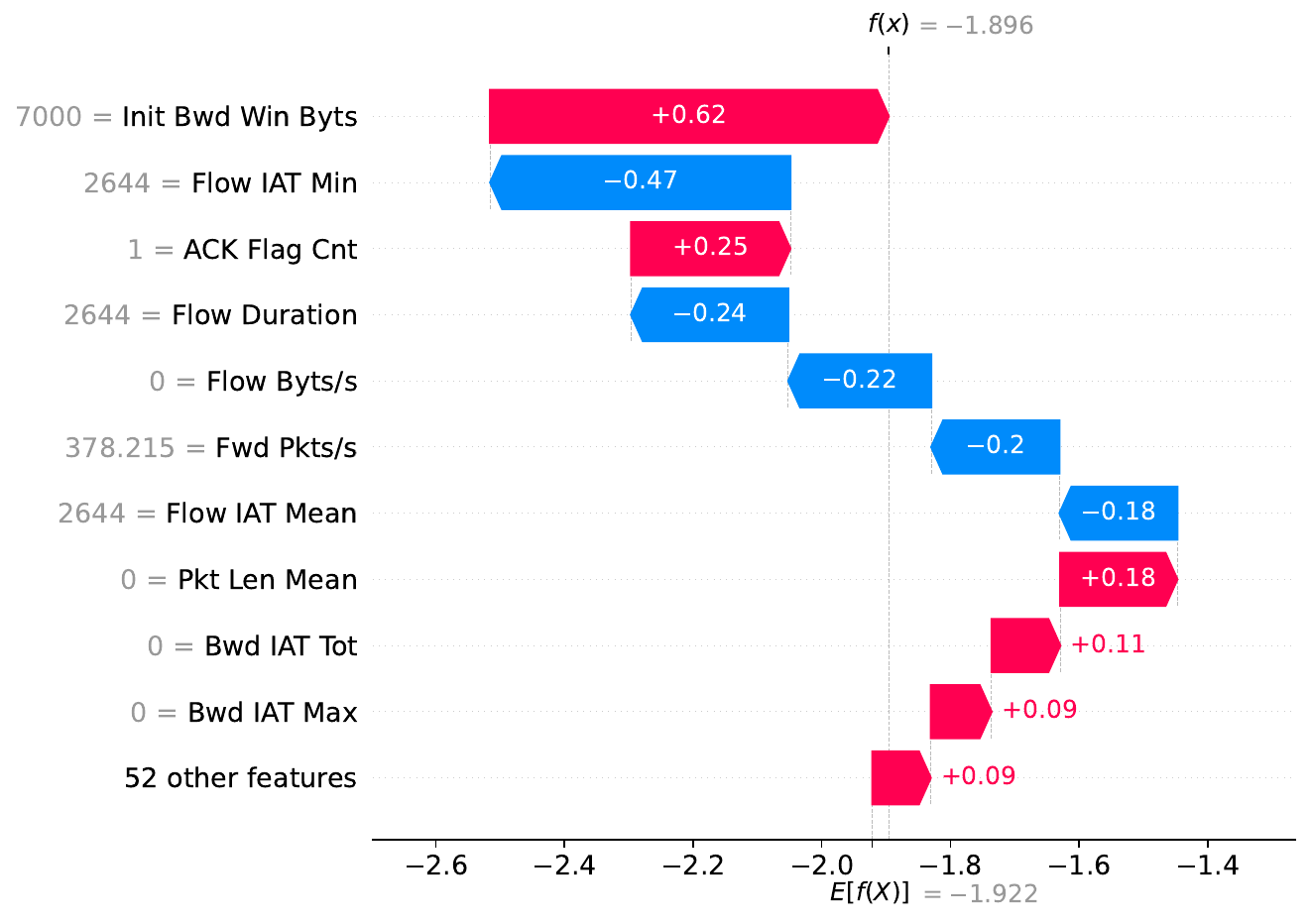} &
\includegraphics[width=0.5\linewidth]{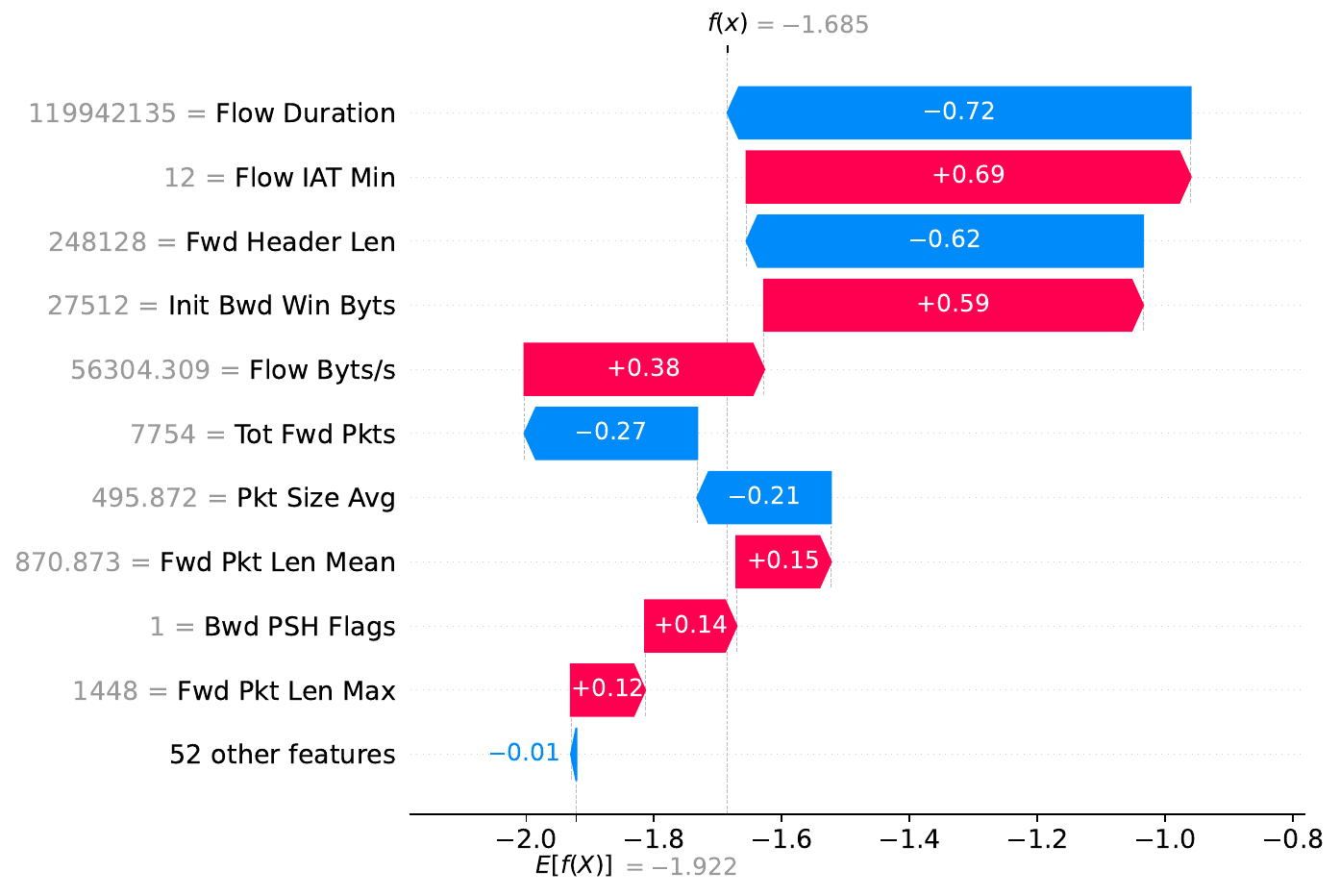} \\ 
(e) Share in Training.   & (f) Share in Testing. \\

\end{tabular}
\caption{SHAP values using XGB model while training and testing for IoTCam sample while flow classifier.}
\label{fig:SHAP_traintestplots_4Class_appendix}
\end{figure*}

\begin{figure*}[h]
\centering
\begin{tabular}{cc}
\includegraphics[width=0.45\linewidth]{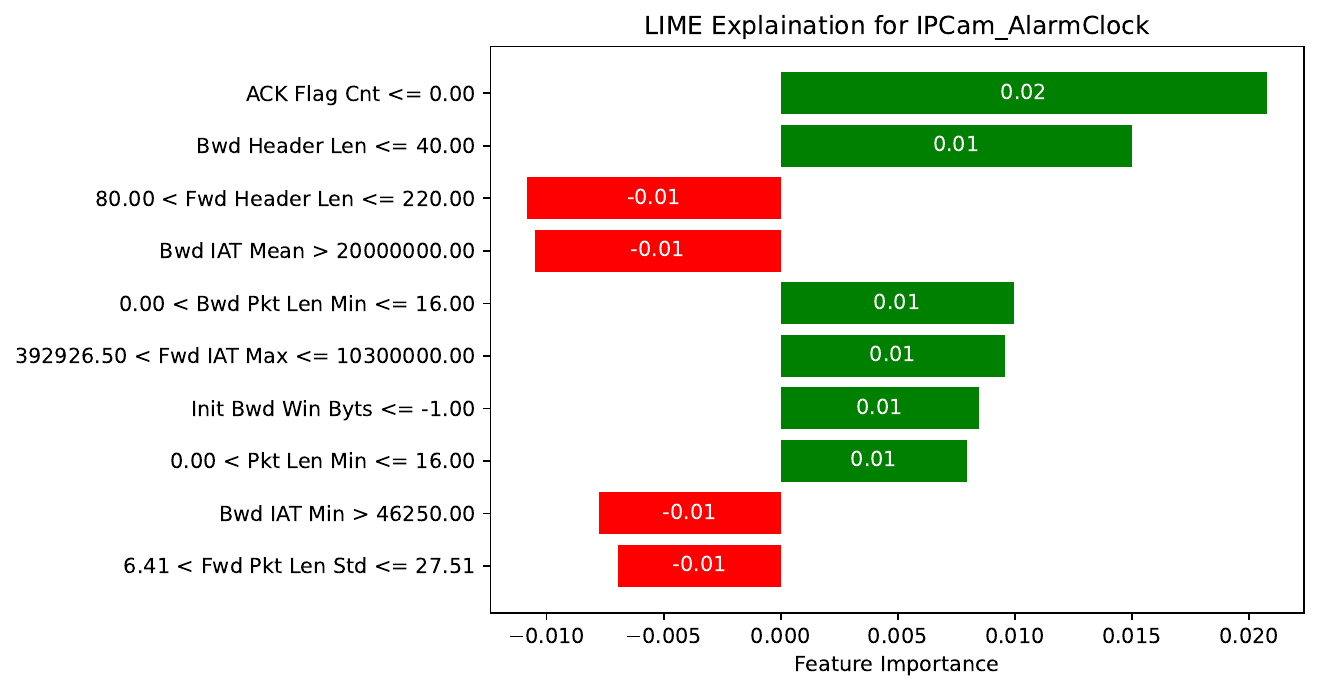} &
\includegraphics[width=0.45\linewidth]{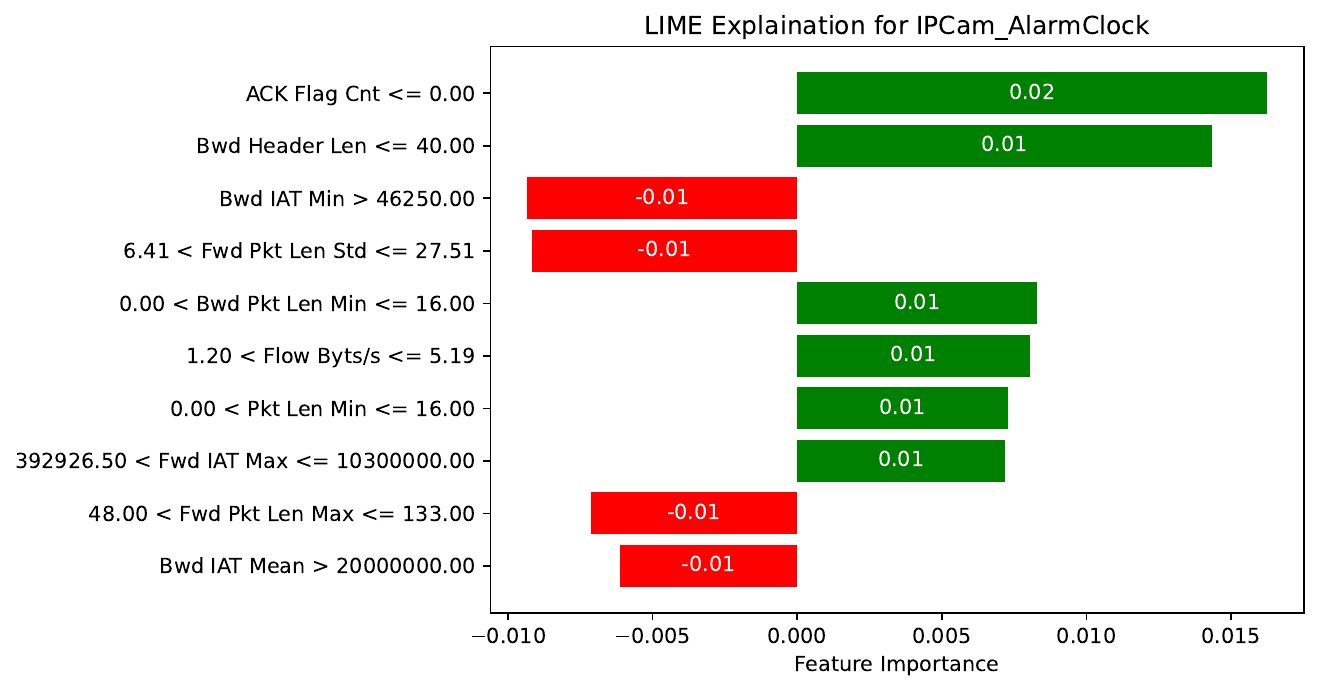} \\ 
(a) AlarmClock in Training & (b) AlarmClock in Testing\\

\includegraphics[width=0.45\linewidth]{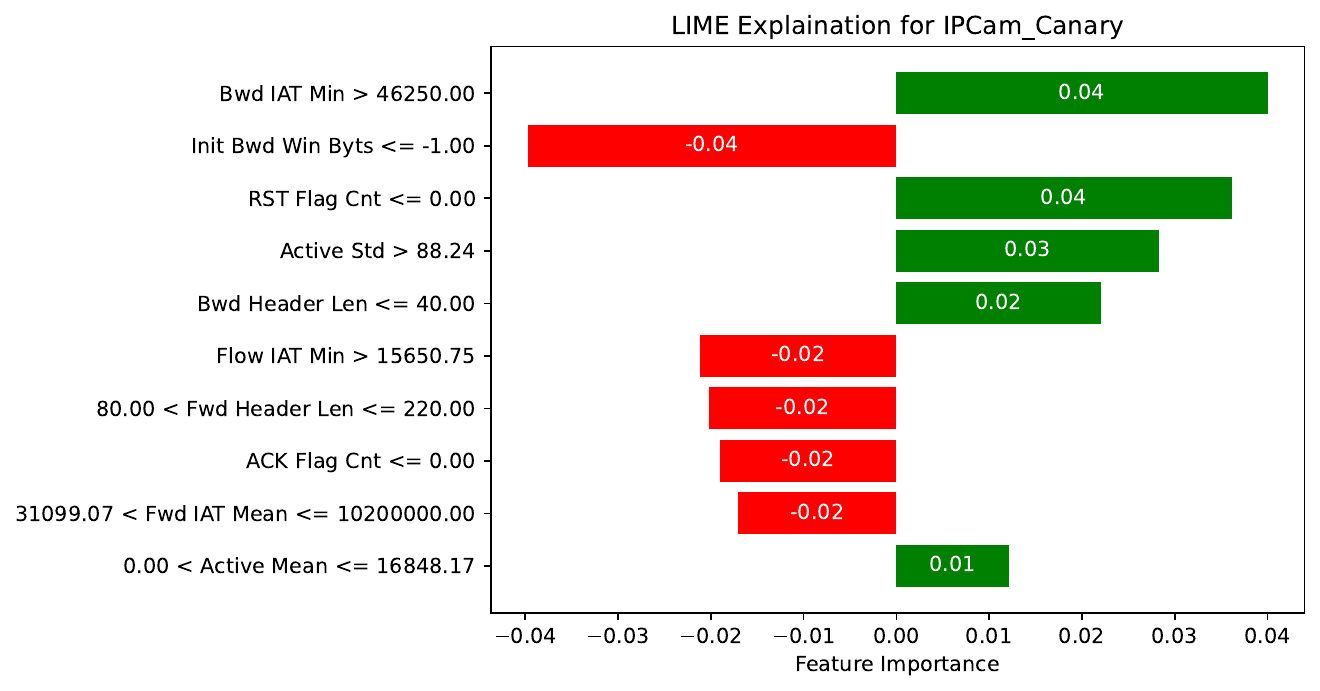} & 
 \includegraphics[width=0.45\linewidth]{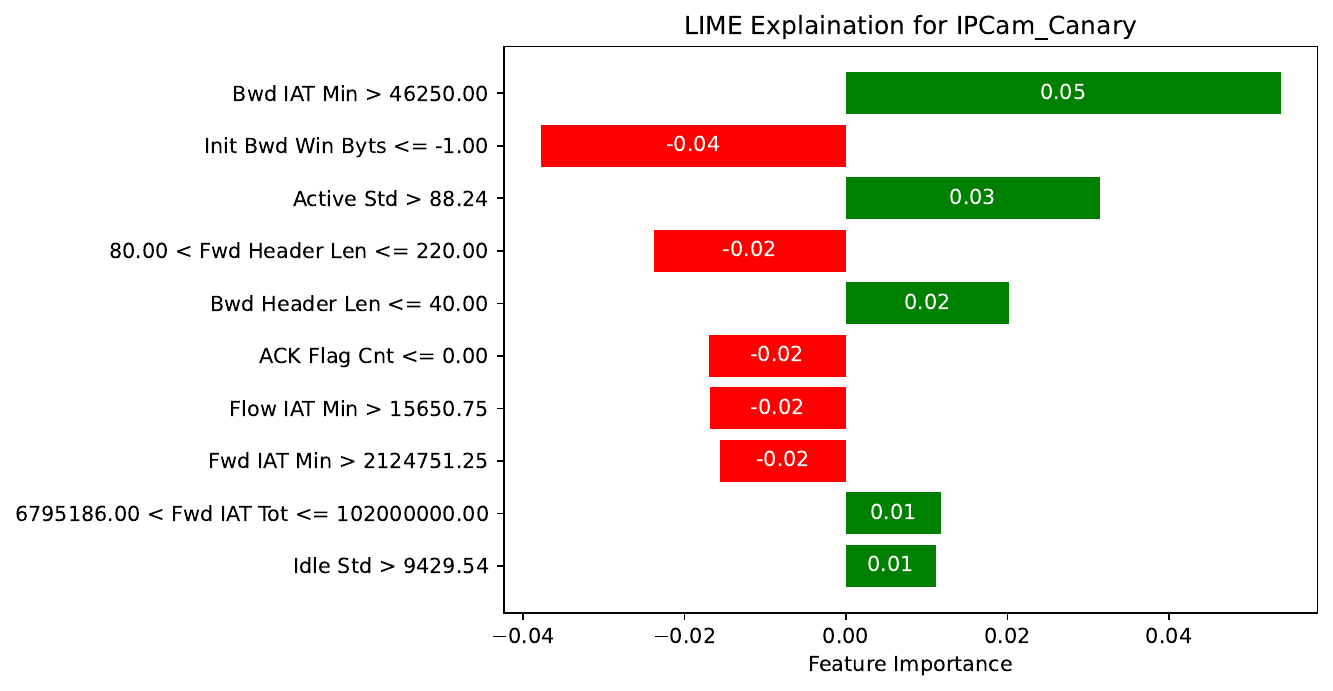} \\
 (c) Canary in Training & (d) Canary in Testing \\
 
 \includegraphics[width=0.45\linewidth]{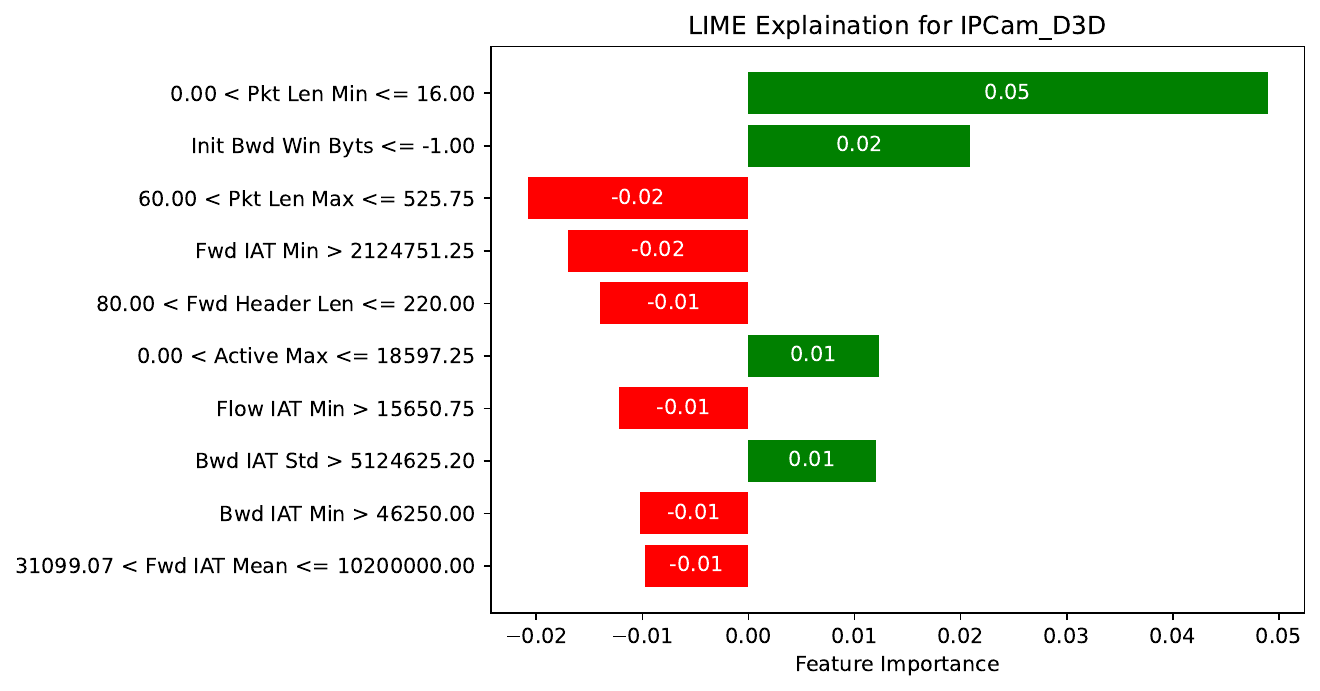} &
 \includegraphics[width=0.45\linewidth]{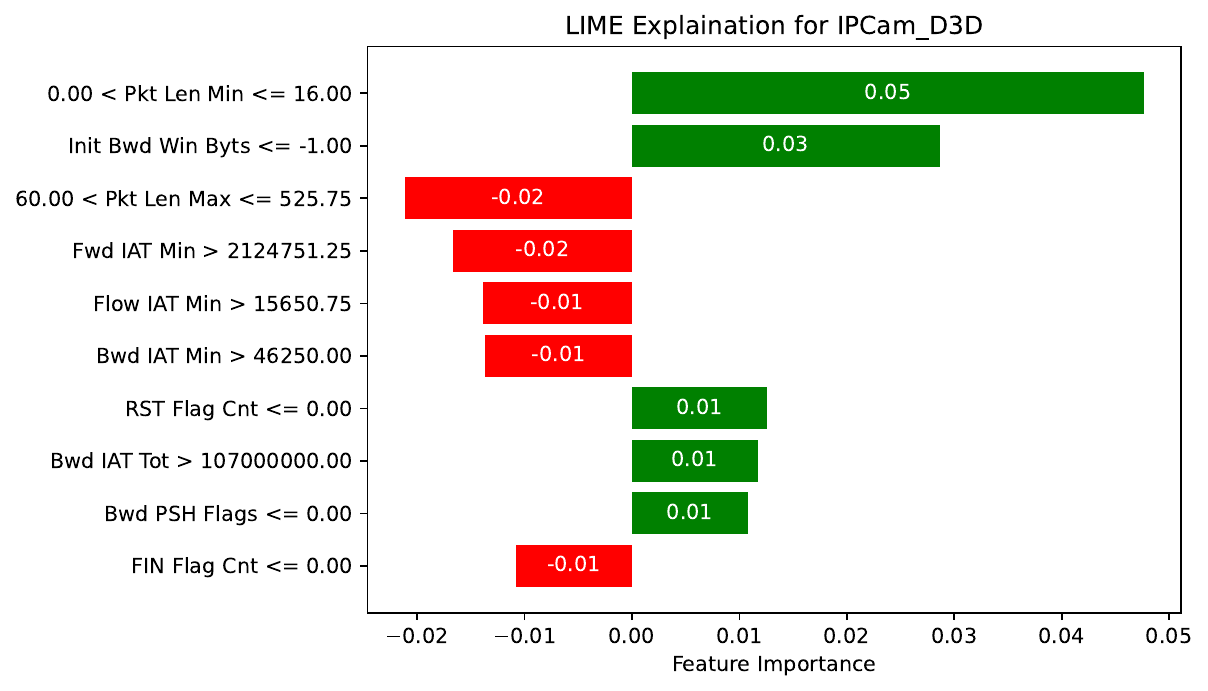} \\
 (e) D3D in Training & (f) D3D in Testing \\

 \includegraphics[width=0.45\linewidth]{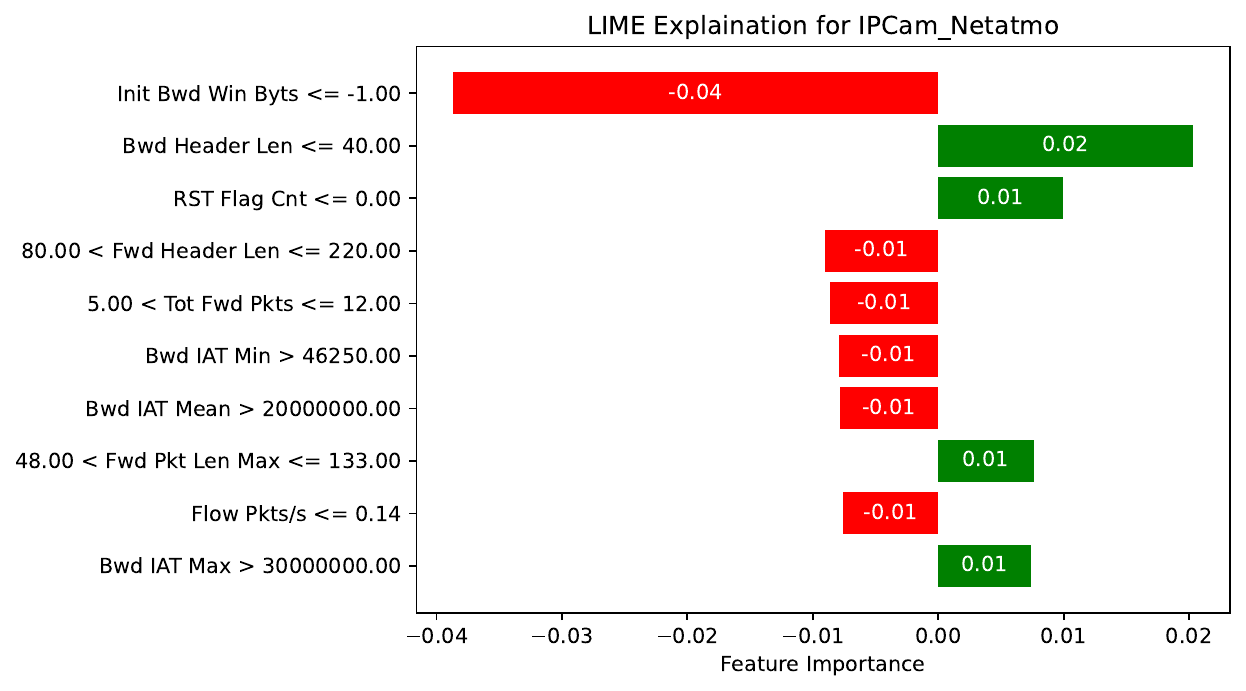} &
 \includegraphics[width=0.45\linewidth]{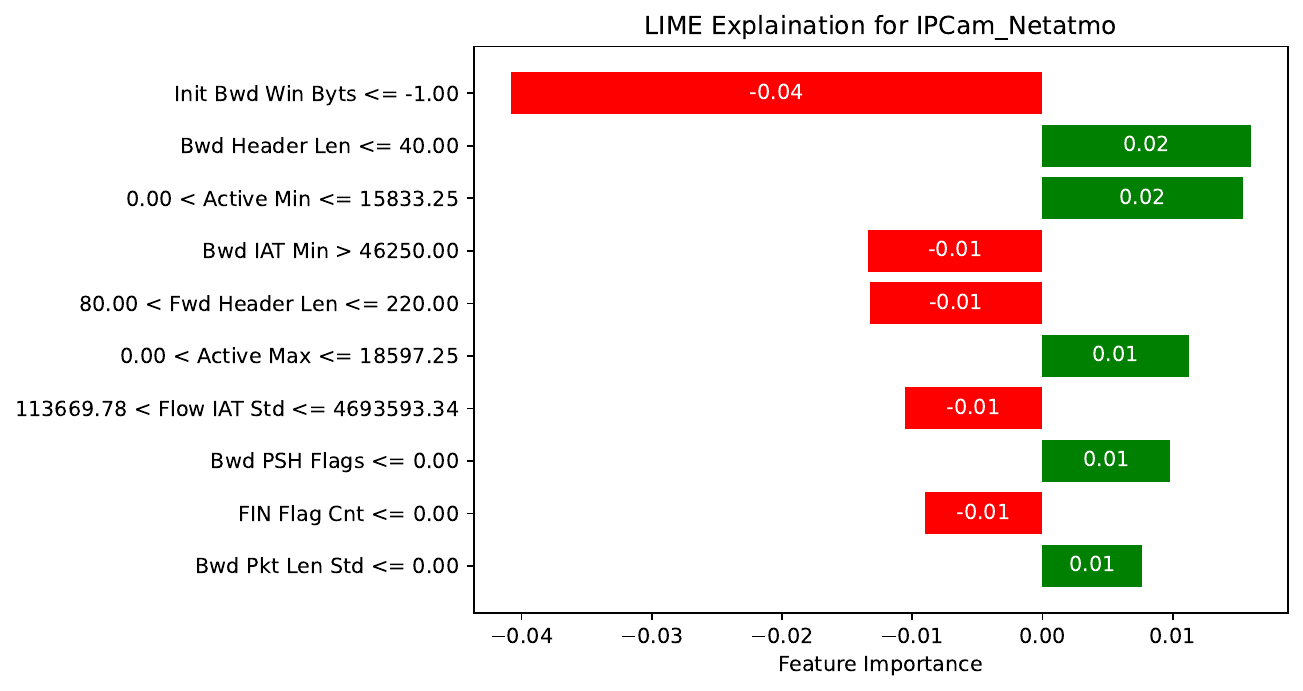} \\
 (g) Netatmo in Training & (h) Netatmo in Testing \\

\includegraphics[width=0.45\linewidth]{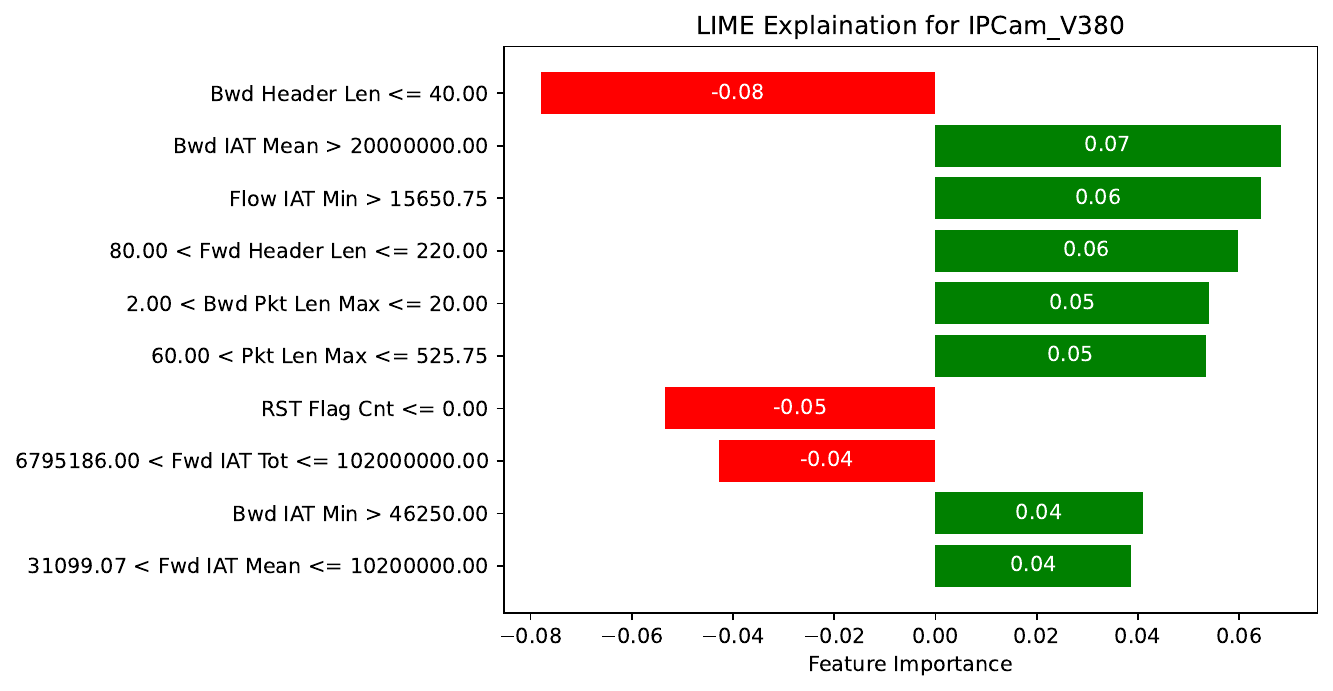} &
 \includegraphics[width=0.45\linewidth]{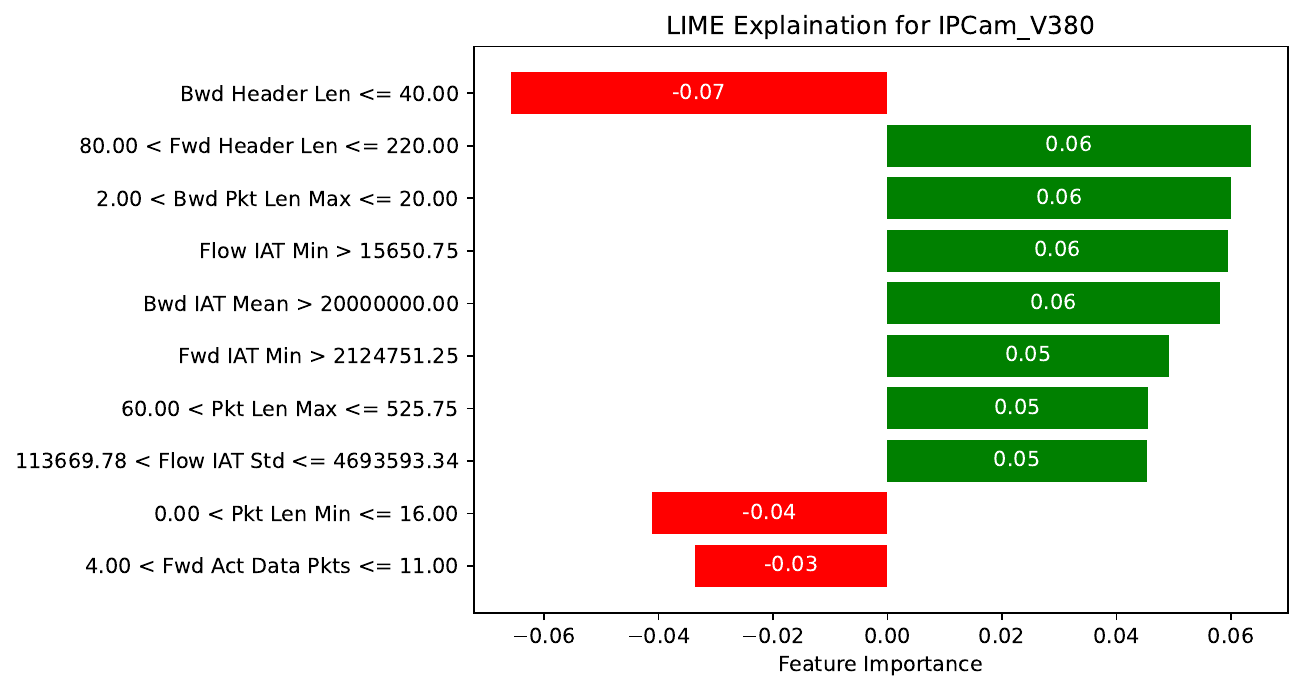} \\
 (i) V380 in Training & (j) V380 in Testing \\

\end{tabular}
\caption{LIME explaination using XGB model while training and testing of Ezviz sample for SmartCam Classifier}
   \label{fig:LIME_traintestplots_6Class_appendix}
\end{figure*}

\end{appendices}

\end{document}